\documentclass[letterpaper]{emulateapj}

\usepackage{psfig}
\usepackage{natbib}
\usepackage{txfonts}
\usepackage{lscape}

\begin{document}
\def\lsun{L_{\sun}}
\def\msun{M_{\sun}}
\def\simle{{\mathop{\stackrel{\sim}{\scriptstyle <}}\nolimits}}
\def\simgr{{\mathop{\stackrel{\sim}{\scriptstyle >}}\nolimits}}
\def\lesim{{\mathop{\stackrel{\scriptstyle <}{\sim}}\nolimits}}
\shorttitle{Spitzer IRDCs}
\shortauthors{Ragan et al.}

\title{Detection of Structure in Infrared Dark Clouds with Spitzer:  Characterizing Star Formation in the Molecular Ring}

\author{Sarah~E.~Ragan\altaffilmark{1}, Edwin~A.~Bergin\altaffilmark{1}, 
	Robert~A.~Gutermuth\altaffilmark{2}}

\altaffiltext{1}{Department of Astronomy, University of Michigan, 830 Dennison Building, 500 Church Street, Ann Arbor, MI, 48109 USA}
\altaffiltext{2}{Department of Astronomy, Smith College, Northampton, MA, 01063 USA}

\begin{abstract}
We have conducted a survey of a sample of infrared-dark clouds (IRDCs) with the {\em Spitzer Space Telescope} in order to explore their mass distribution.  We present a method for tracing mass using dust absorption against the bright Galactic background at 8~$\mu$m.  The IRDCs in this sample are comprised of tens of clumps, ranging in sizes from 0.02 to 0.3~pc in diameter and masses from 0.5 to a few 10$^3\msun$, the broadest dynamic range in any clump mass spectrum study to date.  Structure with this range in scales confirms that IRDCs are the the precursors to stellar clusters in an early phase of fragmentation.  Young stars are distributed in the vicinity of the IRDCs, but the clumps are typically not associated with stars and appear pre-stellar in nature.  We find an IRDC clump mass spectrum with a slope of $\alpha$=1.76$\pm$0.05 for masses from 30$\msun$ to 3000$\msun$.  This slope is consistent with numerous studies, culled from a variety of observational techniques, of massive star formation regions and is close to the mass function of Galactic stellar clusters and star clusters in other galaxies.  We assert that the shape of the mass function is an intrinsic and universal feature of massive star formation regions, that are the birth sites of stellar clusters.  As these clouds evolve and their constituent clumps fragment, the mass spectrum will steepen and eventually assume the form of the core mass function that is observed locally.
\end{abstract}


\section{Introduction}
\label{intro}

Our understanding of star formation has grown primarily from the study of local regions forming low-mass stars in relative isolation.  Nearby regions such as Taurus \citep[e.g.][]{Goldsmith_Taurus}, $\rho$ Ophiuchus \citep[e.g.][]{Young_oph, Johnstone_Oph}, Perseus \citep[e.g.][]{Enoch2008, Jorgensen_perseus,Kirk_perseus}, Serpens \citep[e.g.][]{ts98, Harvey_serpensIRAC, Enoch2008}, the Pipe nebula \citep[e.g.][]{Lombardi_pipe, Muench_pipe}, and Orion \citep[e.g.][]{Li_orion, Johnstone_OrionB} have been studied to great lengths using a variety of techniques including dust emission, extinction mapping, and molecular line emission.  This past decade of research has shown that star formation regions are assembled hierarchically.  Within molecular clouds (tens of parsecs in size, containing 10$^4$-10$^5~\msun$), we adopt the nomenclature used by \citet{BerginTafalla_ARAA2007} distinguishing ``clouds'' (10$^3-10^4~\msun$, 10$^0-10^1$~pc), ``clumps'' (10-10$^3~\msun$, 10$^{-1}$-10$^0$~pc), and ``cores'' (10$^{-1}$-10$^1$~$\msun$, 10$^{-2}$-10$^{-1}$~pc).

In studies of nearby regions, it is possible to resolve pre-stellar cores with single dish observations \citep[e.g.][]{Johnstone_Oph}.  This permits the examination of the properties of the fragmentation of the natal molecular clouds into smaller components.  It is then straightforward to construct a mass function of cores and a mass function for individual cores can then be constructed.  The core mass distributions typically derived from dust emission studies are found to be strikingly similar to the mass spectra of stars, implying that the masses of stars are a direct result of the way in which the natal molecular cloud fragments.  In contrast, when CO line emission is used as a mass probe for cores \citep{Kramer_CMF}, a more top-heavy distribution results.  

While these studies have brought us a deep understanding of isolated low-mass star formation, this is not the complete picture for star formation in the Galaxy.  The necessary ingredient for star formation is dense molecular gas.  However, the H$_2$ distribution in the Milky Way is not uniform.  The primary reservoir is in the Molecular Ring, which resides at 4~kpc from the Galactic Center and contains $\sim$70\% of the molecular gas inside the solar circle \citep{Jackson_GRS}.  Thus, the Molecular Ring is the heart of Galactic star formation.  Indeed, as \citet{Robinson1984} show, the peak of Galactic far-infrared emission originates from this region.

In local clouds, most of the recent progress in our understanding of the early stages of star formation have come from studies of pre-stellar objects.  Populations of starless cores have been identified in numerous local regions, and they are universally cold and quiescent, exhibiting thermal line widths \citep{BerginTafalla_ARAA2007}.  Recently, a population of cold, dense molecular clouds within the Molecular Ring were detected against the bright Galactic mid-infrared background \citep[from 7 to 25~$\mu$m;][]{egan_msx, carey_msx}.  These clouds are opaque to mid-infrared radiation and show little or no typical signs of star formation, such as association with IRAS point sources.  Initial studies demonstrated that these objects, termed infrared-dark clouds (IRDCs) are dense (n(H$_2$) $> 10^5$ cm$^{-3}$), cold (T $<$ 20K) concentrations of 10$^3$ - 10$^5~\msun$ of molecular gas.  Based upon the available mass for star formation, infrared-dark clouds are likely the sites of massive star formation.

Since their discovery, further studies of infrared-dark clouds have established their place as the precursors to clusters.  A number of studies have detected the presence of deeply embedded massive protostars using sub-millimeter probes \citep{Beuther_protostars_IRDC, Rathborne_2007_protostars, Pillai_G11}, which confirms that IRDCs are the birth-sites of massive stars.  Detailed molecular surveys show that molecules such as NH$_3$ and N$_2$H$^+$ trace the dense gas extremely well \citep{ragan_msxsurv, Pillai_ammonia}, as seen in local dense prestellar cores \citep{Bergin2002}.  Furthermore, the molecular emission corresponding to the absorbing structure of infrared-dark clouds universally exhibit non-thermal linewidths on par with massive star formation regions.  Other studies have uncovered the presence of masers \citep{Beuther2002_masers, Wang_ammonia} and outflows \citep{Beuther_IRDCoutflow}, known indicators of ongoing embedded star formation.  Already, the evidence shows that these are the sites where massive stars and star clusters will form or are already forming.  In order to understand massive star formation, and thus Galactic star formation, it is crucial to understand the structure and evolution of IRDCs.

Studies of infrared-dark clouds to date have left the fundamental properties of cloud fragmentation go relatively unexplored.  \citet{rathborne2006} showed that IRDCs exhibit structure with median size of $\sim$0.5~pc, but observations of IRDCs with the Spitzer Space Telescope, which we describe in $\S$\ref{obs}, reveal that there exists structure well below this level.  We characterize the environment in $\S$\ref{env} and highly structured nature of infrared-dark clouds in $\S$\ref{clumps} by utilizing the high-resolution imaging capabilities of the Spitzer.  In $\S$\ref{mf}, we analyze the IRDC absorbing structure, derive the clump mass function, and put the results in the context of previous studies.  We find the mass function to be shallower than Salpeter initial mass function (IMF) \citep{Salpeter_imf} and more closely aligned with that observed using CO in massive star forming regions.  Given the strong evidence for fragmentation and star formation characteristics of these objects, we suggest they are in the initial stages of fragmentation.  The conclusions as well as the broad impact of these results are discussed in $\S$\ref{conclusion}. The results of this study provide an important foundation for further studies of IRDCs with the instruments of the future, allowing us to probe the dominant mode of star formation in the Galaxy, which may be fundamentally different from the processes that govern local star formation.

\section{Observations \& Data Reduction}
\label{obs}

\subsection{Targets}

Searching in the vicinity of ultra-compact HII (UCHII) regions \citep{wc89} for infrared-dark cloud candidates, \citet{ragan_msxsurv} performed a survey of 114 candidates in N$_2$H$^+$(1-0), CS(2-1), and C$^{18}$O(1-0) with the FCRAO.  In order to study substructure with Spitzer, we have selected a sample of targets from the \citet{ragan_msxsurv} sample  which are compact, typically $2\arcmin \times 2\arcmin$ (or $2 \times 2$~pc at 4~kpc), and opaque, providing the starkest contrast at 8~$\mu$m (MSX Band A) with which to examine the absorbing structure.  The selected objects also exhibit significant emission in transitions of CS and N$_2$H$^+$ that are known to trace high-density gas, based on their high critical densities.  By selecting objects with strong emission in these lines, we ensure that their densities are  $>$10$^4$ cm$^{-3}$ and their temperatures are less than 20~K.  Under these conditions in local clouds, N$_2$H$^+$ is strongest when CO is depleted in the pre-stellar phase \citep{bl97}, hence a high N$_2$H$^+$/CO ratio guided our attempt to select the truly ``starless'' dark clouds in the IRDC sample.  Our selection criteria are aimed to isolate earliest stages of star formation in local clouds and give us the best hope of detecting massive starless objects. The eleven IRDCs observed are listed in Table~1 with the distances derived in \citet{ragan_msxsurv} using a Milky Way rotation curve model \citep{Fich:1989} assuming the ``near'' kinematic distance.  The listed uncertainties in Table~1 arise from the $\pm$14\% maximal deviation inherent in the rotation curve model. 

\vspace{0.5in}

\subsection{Spitzer Observations \& Data Processing}

Observations of this sample of objects were made on 2005 May 7 -- 9 and September 15 -- 18 with IRAC centered on the coordinates listed in Table~1.  Each region was observed 10 times with slightly offset single points in the 12s high dynamic range mode.  All four IRAC bands were observed over 7$' \times$ 7$'$ common field-of-view.  MIPS observations were obtained on 2005 April 7 -- 10 of the objects in this sample.  Using the ``large" field size, each region was observed in 3 cycles for 3s at 24~$\mu$m.  MIPS observations cover smaller 5.5$' \times$ 5.5$'$ fields-of-view but big enough to contain the entire IRDC.  Figures~\ref{fig:g0585}$-$\ref{fig:g3744} show each IRDC field in all observed wavebands.  The absorbing structures of the IRDCs are most prominent at 8~$\mu$m and 24~$\mu$m.

We used IRAC images processed by the Spitzer Science Center (SSC) using pipeline version S14.0.0 to create basic calibrated data (BCD) images.  These calibrated data were corrected for bright source artifacts (``banding'', ``pulldown'', and ``muxbleed''), cleaned of cosmic ray hits, and made into mosaics using Gutermuth's WCS-based IRAC post-processing and mosaicking package \citep[see][for further details]{Gutermuth_ngc1333}.  

Source finding and aperture photometry were performed using Gutermuth's PhotVis version 1.10 \citep{Gutermuth_ngc1333}.  We used a 2.4$\arcsec$ aperture radius and a sky annulus from 2.4$\arcsec$ to 6$\arcsec$ for the IRAC photometry.  The photometric zero points for the [3.6], [4.5], [5.8], and [8.0] bands were 22.750, 21.995, 19.793, and 20.187 magnitudes, respectively.  For the MIPS 24~$\mu$m photometry, we use a 7.6$\arcsec$ aperture with 7.6$\arcsec$ to 17.8$\arcsec$ sky annuli radii and a photometric zero point of 15.646 magnitude.  All photometric zero points are calibrated for image units of DN and are corrected for the adopted apertures.

To supplement the Spitzer photometry, we incorporate the source photometry from the Two-Micron All Sky Survey (2MASS) Point Source Catalog (PSC).  Source lists are matched for a final catalog by first matching the four IRAC band catalogs using Gutermuth's WCSphotmatch utility, enforcing a 1$\arcsec$ maximal tolerance for positive matches.  Then, the 2MASS sources are matched with tolerance 1$\arcsec$ to the mean positions from the first catalog using the same WCS-based utility.  Finally, the MIPS 24~$\mu$m catalog is integrated with matching tolerance 1.5$\arcsec$.

\section{Stellar Content \& IRDC Environment}
\label{env}

The tremendous sensitivity of Spitzer has given us the first abilty to characterize young stellar populations in detail.  Before the Spitzer era, IRAS led the effort in identifying the brightest infrared point sources in the Galaxy.  Only one object in this sample, G034.74$-$0.12 (Figure~\ref{fig:g3474}) has an IRAS point source (18526+0130) in the vicinity.  Here, with Spitzer, we have identified tens of young stellar objects (YSOs) in the field of each IRDC.

\vspace{-0.05in}

\subsection{Young Stellar Object Identification \& Classification}

With this broad spectral coverage from 2MASS to IRAC to MIPS, we apply the robust critieria described in \citet{Gutermuth_ngc1333} to identify young stellar objects (YSOs) and classify them.  Table~2 lists the J, H, K$_s$, 3.6, 4.5, 5.8, 8.0 and 24~$\mu$m photometry for all stars that met the YSO criteria, and we note the classification as Class I (CI), Class II (CII), embedded protostars (EP), or transition disk objects (TD).  A color-color diagram displaying these various classes of YSOs 


\begin{figure}
\begin{center}
\includegraphics[scale=0.9]{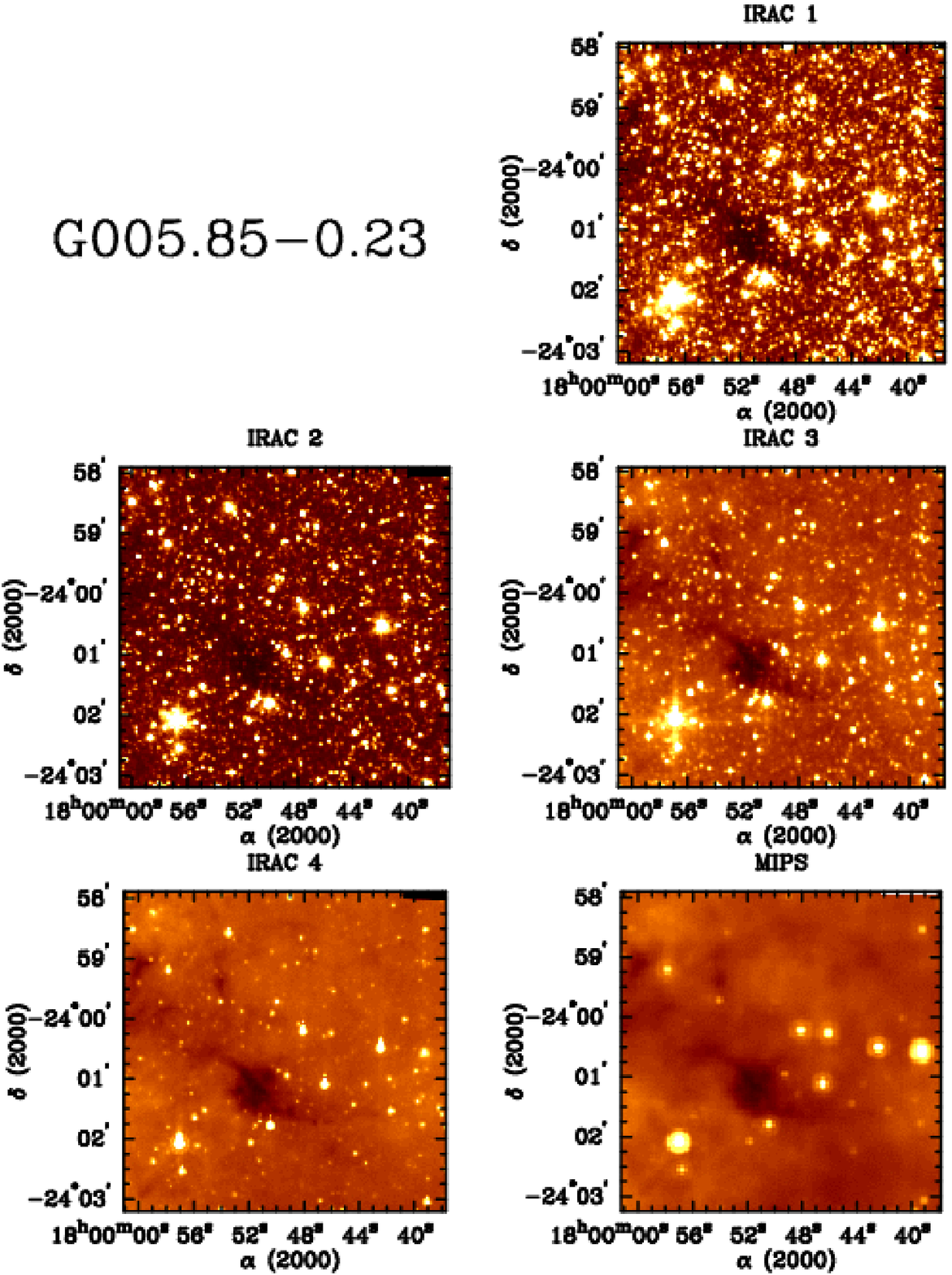}
\end{center}
\caption{G005.85$-$0.23: {\it Top Row Right: 3.6$\micron$.  Middle Row Left: 4.5$\micron$.  Middle Row Right: 5.8$\micron$.  Bottom Row Left: 8$\micron$.  Bottom Row Right: 24$\micron$.}}
\label{fig:g0585}
\end{figure}

\clearpage

\begin{figure}
\begin{center}
\includegraphics[scale=0.9]{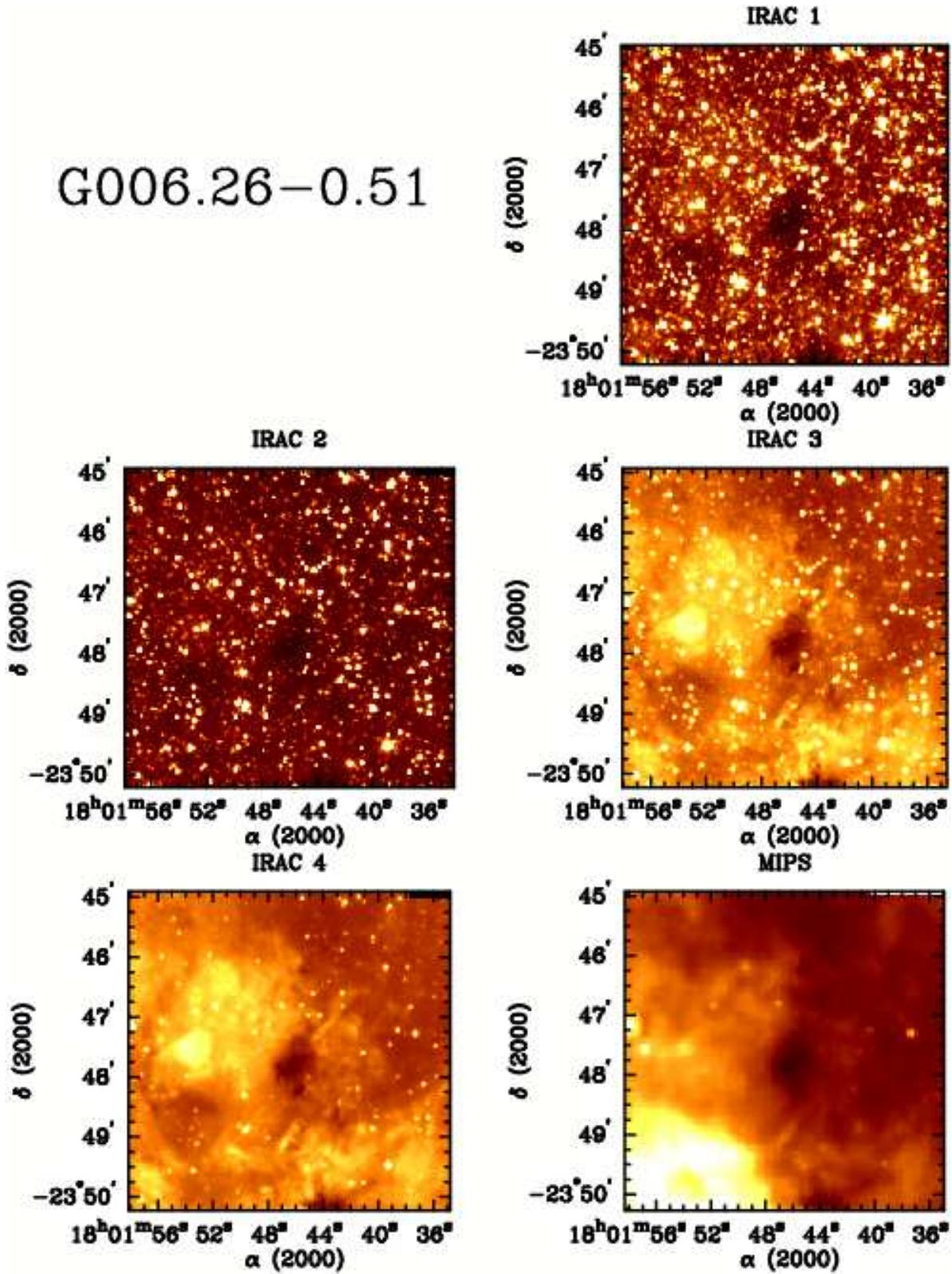}
\end{center}
\caption{G006.26$-$0.51: Wavelengths as noted in Figure~1.}
\label{fig:g0626}
\end{figure}

\clearpage

\begin{figure}
\begin{center}
\includegraphics[scale=0.9]{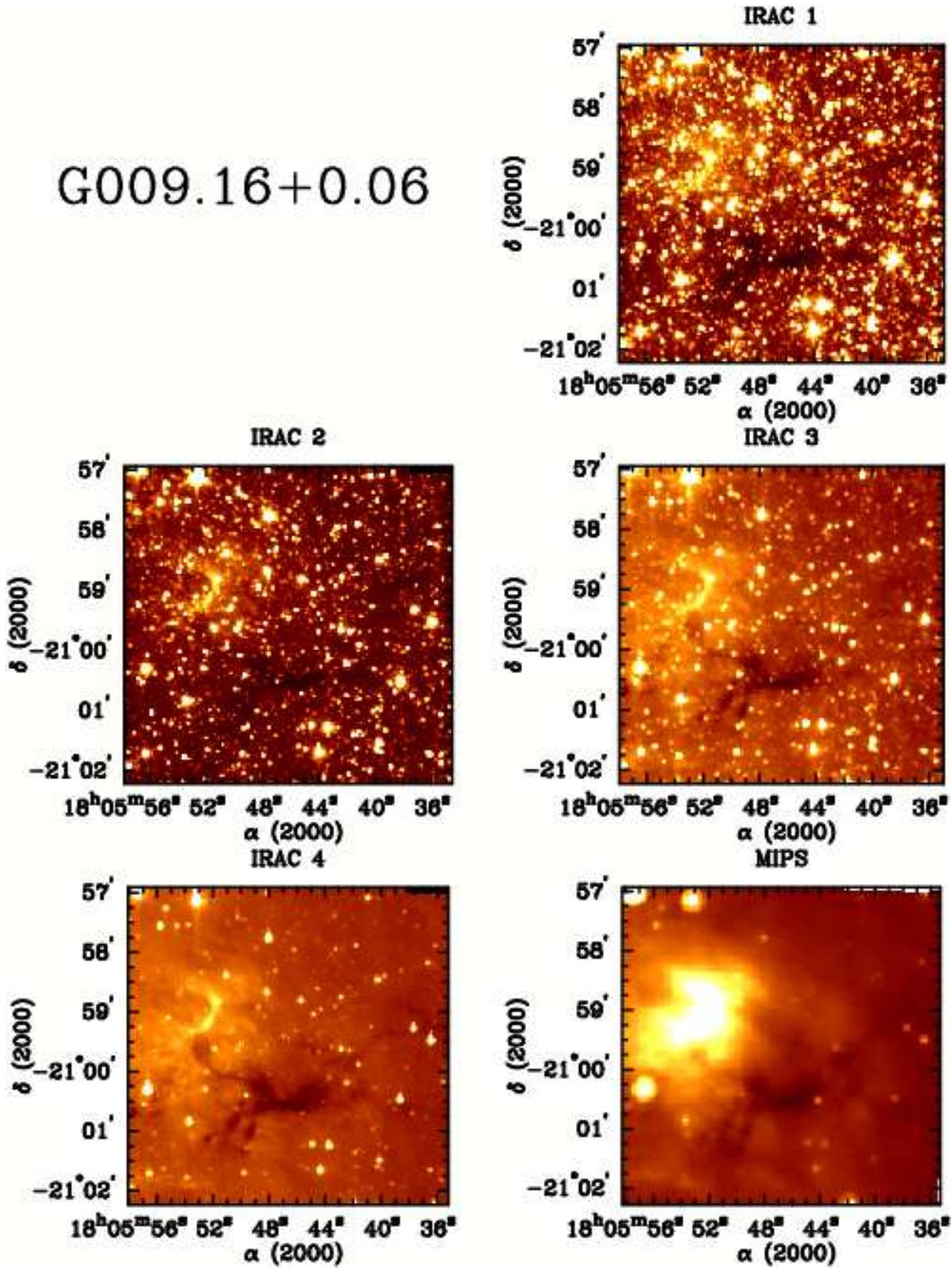}
\end{center}
\caption{G009.16$+$0.06: Wavelengths as noted in Figure~1.}
\label{fig:g0916}
\end{figure}

\clearpage

\begin{figure}
\begin{center}
\includegraphics[scale=0.9]{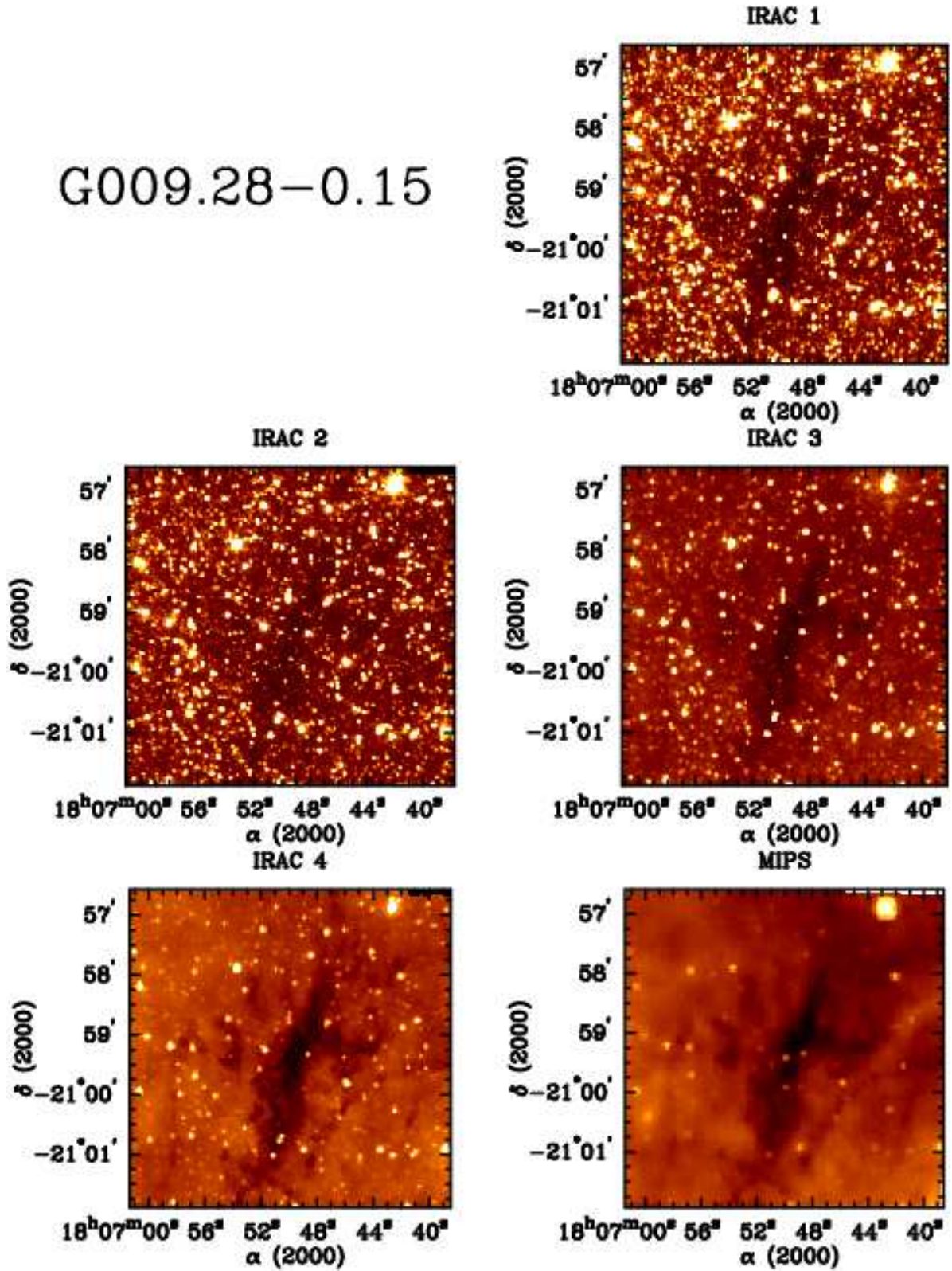}
\end{center}
\caption{G009.28$-$0.15: Wavelengths as noted in Figure~1.}
\label{fig:g0928}
\end{figure}

\clearpage

\begin{figure}
\begin{center}
\includegraphics[scale=0.9]{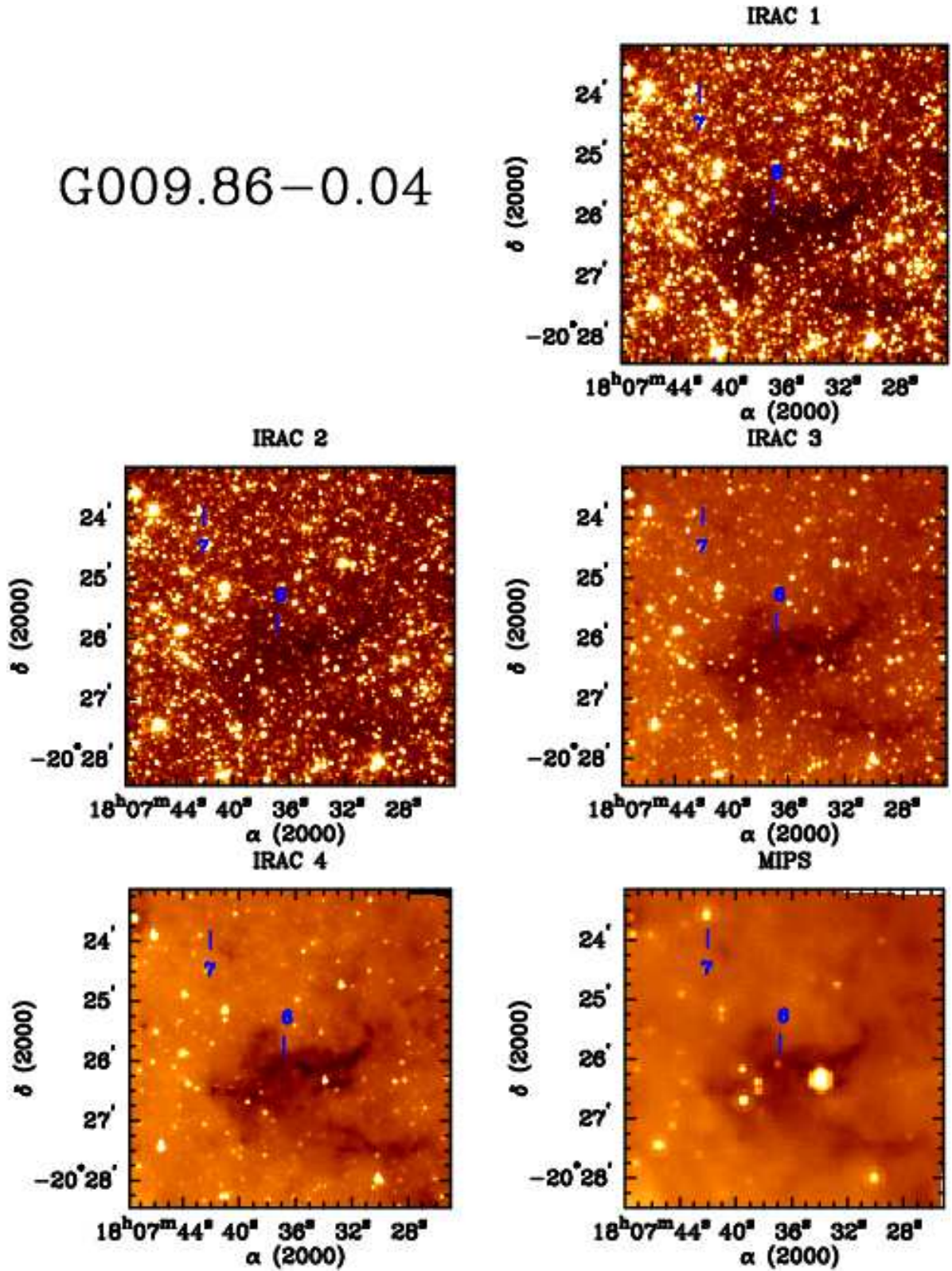}
\end{center}
\caption{G009.86$-$0.04: Wavelengths as noted in Figure~1. - Embedded Objects (indices 6 and 7 in Table~2 under source G009.86$-$0.04) are labeled.  Source G009.86$-$0.04 index 6 is only detectable at 24~$\mu$m and lies right at the heart of the dust absorption.}
\label{fig:g0986}
\end{figure}

\clearpage

\begin{figure}
\begin{center}
\includegraphics[scale=0.9]{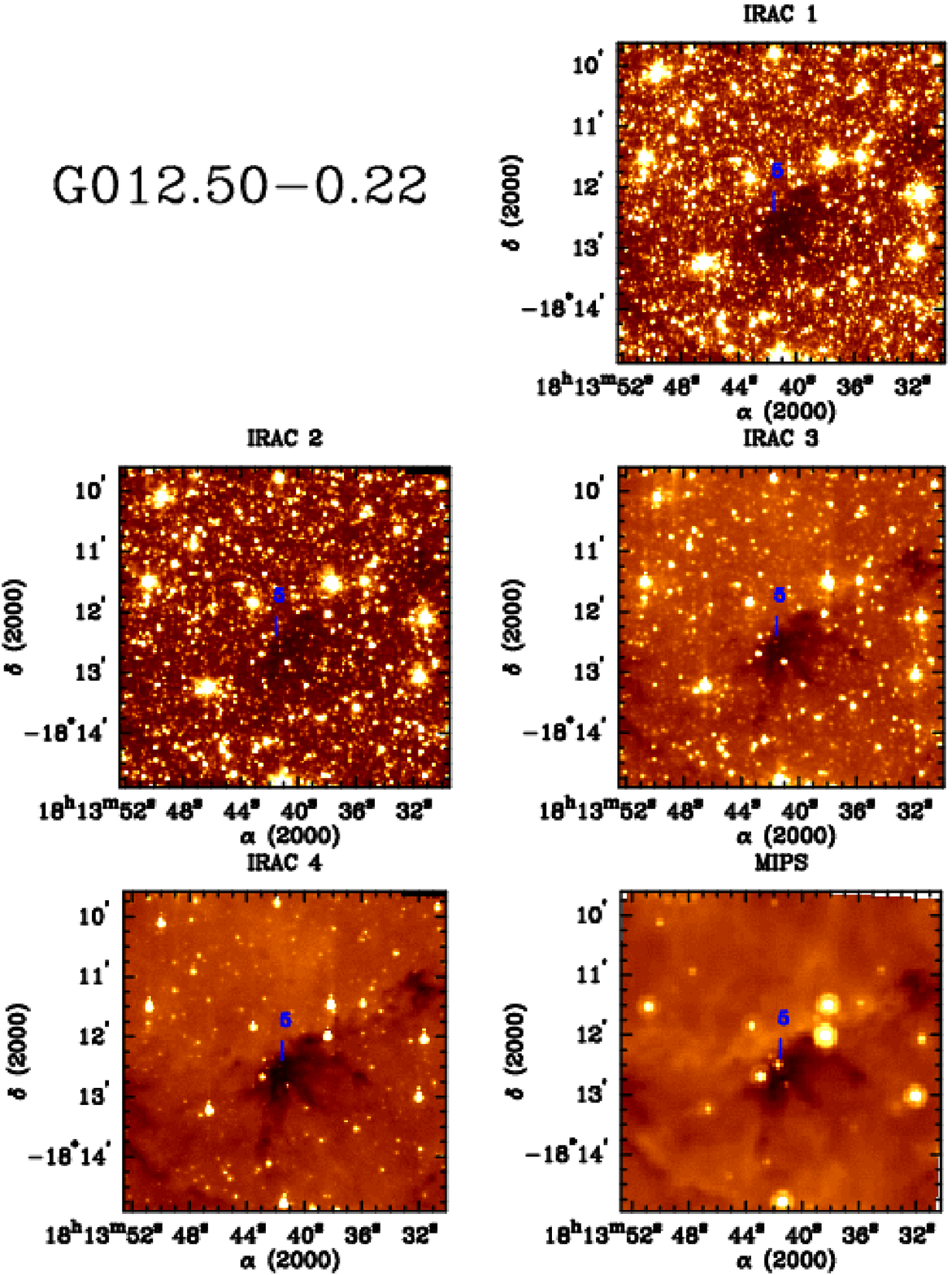}
\end{center}
\caption{G012.50$-$0.22: Wavelengths as noted in Figure~1. - Embedded Object (index 5 in Table~2 under source G012.50-0.22) is labeled.}
\label{fig:g1250}
\end{figure}

\clearpage

\begin{figure}
\begin{center}
\includegraphics[scale=0.9]{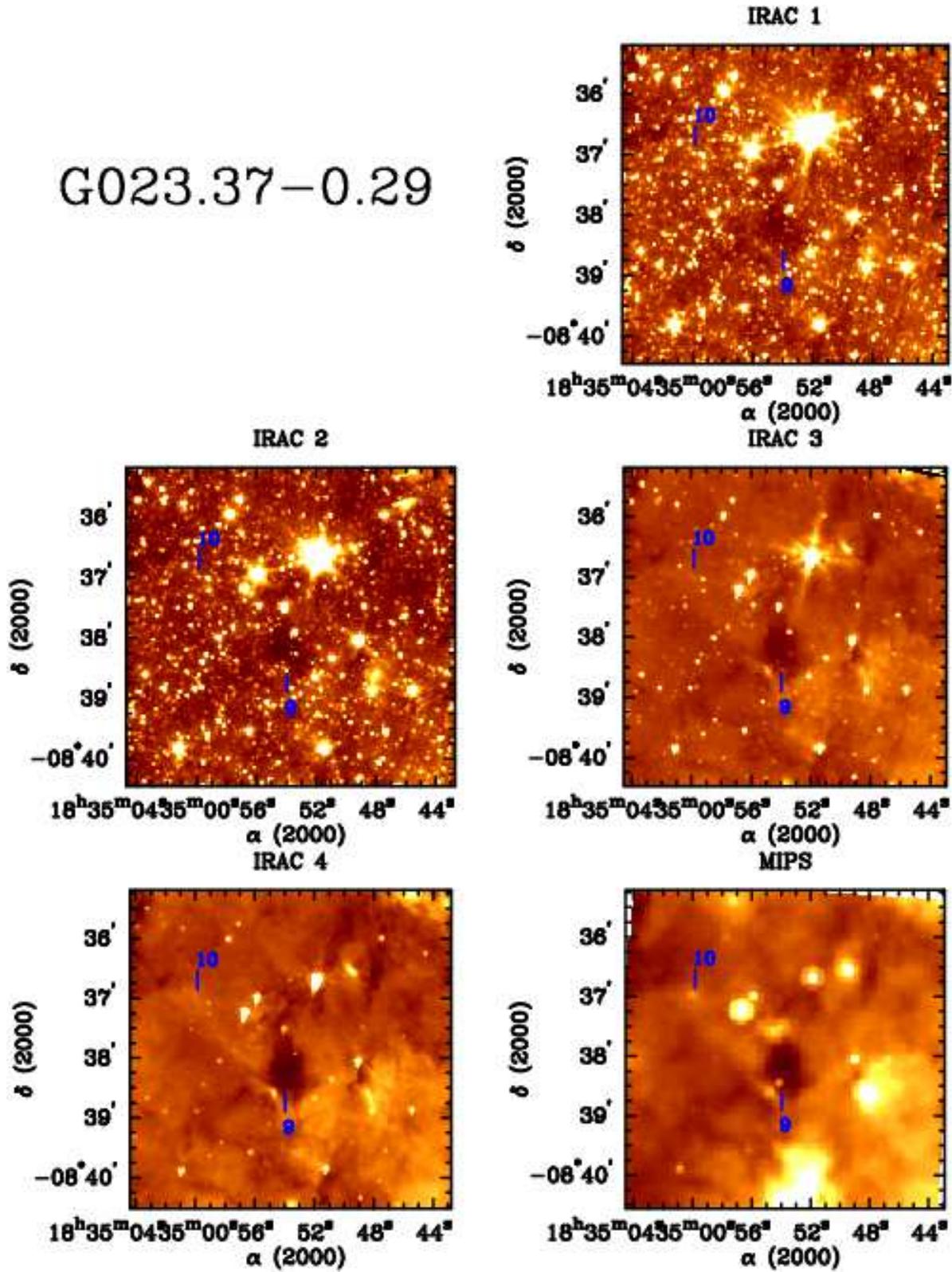}
\end{center}
\caption{G023.37$-$0.29: Wavelengths as noted in Figure~1. - Embedded Objects (indices 9 and 10 in Table~2 under source G023.37$-$0.29) are labeled.}
\label{fig:g2337}
\end{figure}

\clearpage

\begin{figure}
\begin{center}
\includegraphics[scale=0.9]{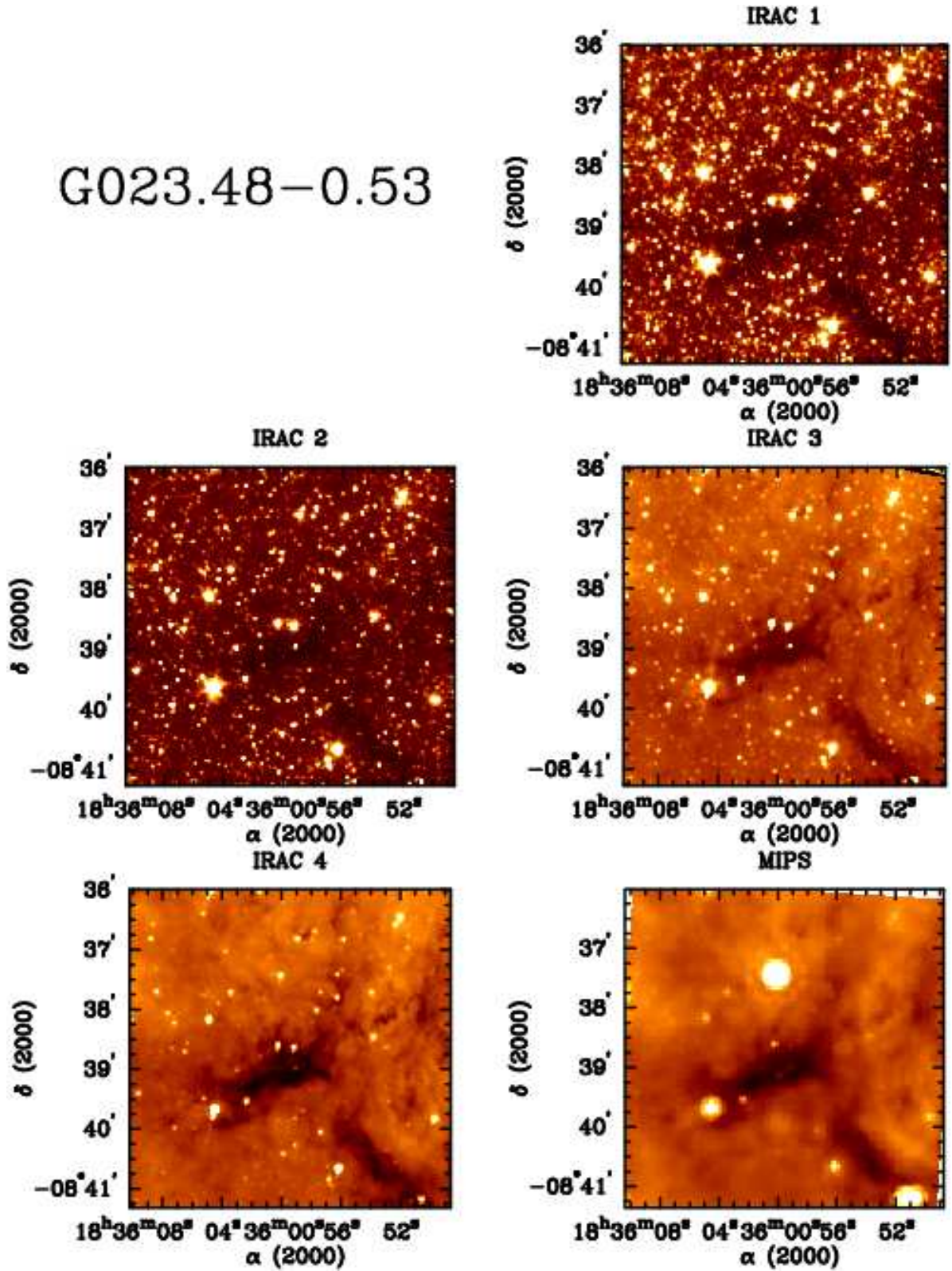}
\end{center}
\caption{G023.48$-$0.53: Wavelengths as noted in Figure~1.}
\label{fig:g2348}
\end{figure}

\clearpage

\begin{figure}
\begin{center}
\includegraphics[scale=0.9]{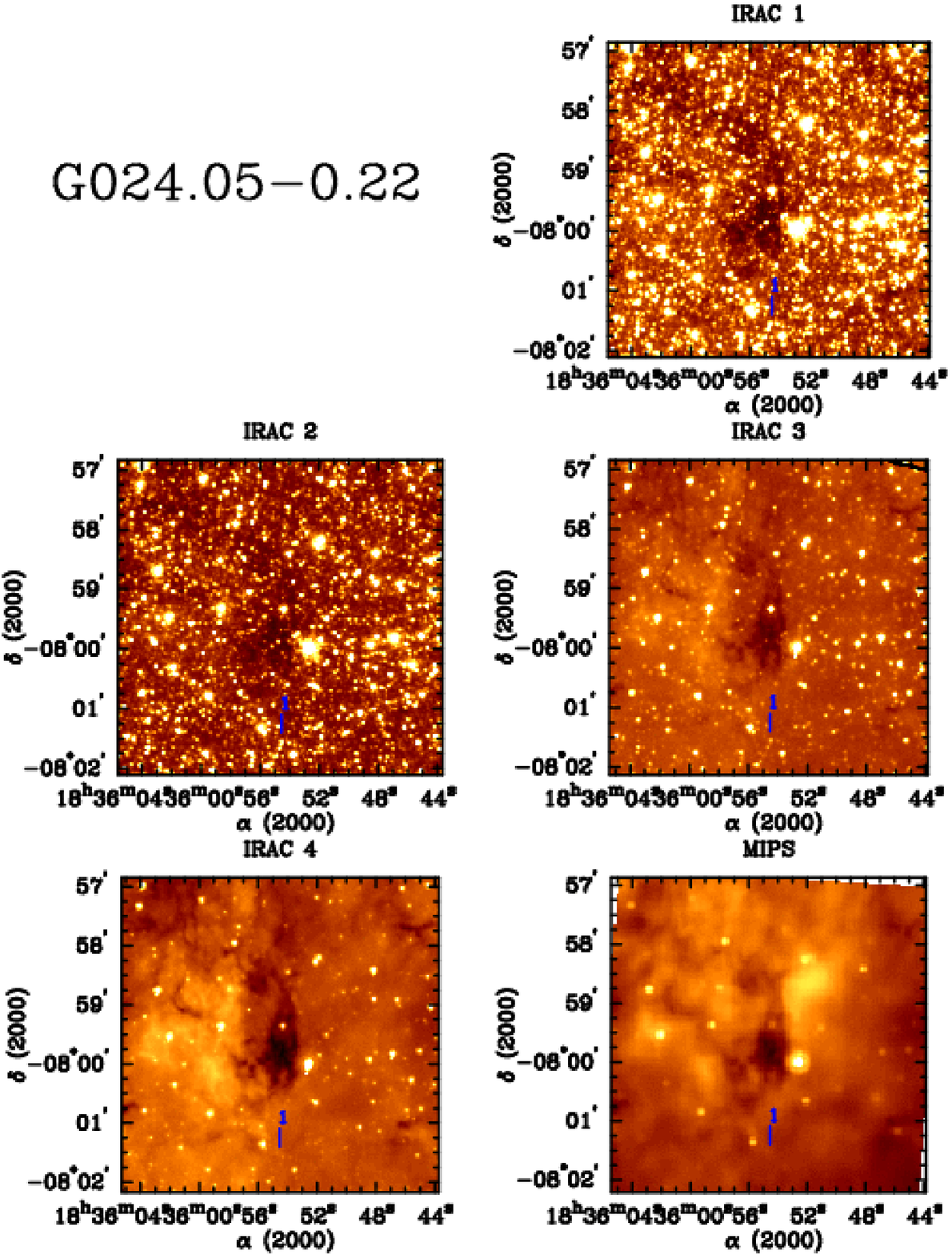}
\end{center}
\caption{G024.0$5-$0.22: Wavelengths as noted in Figure~1. - Embedded Object (index 1 in Table~2 under source G24.05$-$0.22) is labeled.}
\label{fig:g2405}
\end{figure}

\clearpage

\begin{figure}
\begin{center}
\includegraphics[scale=0.9]{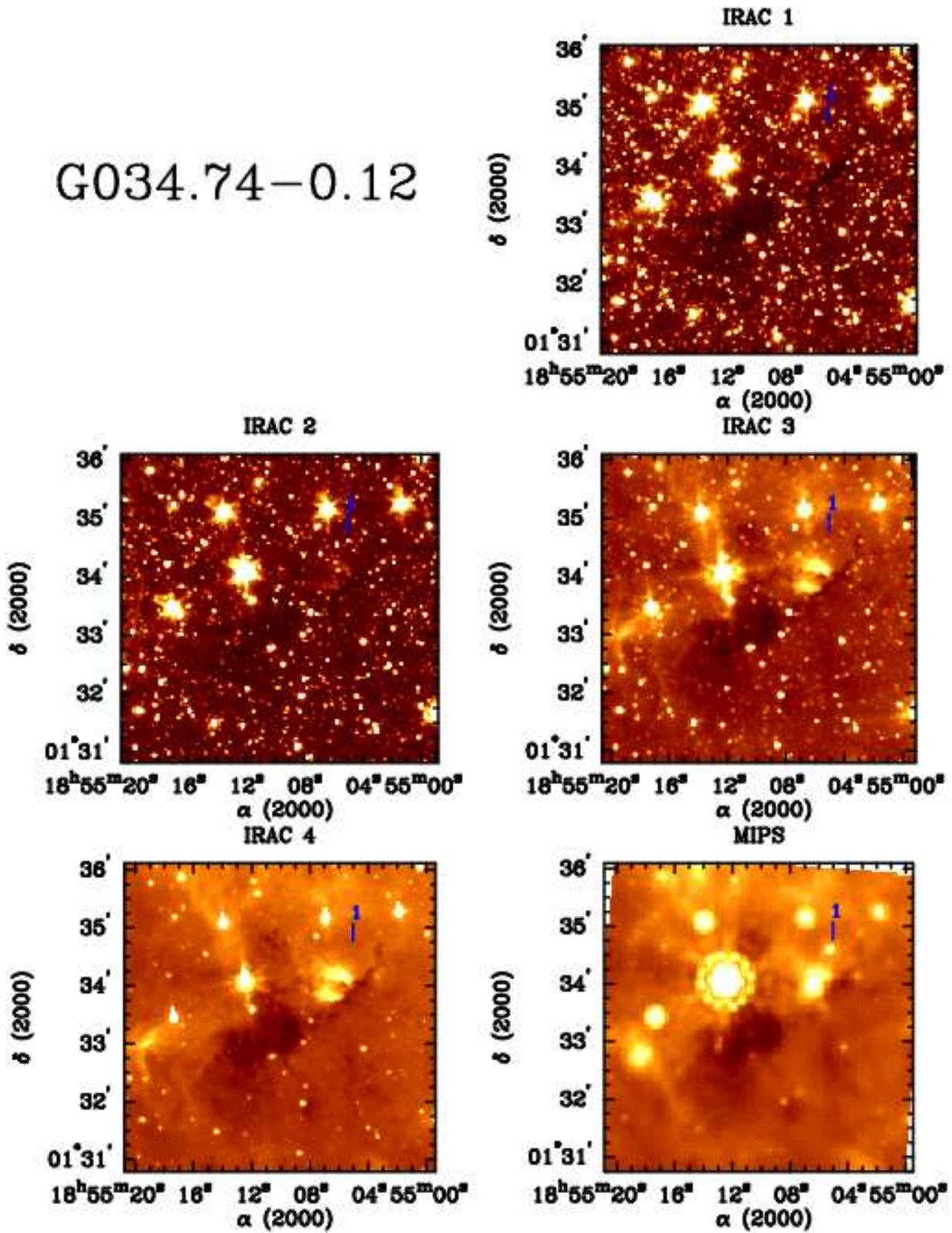}
\end{center}
\caption{G034.74$-$0.12: Wavelengths as noted in Figure~1. - Embedded Object (index 5 in Table~2 under source G034.74$-$0.12.) is labeled.}
\label{fig:g3474}
\end{figure}

\clearpage

\begin{figure}
\begin{center}
\includegraphics[scale=0.9]{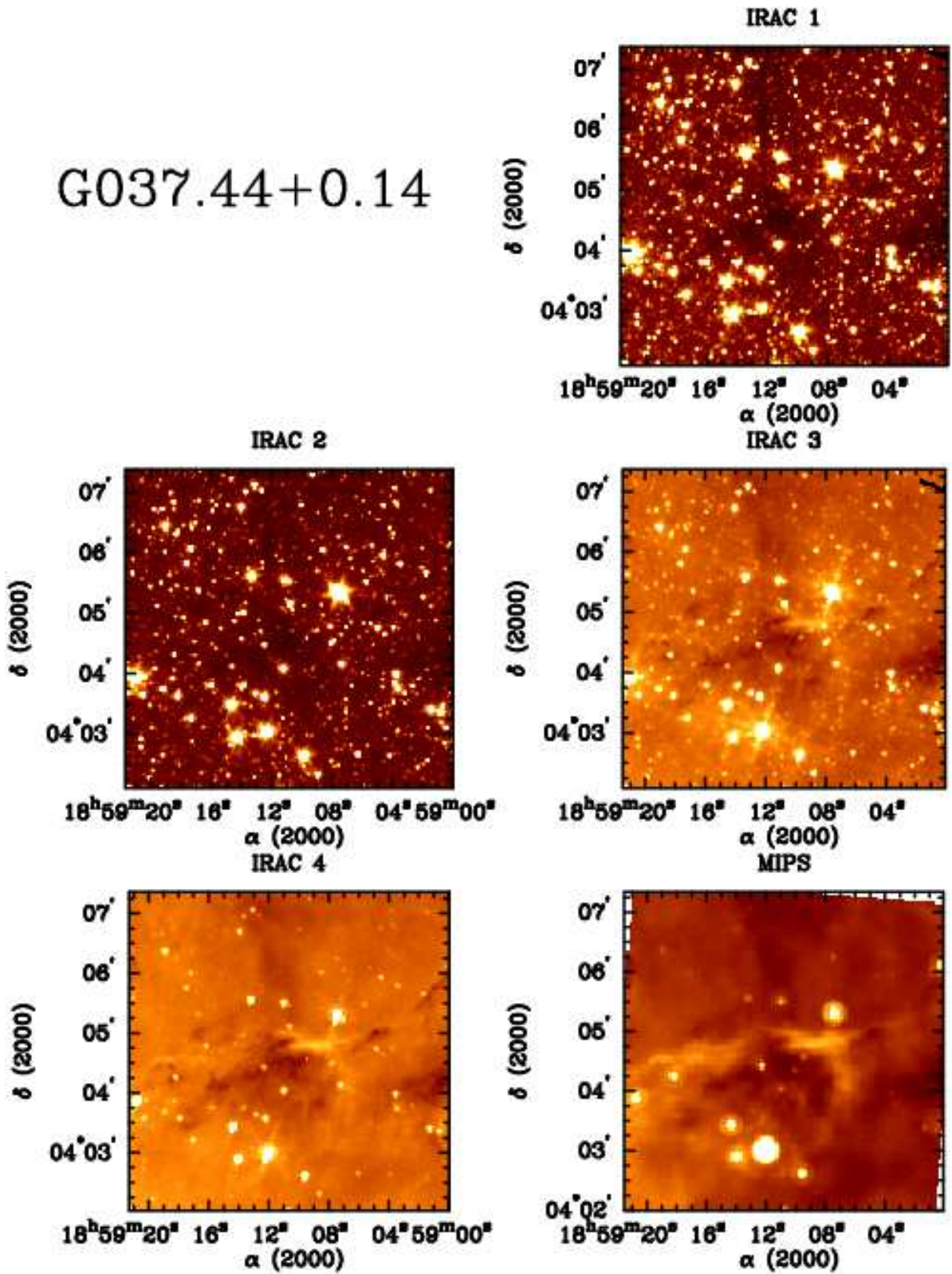}
\end{center}
\caption{G037.44$+$0.14: Wavelengths as noted in Figure~1.}
\label{fig:g3744}
\end{figure}

\clearpage

\begin{figure}
\hspace{-0.85in}
\includegraphics[scale=0.8]{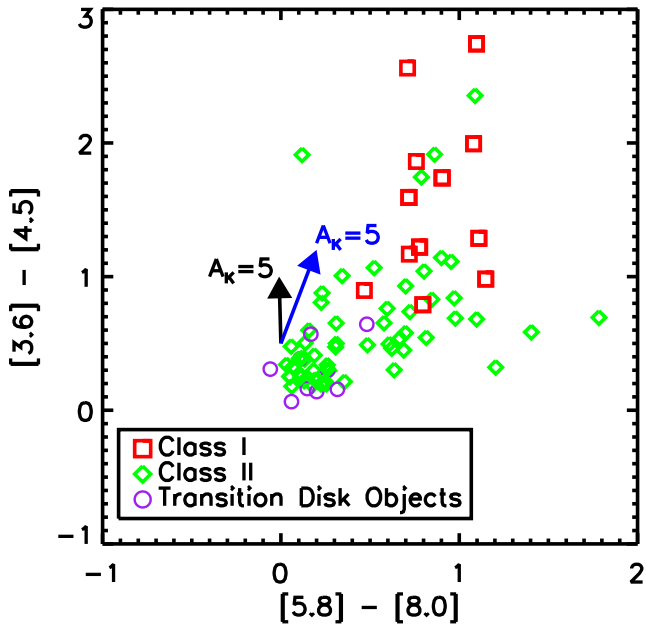}
\caption{\footnotesize{IRAC four color plot for all objects in the IRDC sample for all objects with photometry in all four bands that had errors less than 0.2 magnitudes.  Class I protostars are marked with red squares, green circles mark the more-evolved Class II sources, and transition/debris disk objects are marked with purple circles.  The deeply embedded objects identified with this analysis did not have sufficient detections in IRAC bands to appear on the color-color plots.  The extinction law from \citet{Flaherty2007} indicated by the black arrow, and the extinction law from \citet{Indebetouw2005} is plotted as the blue arrow.}}
\label{fig:colorcolor}
\end{figure}

\begin{figure*}
\begin{center}
\includegraphics[angle=270,scale=0.65]{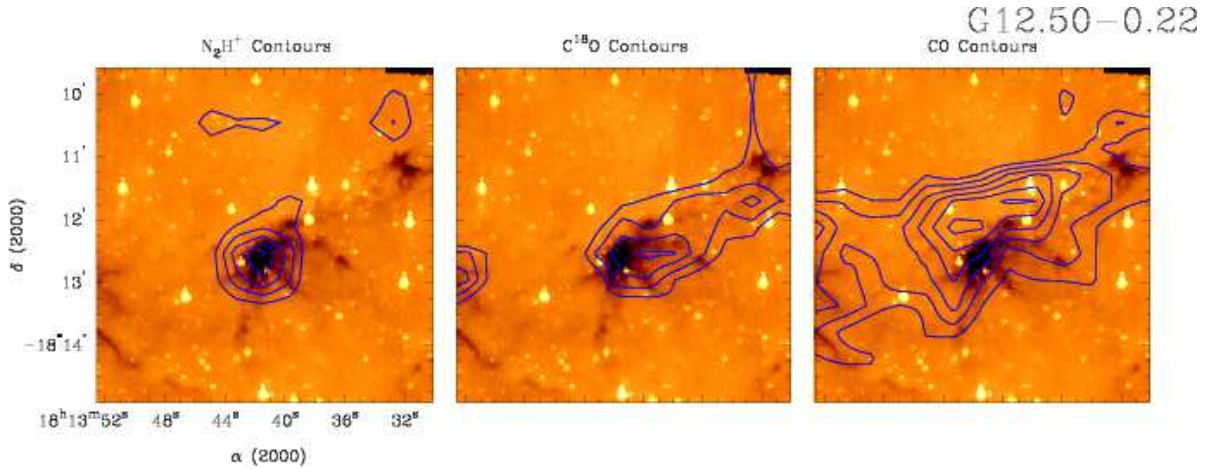}
\caption{\footnotesize{FCRAO molecular line contours of N$_2$H$^+$~(1-0) (left), C$^{18}$~(1-0) (center) and $^{12}$CO~(1-0) (right) plotted over the {\em Spitzer} 8~$\mu$m image of G012.50$-$0.22.  The critical density of the molecular transition decreases from left to right.}}
\label{fig:g1250_env}
\end{center}
\end{figure*}

\noindent in the entire sample is shown in Figure~\ref{fig:colorcolor}\footnote{No embedded protostar was detected in all four IRAC bands, so none are plotted in Figure~\ref{fig:colorcolor}.}.  The extinction laws from both \citet{Flaherty2007} and \citet{Indebetouw2005} are plotted to show the effect of five magnitudes of visual extinction.  The objects associated with these IRDCs are a great distance from us and in the plane of the Galaxy, so they naturally suffer from a great deal of extinction, reddening, and foreground contamination.  Furthermore, the reddening law used in this classification scheme and the measures taken to extricate extragalactic contaminants may be inaccurate due to the great distance to IRDCs, as the criteria were originally designed to suit local regions.  This may result in misclassification of sources. For example, a highly reddened Class I object might appear as an embedded protostar.  Nonetheless, if these objects are indeed protostars, it is likely that they are associated with the IRDC.

In Table~3, we summarize the number of each class of YSO in each IRDC field.  We note the number of these YSOs that are spatially coincident with the absorbing IRDC clumps (see $\S$\ref{structure}).  Only $\sim$10\% of the YSOs are associated with the dense gas.  The rest appear to be a distributed population of stars surrounding the IRDC.  This may be because any star directly associated with the IRDC is too heavily obscured to be detected even with the deep Spitzer observations we undertook.  Our observations are sensitive to 1-3~$\msun$, 1Myr-old pre-main sequence stars \citep{Baraffe1998}, or 1~$\lsun$ Class 0 protostar at 4~kpc with no extinction \citep{whitney_protostar}.  With extinction, which can reach 1-2 magnitudes in the {\em Spitzer} bands, embedded YSOs up to 3-4~$\msun$ might be present, but hidden from our view.  Another possible reason for the lack of YSOs detected coincident with the dense gas is that the IRDC itself could be in a stage prior to the onset of star formation, and the surrounding stars that are observed have disrupted their natal molecular gas.

Table~4 lists all of the objects identified as embedded objects that are spatially coincident with an IRDC.  We list the flux density at each Spitzer wavelength and an estimate of the mid-infrared luminosity derived from integrating the spectral energy distribution, which is dominated by emission at 24~$\mu$m.  In the likely event that the embedded objects are extincted, these mid-infrared luminosities will be underestimated.  Taking the average extinction estimations, which can be derived most reliably from the measurements of Class II objects, A$_K$ ranges from 1 to 3, which, if the extinction law \citet{Flaherty2007} is applied, corresponds to A$_{24}$ of 0.5 to 1.6.  As a check, we use a second method to estimate the extinction: based on average values of the optical depth we measure in the IRDCs, we confirm that A$_{24}\sim$1 is typical in these objects.  Given the uncertain extinction properties, and the fact that a large portion of these embedded sources' luminosity will emerge at longer wavelengths not observed here, the luminosities presented in Table~4 are lower limits.  Stars with luminosities in this range, according to \citet{Robitaille2006}, arise from stars ranging from 0.1 to 2~$\msun$, but are likely much greater.

\subsection{Nebulosity at 8 and 24~$\mu$m}

Four IRDCs in our sample (G006.26$-$0.51, Figure~\ref{fig:g0626}; G009.16$+$0.06, Figure~\ref{fig:g0916}; G023.37$-$0.29 Figure~\ref{fig:g2337}; G034.74$-$0.12, Figure~\ref{fig:g3474}) exhibit bright emission nebulosity in the IRDC field at 8 and 24~$\mu$m.  These regions tend to be brightest in the thermal infrared (e.g. 24~$\mu$m) but show some emission at 8~$\mu$m, which suggests they are sites of high mass star cluster formation.  To test whether the apparent active star formation is associated with the IRDC in question, or if it is in the vicinity, we correlate each instance of a bright emission with the molecular observations of the object obtained by \citet{ragan_msxsurv}.  The molecular observations provide velocity information which, due to Galactic rotation, aid in estimating the distance to the mid-infrared emission \citep{Fich:1989}.  This distance compared with the distance to the IRDC enables us to discern whether the IRDC and young cluster are at the same distance or one is in the foreground or background.

In the case of G006.26$-$0.51 (Figure~\ref{fig:g0626}), we detect infrared emission at 24~$\mu$m east of the IRDC.  This is spatially coincident and has similar morphology to C$^{18}$O~(1-0) emission emitting at a characteristic velocity of 17~km~s$^{-1}$ \citep{ragan_msxsurv}, corresponding to a distance of about 3$\pm$0.5~kpc.  The IRDC has a velocity of 23~km~s$^{-1}$, which gives a distance of 3.8~kpc, but with an uncertainty of over 500~pc (see Table~1 and \citet{ragan_msxsurv}).  Given the errors inherent in the distance derivation from the Galactic rotation curve, we cannot conclusively confirm or rule out association.  G009.16$+$0.06 (Figure~\ref{fig:g0916}), has neither distinct velocity component evident in the molecular observations nor does the molecular emission associated with the IRDC overlap with the 24~$\mu$m emission.  Embedded clusters should be associated with molecular emission especially C$^{18}$O which is included in the FCRAO survey.  Associated emission for this object likely lies outside the bandpass of the FCRAO observations and is at a greater or lesser distance than IRDC.  The 24~$\mu$m image of G023.37$-$0.29 (Figure~\ref{fig:g2337}) shows bright emission to the south of the IRDC and another region slightly south and west of the IRDC.  This emission is not prominent in the IRAC images, suggesting that this is potentially an embedded star cluster.  Molecular observations show strong emission peaks in both CS~(2-1) and N$_2$H$^+$~(1-0) in the vicinity of the IRAC 8~$\mu$m and MIPS 24~$\mu$m emission.  However, there are three distinct velocity components evident in the observed bandpass, none of which is more spatially coincident with the 24~$\mu$m emission than the others.  Unfortunately, the spatial resolution of the FCRAO survey is insufficient for definitive correlation.  Finally, in G34.74$-$0.12 (Figure~\ref{fig:g3744}), no molecular emission is distinctly associated with the nebulosity; the most likely scenario for this object is that the associated molecular emission lies outside the bandpass of the FCRAO observation and, therefore, is not associated.  

\subsection{Summary of Stellar Content and IRDC Environment}

We have characterized the star formation that is {\em possibly} associated with the IRDCs to the extent that the Spitzer and millimeter data allow.  The YSO population is distributed, and only a handful of objects identified are directly spatially associated with the IRDC.  More explicitly, in this sample, half (5/11) of the sample shows no clear evidence for {\em embedded} sources in the dense absorbing gas, and instead appear populated sparsely with young protostars, the photometric properties of which are given in Table~2, and the overall IRDC star content is summarized in Table~3.  Among those embedded objects correlated with the absorbing structure at 8~$\mu$m, which are summarized in Table~4, we find a marked lack of luminous sources ($>$5~$\lsun$) at these wavelengths.  There may be significant extinction at 24~$\mu$m, in which case we would underestimate their luminosity.  Further, even in IRDCs with embedded protostars, most of the cloud core mass is not associated with an embedded source.  It is our contention that most of the IRDC mass does not harbor significant massive star formation, and, hence IRDCs are in an early phase of cloud evolution.  

Bright emission nebulosity is evident at 8~$\mu$m and 24~$\mu$m in four fields, presumably due to the presence of high mass stars or a cluster.  If the IRDC were associated with the nebulosity, it would be a strong indication that the IRDCs have massive star formation occurring already in the vicinity.  Molecular data give no definitive clues that these regions are associated with the IRDCs.  

Most studies including this one focus primarily on the dense structures that comprise infrared-dark clouds, yet their connection to the surrounding environment has not yet been discussed in the literature.  While it is clear that some star formation is directly associated with the dense material, star formation is also occurring beyond the extent of the IRDC as it appears in absorption.  Figure~\ref{fig:g1250_env} shows molecular line contours from \citet{ragan_msxsurv} over the {\em Spitzer} 8~$\mu$m image.  N$_2$H$^+$, a molecule known to trace very dense gas, corresponds exclusively to the dark cloud.  On the other hand, C$^{18}$O and, to a greater extent $^{12}$CO, show a much more extended structure, which suggests that the infrared-dark cloud resides within a greater molecular cloud complex.  For all of the objects in our sample, the $^{12}$CO emission was present at the edge of the map (up to 2$'$ away from the central position), so it is likely that the emission, and therefore the more diffuse cloud that it probes, extends beyond the mapped area.  Thus, the full extent of the surrounding cloud is not probed by our data.

\section{Tracing mass with dust absorption at 8~$\micron$}
\label{clumps}

Each infrared-dark cloud features distinct absorbing structures evident at all Spitzer wavelengths, but they are most pronounced at 8~$\mu$m and 24~$\mu$m due to strong background emission from polycyclic aeromatic hydrocarbons (PAHs) and small dust grains in the respective bandpasses \citep{Draine_dustreview}.  The IRDCs in this sample exhibit a range of morphologies and surrounding environments.  Figures~\ref{fig:g0585}-\ref{fig:g3744} shows a morphological mix of filamentary dark clouds (e.g. G037.44$+$0.14, Figure~\ref{fig:g3744}) and large ``round'' concentrations (e.g. G006.26$-$0.51, Figure~\ref{fig:g0626}). Remarkably, these detailed structures correspond almost identically between the 8~$\mu$m and 24~$\mu$m bands, despite the fact that the source of the background radiation arises from separate mechanisms.  At 8~$\mu$m emission from PAHs dominate on average, and at 24~$\mu$m, the bright background is due to the thermal emission of dust in the Galactic plane.  Considering this scenario, it is unlikely that we are mistaking random background fluctuations for dense, absorbing gas with the appropriate characteristics to give rise to massive star and cluster formation.

\begin{figure*}
\begin{center}
\includegraphics[angle=270,scale=0.6]{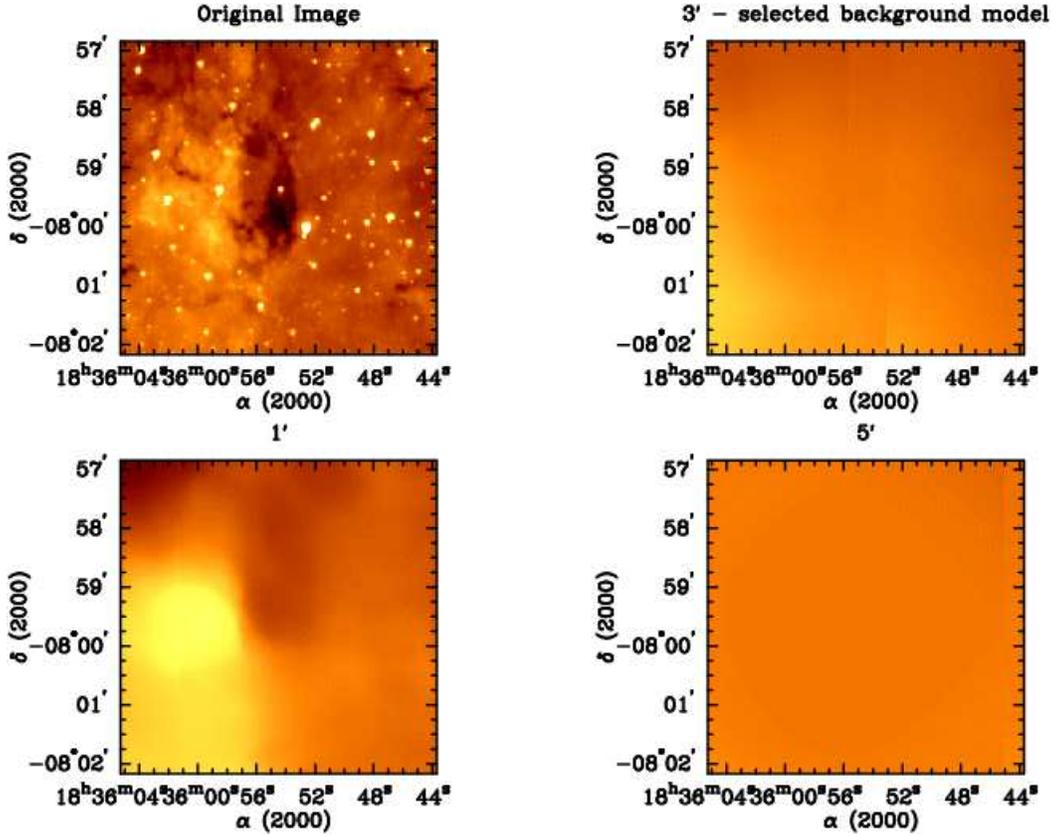}
\end{center}
\caption{\footnotesize{{\it Upper left:} Original IRAC 8~$\mu$m image of G024.05-0.22.  {\it Upper right:} Background model using the spatial median filtering technique with a 3$\arcmin$ radius.  The dark cloud is virtually eliminated from the background, but still accounts for the large-scale variations. {\it Lower left:}  Same as upper right panel, except that a 1$\arcmin$ radius is used, which models the dark cloud as part of the background.  {\it Lower right:} Same as upper right and lower left panels, except that a 5$\arcmin$ radius is used, which misses the background variation and is almost a constant value.}}
\label{fig:bgmethod}
\end{figure*}

\subsection{Modeling the Foreground and Background}
\label{bg}

In the Galactic plane, the 8~$\mu$m background emission varies on scales of a few arcminutes.  To accurately estimate structures seen in absorption, we account for these variations using a spatial median filtering technique, motivated by the methods used in \citet{Simon2006}.  For each pixel in the IRAC image, we compute the median value of all pixels within a variable radius and assign that value to the corresponding pixel in the background model.  Figure~\ref{fig:bgmethod} illustrates an example of several trials of this method, including models with 1$\arcmin$, 3$\arcmin$, and 5$\arcmin$ radius of pixels included in a given pixel's median calculation.  We select the size of the filter to be as small as possible such that the resulting map shows no absorption as background features.  If the radius is too small, most of the included pixels will have low values with few representing the true background in the areas where absorption is concentrated (lower left panel of Figure~\ref{fig:bgmethod}).  The background variations are also not well-represented if we select a radius too large (lower right panel of Figure~\ref{fig:bgmethod}).  Based on our analysis, the best size for the filter is 3$\arcmin$.  The observed 8~$\mu$m emission is a combination of both background and foreground contributions.  

\begin{equation}
\int I^{estimate} d\lambda = \int I_{BG}^{true} d\lambda + \int I_{FG} d\lambda
\label{eq1}
\end{equation}

\noindent where $\int I^{estimate}~d\lambda$ is the intensity that we measure from the method described above, $\int I_{BG}^{true}~d\lambda$ is the true background intensity, which can only be observed in conjunction with $\int I_{FG}~d\lambda$, the foreground intensity, all at 8~$\mu$m.  The relative importance of the foreground emission is not well-known. For simplicity, we assume the foreground can be approximated by constant fraction, $x$, of the emission across each field.

\begin{equation}
\int I_{FG}~d\lambda = x \int~I_{BG}^{true}~d\lambda
\end{equation}

\begin{figure}
\hspace{-0.3in}
\includegraphics[scale=0.55]{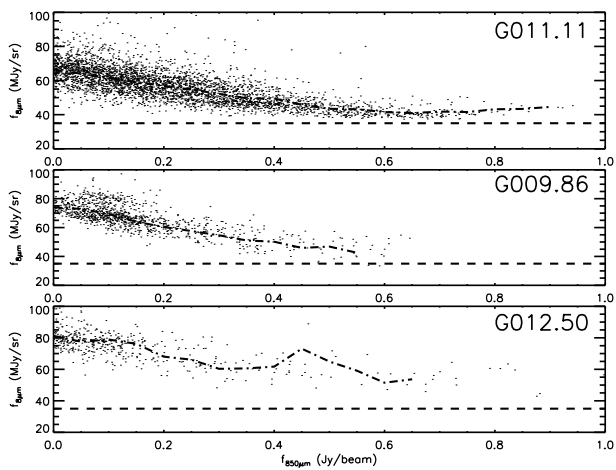}
\caption{\footnotesize{Spitzer 8~$\mu$m vs. SCUBA 850$\mu$m flux for IRDC G011.11$-$0.12, G009.86$-$0.04, and G012.50$-$0.22.  The horizontal dashed line marks where the 8~$\mu$m flux density reaches a minimum in G011.11$-$0.12, which is also indicated for the two other IRDCs with available SCUBA data.  This flux density serves as an estimate of the foreground emission at 8~$\mu$m.  The dash-dotted line indicates the mean 8~$\mu$m emission.}}
\label{fig:g1111_fluxplot}
\end{figure}

One way to estimate the foreground contribution has already been demonstrated by \citet{Johnstone_G11}.  The authors compare observations of IRDC G011.11$-$0.12 with the {\em Midcourse Science Experiment (MSX)} at 8~$\mu$m and the Submillimeter Common-User Bolometer Array (SCUBA) on the James Clerk Maxwell Telescope (JCMT) at 850~$\mu$m (see their Figure~3) and use the point at which the 8~$\mu$m integrated flux is at its lowest at high values of 850~$\mu$m flux for the foreground estimate.  The top panel Figure~\ref{fig:g1111_fluxplot} shows a similar plot to Figure~3 in \citet{Johnstone_G11}, except our integrated 8~$\mu$m flux is measured with {\em Spitzer} and presented here in units of MJy/sr.  SCUBA 850$\mu$m data for two of the IRDCs in this sample (G009.86$-$0.04 and G012.50$-$0.22) are available as part of the legacy data release \citep{SCUBA_legacy} and are included in this plot.  Just as \citet{Johnstone_G11} point out, we see a clear trend: where 8~$\mu$m emission is low along the filament, the 850$\mu$m flux is at its highest.  In the case of G011.11$-$0.12, where the SCUBA data are of the highest quality, we take the minimum 8~$\mu$m flux density to be an estimate of the foreground contribution.  Assuming this trend is valid for our sample of IRDCs, we use the 8~$\mu$m emission value measured at the dust opacity peak in each source as our estimation of the foreground level for that object (for the remainder of this paper, we will refer to this method as foreground estimation method ``A'').  Given these considerations, we find values for $x$ to range between 2 and 5.  Up to 20\% of this foreground contamination is likely due to scattered light in the detector (S.T. Megeath, private communication).  We assume constant foreground flux at this level.  As an alternative foreground estimate, we also test a case in which we attribute half of the model flux to the background and half to the foreground.  This is equivalent to choosing a value of $x$ of 1, and based on Figure~\ref{fig:g1111_fluxplot}, is also a reasonable estimate.  This method will be referred to as foreground estimation method ``B.''  For most of the following figures and discussion, we use estimation method A and refer the results from method B in the text when applicable.

With an estimation of the foreground contribution, the absorption can be quantitatively linked to the optical depth of the cloud.  The measured integrated flux, $\int I_m d\lambda$ at any point in the image, including contributions from both the foreground and background, can then be expressed as 

\begin{equation}
\int I_m d\lambda =\int I_{BG}^{true}e^{-\tau_8} d\lambda+ \int I_{FG} d\lambda
\end{equation}

\noindent where $\tau_8$ is the optical depth of the absorbing material.  For the subsequent calculations, we use the average intensity, assuming uniform transmission over the IRAC channel 4 passband, and average over the extinction law \citep[][see Section~\ref{structure}]{weingartner_draine01} in this wavelength region in order to convert the optical depth into a column density (see discussion in next section).  We note that we make no attempt to correct for the spectral shape of the the dominant PAH emission feature in the 8~$\mu$m Spitzer bandpass, which we assume dominates the background radiation.  In addition, clumpy material that may be optically thick and is not resolved by these observations will cause us to underestimate the column density.  These factors could introduce an uncertainty in the conversion of optical depth to column density.  Still, we will show in Section~\ref{columnprobe} that dust models compare favorably to our estimation of the dust absorption cross section, lending credence to our use of $\tau$ as a tracer of column density.

\subsection{Identification of Structure}
\label{structure}

\begin{figure}
\begin{center}
\includegraphics[scale=0.5]{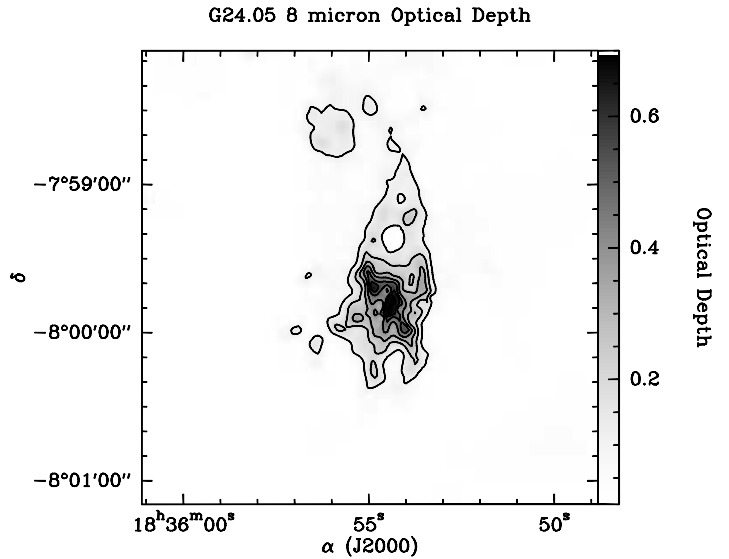}
\end{center}
\caption{\footnotesize{G024.05$-$0.22. 8~$\mu$m optical depth with contours highlighting the structures.}}
\label{fig:opaccontours}
\medskip
\end{figure}

Figure~\ref{fig:opaccontours} shows a map of optical depth G024.05$-$0.22.  This provides an example of the the absorbing substructure in one of the IRDCs in our sample.  Owing to the high spatial resolution of Spitzer at 8~$\mu$m (1 pixel = 0.01~pc at 4~kpc, accounting for oversampling), we see substructures down to very small scales ($\sim$0.03~pc) in {\em all} IRDCs in our sample.  

In order to identify independent absorbing structures in the 8~$\mu$m optical depth map, we employed the {\tt clumpfind} algorithm \citep{williams_clumpfind}.  In the two-dimensional version, {\tt clfind2d}, the algorithm calculates the location, size, and the peak and total flux of structures based on specified contour levels.  We use the Spitzer PET\footnote{http://ssc.spitzer.caltech.edu/tools/senspet/} to calculate the sensitivity of the observations, i.e. to what level the data permit us to discern true variations from noise fluctuations.  At 8~$\mu$m, the observations are sensitive to 0.0934 MJy/sr which, on average, corresponds to an optical depth sensitivity (10-$\sigma$) of $\sim$0.02.  While the clumps take on a variety of morphologies, since {\tt clumpfind} makes no assumptions about the clump shapes, we approximate the clump ``size'' by its effective radius, 

\begin{equation}
r_{eff}=\sqrt{\frac{n_{pix}~A_{pix}}{\pi~f_{os}}}
\end{equation}

\noindent where n$_{pix}$ is the number of pixels assigned to the clump by {\tt clumpfind}, and A$_{pix}$ is the area subtended by a single pixel.  The correction factor for oversampling, $f_{os}$ accounts for the fact that the {\em Spitzer Space Telescope} has an angular resolution of 2.4$''$ at 8~$\mu$m, while the pixel scale on the IRAC chip is 1.2$''$, resulting in oversampling by a factor of 4.  

The number and size of structures identified with {\tt clumpfind} varies depending on the number of contouring levels between the fixed lower threshold, which is set by the sensitivity of the observations, and the highest level set by the deepest absorption.  We set the lowest contour level to 10$\sigma$ above the average background level.  In general, increasing the number of contour levels serves to increase the number of clumps found.   In all cases, we reach a number of levels where the addition of further contouring levels results in no additional structures.  We therefore select the number of contour levels at which the number of clumps levels off, i.e. when the addition of more contour levels reveals no new clumps.  We also remove those clumps found at the image edge or bordering a star, as the background estimation is likely inaccurate and/or at least a portion of the clump is probably obscured by the star, rendering any estimation of the optical depth inaccurate.

\begin{figure}
\begin{center}
\includegraphics[angle=270, scale=0.4]{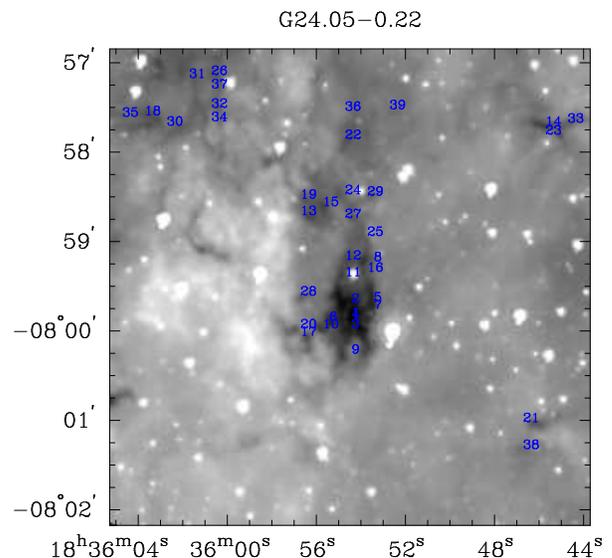}
\end{center}
\caption{\footnotesize{G024.05$-$0.22. Results of the {\tt clumpfind} algorithm plotted over Spitzer 8~$\mu$m image.  Absorption identified as a ``clump'' is denoted by a number.}
\label{fig:numberedclumps}}
\end{figure}

Using {\tt clumpfind}, each IRDC broke down into tens of clumps, ranging in size from tens to hundreds of pixels per clump.  The average clump size is 0.04~pc.  Typically, there is one or two central most-massive clumps and multiple smaller clumps in close proximity.  In some instances, clumps are strung along a filamentary structure, while in other cases, clumps are radially distributed about a highly-concentrated center.  Figure~\ref{fig:numberedclumps} shows an example of how the clumps are distributed spatially in G024.05$-$0.22 as {\tt clumpfind} identifies them.   

With reliable identification of clumps, we next calculate individual clump masses.  As described, {\tt clumpfind} gives total optical depth measured at 8~$\mu$m, $\tau_{8,tot}$, within the clump boundary, its size and position.  This can be directly transformed into $N(H)_{tot}$ via the relationship 

\begin{equation}
\label{colequation}
N(H)_{tot} = {\frac{\tau_{8,tot}}{\sigma_8~f_{os}}}
\end{equation}

\noindent where $\sigma_8$ is the dust absorption cross section at 8~$\mu$m.  We derive an average value of $\sigma_8$ over the IRAC channel 4 bandpass using dust models that take into account higher values of R$_V$ corresponding to dense regions in the ISM.  Using \citet{weingartner_draine01}, we use $R_V$ = 5.5, case B values, which agree with recent results from \citet{Indebetouw2005}.  We find the value of $\sigma_8$ to be 2.3$\times$10$^{-23}$cm$^2$.  

The column density can then be used with the average clump size and the known distance to the IRDC, assuming all clumps are at approximately the IRDC distance, to find the clump mass.  The mass of a clump is given as 

\begin{equation}
M_{clump} = 1.16 m_H N(H)_{tot} A_{clump} 
\end{equation}

\noindent where m$_H$ is the mass of hydrogen, N(H)$_{tot}$ is the total column density of hydrogen, the factor 1.16 is the correction for helium and A$_{clump}$ is the area of the clump.  Table~6 gives the location, calculated mass and size of all the clumps identified with {\tt clumpfind}.  We also note which clumps are in the vicinity of candidate young stellar objects (Table~2) or foreground stars, thereby subjecting the given clump properties to greater uncertainty.  On average (for foreground estimation method A), 25\% of clumps border a field star, and these clumps are flagged and not used in the further analysis.  In each infrared-dark cloud, we find between 3000$\msun$ and 10$^4\msun$ total mass in clumps, and typically $\sim$15\% of that mass is found in the most massive clump.  

We perform the same analysis on the maps produced with foreground estimation method B. The foreground assumption in this case leads to lower optical depths across the map.  Due to the different dynamic range in the optical depth map, {\tt clumpfind} does not reproduce the clumps that are found with method A exactly.  The discrepancy arises in how {\tt clumpfind} assigns pixels in crowded regions of the optical depth map, so while at large the same material is counted as a clump, the exact assignment of pixels to specific clumps varies somewhat.  On average, the clumps found in the ``method B'' maps tend to have lower masses by a factor of 2, though the sizes do not differ appreciably from those found with foreground estimation method A.  

\begin{figure*}
\begin{center}
\includegraphics[angle=270,scale=0.6]{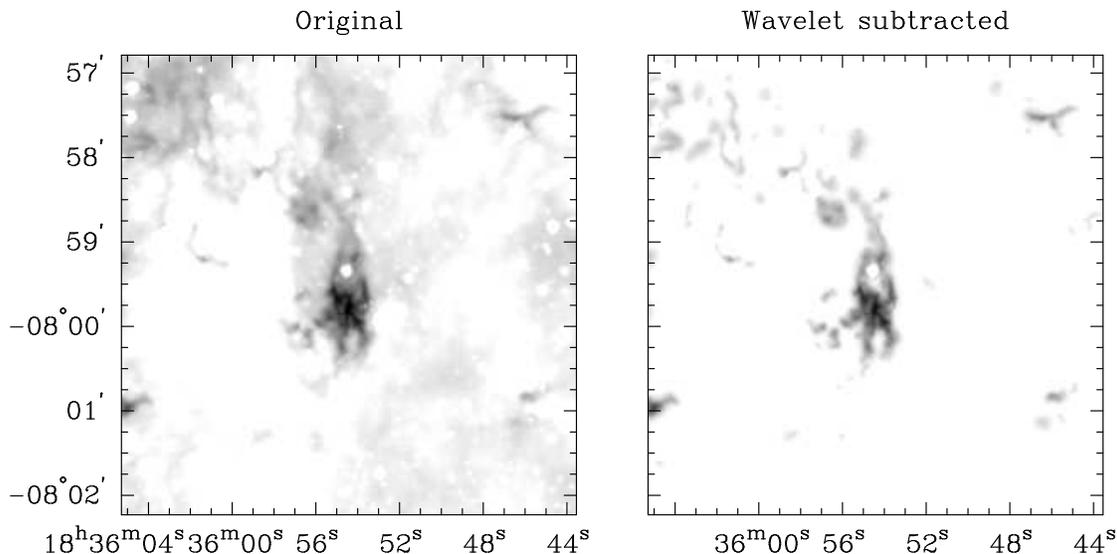}
\end{center}
\caption{\footnotesize{Original optical depth image of G024.05$-$0.22 (left) and the wavelet subtracted image (right) of the same region.}}
\label{fig:wavelet}
\end{figure*}

\subsection{Resolving Inaccuracy in Clump Mass Calculation}
\label{masserr}

The clumps identified in this fashion include a contribution from the material in the surrounding envelope.  As a result, a portion of the low-mass clump population may not be detected, and the amount of material in a given clump may be overestimated.  To examine this effect, we use the {\tt gaussclumps} algorithm \citep{StutzkiGusten1990} to identify clumps while accounting for the contribution from the cloud envelope.  This method was designed to decompose three-dimensional molecular line observations by deconvoloving the data into clumps fit by Gaussians.  To use the algorithm here without altering the code, we fabricated a data cube by essentially by mimicking a third (velocity) dimension, thus simulating three-dimensional clumps that were all centered in velocity on a single central plane.  \citet{mookerjea2004} and \citet{Motte2003} have used similar techniques to simulate a third dimension to their dust continuum data sets.  The {\tt gaussclumps} algorithm inherently accounts for an elevated baseline level, which can be used to approximate the envelope.  Applied to our data set, {\tt gaussclumps} finds that 15-50\% of the material is in the envelope.   Further discussion of the envelope contribution, including its effect on the mass function, is given in $\S$\ref{envelope}.

The {\tt clumpfind} and {\tt gaussclumps} methods result in nearly one-to-one clump identification in the central region of the IRDC.  However, because the contribution from the cloud envelope falls off further away from the central concentration of mass in the IRDC, {\tt gaussclumps} fails to find low-mass clumps on the outskirts of IRDCs as successfully as {\tt clumpfind}, despite being statistically valid relative to their local background.  We conclude that {\tt gaussclumps} is not suitable to identify of structure in the outskirts of the IRDCs where the envelope is below the central level.  

Another method commonly employed in the literature to account for the extended structures in which dense cores reside is a ``wavelet subtraction'' technique, which is described in \citet{alves_cmfimf}.  To address the varying levels of background across the optical depth map, we use the wavelet transform of the image to extract the dense cores.  For one IRDC in our sample, G024.05$-$0.22, we (with the help of J. Alves, private communication) perform the wavelet analysis on the optical depth map.  Figure~\ref{fig:wavelet} shows a comparison between the original optical depth map and the wavelet-subtracted map.  With the removal of the ``envelope'' contribution in this fashion, the clumps are up to 90\% less massive on average, and their average size decreases by 25\%, or $\sim$0.02~pc.  

Both using {\tt gaussclumps} and applying wavelet subtraction methods to extract clumps show that the contribution of the cloud envelope is not yet well-constrained quantitatively.  Not only is the cloud envelope more difficult to detect, its structure is likely not as simple as these first order techniques have assumed in modeling it.  As such, for the remainder of the paper, we will not attempt to correct the clump masses on an individual basis, but rather focus our attention on the clump population properties as a whole.  In $\S$\ref{validate}, we employ several techniques to calibrate our mass estimation methods.  We will show in $\S$\ref{mf} that the effect of the envelope is systematic and does not skew the derived relationships, such as the slope of the mass function.  

\subsection{Validating 8~$\micron$ absorption as a Tracer of Mass}
\label{validate}

In previous studies, molecular clouds have been predominantly probed with using the emission of warm dust at sub-millimeter wavelengths.  While there are inherent uncertainties in the conversion of flux density to mass, the emission mechanism is well-understood.  The method described above is a powerful way to trace mass in molecular clouds.  To understand the extent of its usefulness, here we validate dust absorption as a mass tracer by drawing comparisons between it and results using more established techniques.  First, we relate the dust absorption to dust emission as probes of column density.  Second, we use observations of molecular tracers of dense gas not only to further cement the validity of the absorbing structures, but also to place the IRDCs in context with their surroundings.  Finally, we show that the sensitivity of the technique does not have a strong dependence on distance.  

\subsubsection{Probing Column Density at Various Wavelengths}
\label{columnprobe}

As we discussed in $\S$\ref{bg}, there is an excellent correlation between the 8~$\mu$m and 850~$\mu$m flux densities in IRDC G011.11-0.12.  Figure~\ref{fig:g1111_fluxplot} shows the point-to-point correlation between the SCUBA 850~$\mu$m flux density and Spitzer 8~$\mu$m flux density.  This correspondence itself corroborates the use of absorption as a dust tracer.  In addition, the fit to the correlation can confirm that the opacity ratio, $\kappa_8~$/$\kappa_{850}$, is consistent with dust behavior in high density environments.  Relating the 8~$\mu$m flux density 

\begin{equation}
f_{8}=f_{bg}e^{-\kappa_{8}\Sigma(x)}+f_{fg}
\end{equation}

\noindent where $\kappa_8~$ is the 8~$\mu$m dust opacity, $\Sigma(x)$ is the mass column density of emitting material, and $f_{bg}$ and $f_{fg}$ are the background and foreground flux density estimates, respectively (from $\S$\ref{bg}), and the 850~$\mu$m flux density 

\begin{equation}
f_{850}=B_{850}(T_d=13K) \kappa_{850} \Sigma(x) \Omega
\end{equation}

\noindent where $B_{850}$ is the Planck function at 850~$\mu$m evaluated for a dust temperature of 13~K, $\kappa_{850}$ is the dust opacity at 850~$\mu$m and $\Omega$ is the solid angle subtended by the JCMT beam at 850~$\mu$m, one can find a simple relation between the two by solving each for $\Sigma(x)$ and equating them. The opacity ratio, put in terms of the flux density measurements is as follows:

\begin{equation}
\frac{\kappa_{8}}{\kappa_{850}}=\frac{B_{850}~\Omega}{f_{850}} ln \left( \frac{f_{bg}}{f_{8}-f_{fg}}\right)
\end{equation}

\noindent From our data, we confirm this ratio is considerably lower ($\sim$500) in cold, high density environments than in the diffuse interstellar dust as found by \citet{Johnstone_G11}.

We perform another consistency check between our data and dust models.  With maps at both 8 and 24~$\mu$m, both showing significant absorbing structure against the bright Galactic background (albeit at lower resolution at 24~$\mu$m), we can calculate the optical depth of at 24~$\mu$m in the same way we did in Section~\ref{bg}.  The optical depth scales with the dust opacity by the inverse of the column density ($\tau_{\lambda} \propto \kappa_{\lambda} / N(H)$), so the ratio of optical depths is equal to the dust opacity ratio.  We find that the typical ratio as measured by Spitzer in IRDCs is

\begin{equation}
\frac{\kappa_{8}}{\kappa_{24}} = \frac{\tau_8}{\tau_{24}} \sim 1.2
\end{equation}

\noindent which is comparable to 1.6, the \citet{weingartner_draine01} prediction (for R$_V$ = 5.5, case B) and 1-1.2 predicted by \citet{ossenkopf_henning} in the high-density case.  We conclude that the dust properties we derive are consistent with the trends that emerge from models of dense environments typical of infrared-dark clouds.

\begin{figure*}
\begin{center}
\hbox{
\vspace{1.0cm}
\hspace{1.5cm}
\psfig{figure=f19a.ps,angle=270,height=7.0cm}
\hspace{0.4cm}
\psfig{figure=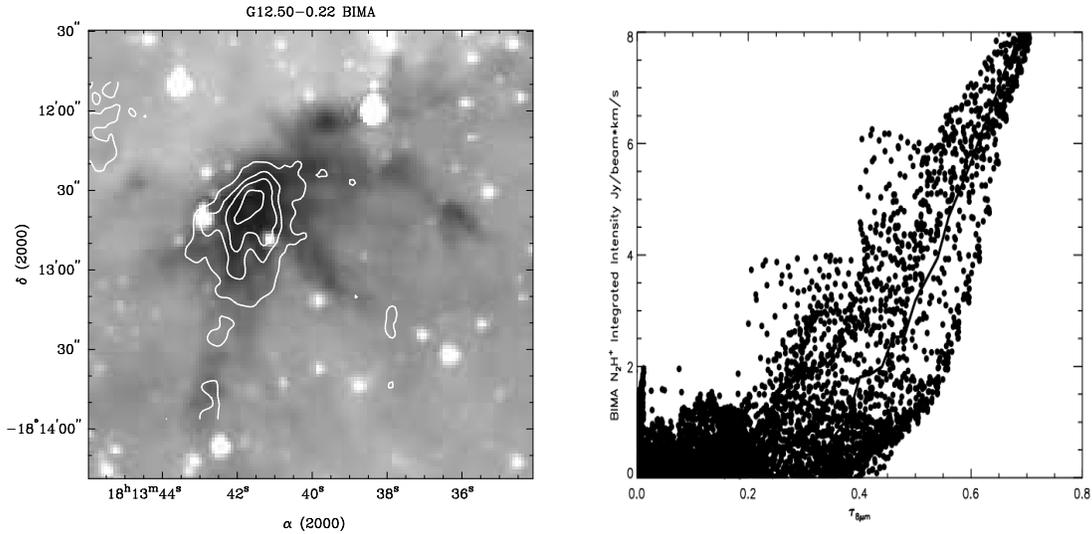,angle=90,height=7.0cm,width=7.0cm}
}
\vspace{-1.0cm}
\end{center}
\caption{\footnotesize{Left: Contours of integrated intensity of N$_2$H$^+$ plotted over the IRAC 8~$\mu$m image of IRDC G012.50$-$0.22.  Right: Point-to-point correlation of the N$_2$H$^+$ integrated intensity and the 8~$\mu$m optical depth.  Points with high integrated intensity but low optical depth correspond to stars, whose presence leads to the underestimation of optical depth in the vicinity.}}
\label{fig:g1250_bimaplot}
\end{figure*}

\begin{figure}
\hspace{-1in}
\includegraphics[scale=0.8]{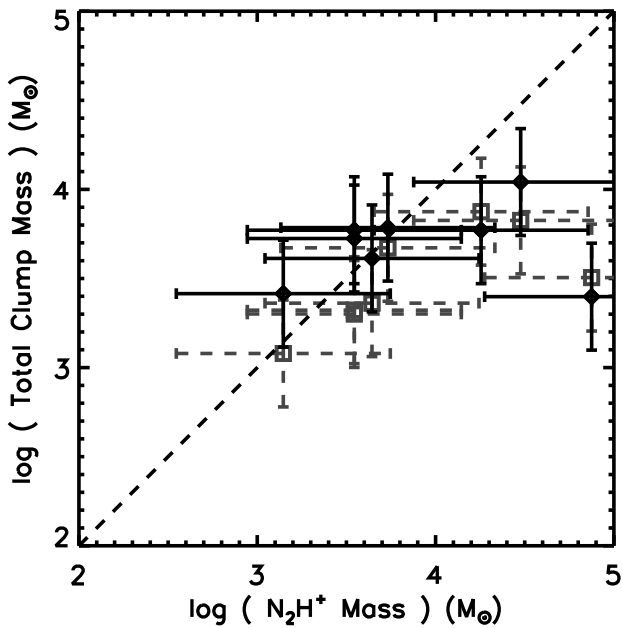}
\caption{\footnotesize{Comparison of the total mass derived from N$_2$H$^+$ maps from \citet{ragan_msxsurv} and total clump mass as derived from dust absorption at 8~$\micron$, where the black diamonds represent the mass using foreground estimation method A and the grey squares show the masses derived using foreground estimation method B (see $\S$\ref{bg}).  Three of the IRDCs in the sample did not have adequate N$_2$H$^+$ detections.  Error bars for 30\% systematic errors in the mass are plotted for the clump mass estimates, and a factor of 5 uncertainty is plotted for the N$_2$H$^+$ mass estimates.  The dashed line shows a one-to-one correspondence for reference.}
\label{fig:n2hpclumps}}
\end{figure}

\subsubsection{Molecular Line Tracers}

Molecular lines are useful probes of dense clouds, with particular molecules being suited for specific density ranges.  For instance, chemical models show that N$_2$H$^+$ is an excellent tracer of dense gas in pre-stellar objects \citep{bl97}.  In support of these models, observations of low-mass dense cores \citep{tafalla_dep, Bergin2002} demonstrate that N$_2$H$^+$ highlights regions of high central density (n$\sim$10$^6$~cm$^{-3}$), while CO readily freezes out onto cold grains (when n~$>~10^4$~cm$^{-3}$), rendering it undetectable in the central denser regions of the cores.  CO is a major destroyer of N$_2$H$^+$, and its freeze-out leads to the rapid rise in N$_2$H$^+$ abundance in cold gas.  When a star is born, the CO evaporates from grains and N$_2$H$^+$ is destroyed in the proximate gas \citep{lbe04}.  Thus, N$_2$H$^+$ is a preferential tracer of the densest gas that has not yet collapsed to form a star in low-mass pre-stellar cores.

While N$_2$H$^+$ has been used extensively as a probe of the innermost regions of local cores, where densities can reach 10$^6$cm$^{-3}$ \citep[e.g.][]{taf04}, this chemical sequence has not yet been observationally proven in more massive star forming regions.  Nonetheless recent surveys \citep[e.g.][]{Sakai2008, ragan_msxsurv} confirm that N$_2$H$^+$ is prevalent in IRDCs, and mapping by \citet{ragan_msxsurv} shows that N$_2$H$^+$ more closely follows the absorbing gas than CS or C$^{18}$O, which affirms that the density is sufficient for appreciable N$_2$H$^+$ emission.  These single dish surveys do not have sufficient resolution to confirm the tracer's reliability on the clump or pre-stellar core scales in IRDCs.  Interferometric observations will be needed to validate N$_2$H$^+$ as a probe of the chemistry and dynamics of individual clumps (Ragan et al., in prep.).  

For one of the objects in our sample, G012.50$-$0.22, we had previous BIMA observations of N$_2$H$^+$ emission with 8$'' \times 4.8''$ spatial resolution.  The BIMA data were reduced using the standard MIRIAD pipeline reduction methods \citep{MIRIAD}.  As in nearby clouds, such as \citet{wmb04}, the integrated intensity of N$_2$H$^+$ relates directly to the dust (measured here in absorption) in this infrared-dark cloud.  Figure~\ref{fig:g1250_bimaplot} illustrates the quality of N$_2$H$^+$ as a tracer of dense gas, both in the N$_2$H$^+$ contours plotted over the 8~$\mu$m {\em Spitzer} image and the point-to-point correlation between the 8~$\mu$m optical depth and the integrated intensity of N$_2$H$^+$.  The points that lie above the average line, with high integrated intensities but low optical depth, are all in the vicinity of a foreground star in the 8~$\mu$m image, which lowers our estimate for optical depth.  In the sample, however, we have shown that the foreground and young stellar population is largely unassociated with the absorption.  

Two trends are apparent in Figure~\ref{fig:g1250_bimaplot}.  First, below $\tau~<~0.25$ there is a lack of N$_2$H$^+$ emission.  This suggests that the absorption may be picking up a contribution from a lower density extended envelope that is incapable of producing significant N$_2$H$^+$ emission.  This issue is discussed in greater detail in $\S$\ref{envelope}.  Alternatively, the interferometer may filter out extended N$_2$H$^+$ emission.  The second trend evident in Figure~\ref{fig:g1250_bimaplot} is that for $\tau~>~0.25$, there is an excellent overall correlation, confirming that mid-infrared absorption in clouds at distances of 2 to 5~kpc is indeed tracing the column density of the {\em dense} gas likely dominated by pre-stellar clumps.

In addition to directly tracing the dense gas in IRDCs, molecular observations can be brought to bear on critical questions regarding the use of absorption against the Galactic mid-infrared background and how best to calibrate the level of foreground emission.  One way to approach this is to use the molecular emission as a tracer of the total core mass and compare this to the total mass estimated from 8~$\mu$m absorption with differing assumptions regarding the contributions of foreground and background (see $\S$~\ref{bg}).  In \citet{ragan_msxsurv} we demonstrated that the distribution of N$_2$H$^+$ emission closely matches that of the mid-infrared absorption (see also $\S$~\ref{clumps}).  This is similar to the close similarity of N$_2$H$^+$ and dust continuum emission in local pre-stellar cores \citep[e.g.][]{BerginTafalla_ARAA2007}.  Thus we can use the mass estimated from the rotational emission of N$_2$H$^+$ to set limits on viable models of the foreground.  In \citet{ragan_msxsurv} we directly computed a mass using an N$_2$H$^+$ abundance assuming local thermodynamic equilibrium (LTE) and using the H$_2$ column density derived from the MSX 8~$\mu$m optical depth.  However, this estimate is highly uncertain as the optical depth was derived assuming no foreground emission, and the N$_2$H$^+$ emission may not be in LTE.  Instead, here, we will use chemical theory and observations of clouds to set limits.

N$_2$H$^+$ appears strong in emission in dense pre-stellar gas due to the freeze-out of CO, its primary destruction route.  Detailed theoretical models of this process in gas with densities in excess of 10$^5$~cm$^{-3}$ \citep{aikawa_be}, as expected for IRDCs, suggest a typical abundance should be $\sim$10$^{-10}$ with respect to H$_2$ \citep{maret_n2, aikawa_be, Pagani2007}.  This value is consistent with that measured in dense gas in several starless cores \citep{tafalla_dep, maret_n2}.  Using this value we now have a rough test of our foreground and background estimates.  For example, in G024.05$-$0.22 we find a total mass of 4100~$\msun$ (foreground estimation method A).  Using the data in \citet{ragan_msxsurv}, we find that the total mass traced by N$_2$H$^+$ is 4400~$\msun$, providing support for our assumptions.  Figure~\ref{fig:n2hpclumps} shows the relationship between the total clump mass derived from absorption and the total mass derived from our low-resolution maps of N$_2$H$^+$ for the eight IRDCs in our sample that were detected in N$_2$H$^+$.  In general, there is good agreement.  We plot a 30\% systematic error in the total clump masses (abscissa) and a factor of 5 in for the total N$_2$H$^+$ mass estimate (ordinate).  In the cases where the estimates differ, the N$_2$H$^+$ mass estimate tends to be greater than the total mass derived from the dust absorption clumps.  This discrepancy likely arises in large part from an under-estimation of N$_2$H$^+$ abundance and/or non-LTE conditions.  All the same, the consistency of the mass estimates, together with the morphological correspondence, reaffirms that the we are probing the dense clumps in IRDCs and that our mass probe is reasonably calibrated.

We find no discernible difference between methods A and B of foreground estimation.  However, we note that both are substantially better than assuming no foreground contribution.  We therefore believe that method A is an appropriate estimate of the foreground contribution (see $\S$\ref{bg}).  

\subsubsection{Effects of Distance on Sensitivity}
\label{sens}

\begin{figure}
\begin{center}
\includegraphics[angle=90,scale=0.4]{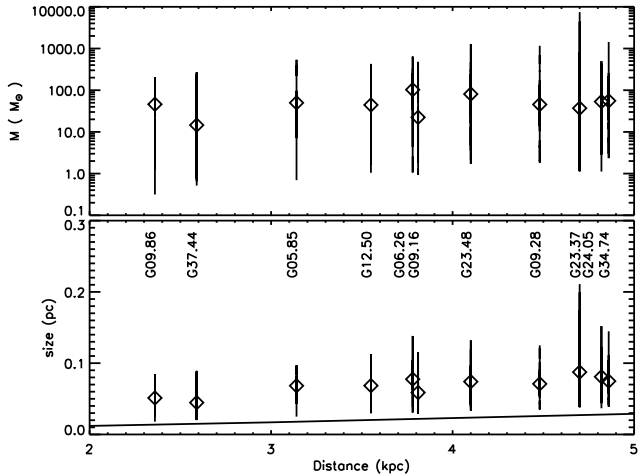}
\end{center}
\caption{\footnotesize{Top: The range in clump mass as a function of distance.  The median clump mass for each IRDC in the sample is indicated with a diamond.  Bottom: The range in clump size as a function of distance.  The median clump size for each IRDC in the sample is indicated with a diamond.  The resolution limit is plotted as a solid line, and it shows the boundary at which {\tt clumpfind} defines a ``clump'' for an object at the distance of the indicated host IRDC.}}
\vspace{0.2in}
\label{fig:masssens}
\end{figure}

Infrared-dark clouds are much more distant than the local, well-studied clouds such as Taurus or $\rho$ Ophiuchus.  As such, a clear concern is that the distance to IRDCs may preclude a well-defined census of the clump population.  The most likely way in which the our survey is incomplete is the under-representation of low-mass objects due to their relatively small size, blending of clumps along the line of sight, or insensitivity to their absorption against the background.  One observable consequence of this effect, assuming IRDCs are a structurally homogeneous class of objects, might be that more distant IRDCs should exhibit a greater number of massive clumps at the expense of the combination of multiple smaller clumps.  Another possible effect is the greater the distance to the IRDC, the less sensitive we become to small clumps, and clumps should appear to blend together (i.e. neighboring clumps will appear as one giant clump).  Due to this effect, we expect that the most massive clumps of the population will be over-represented.  As a test, we examine the distribution of masses and sizes of clumps as a function of IRDC distance, which is shown in Figure~\ref{fig:masssens}.  This sample, with IRDCs ranging in distance from 2.4 to 4.9~kpc away, does not show a strong trend of this nature.  We show the detection limit for clumps to illustrate the very good sensitivity of this technique and that while it does impose a lower boundary on clump detectability, most clumps are not close to this value.  We found no strong dependence of clump mass or size on the distance to the IRDC and conclude that blending of clumps does not have a great effect on the mass sensitivity.

Typical low-mass star forming cores range in size from 0.03 to 0.1~pc \citep{BerginTafalla_ARAA2007}.  If one were to observe such objects 4~kpc, they would only subtend a few arcseconds.  For example, if L1544, a prototypical pre-stellar core, resided at the typical distance to the IRDCs in the sample, it would show sufficient absorption \citep[based on reported column density measurements by][]{bacmann_iso} against the Galactic background, but according to \citet{Williams_2006}, would subtend 3$''$ in diameter at our fiducial 4~kpc distance, which is very close to our detection limit.  In addition, very low mass clumps could blended into any extended low-density material that is included in our absorption measurement.  These effects should limit our sensitivity to the very low-mass end of our clump mass function. 

To first order, we have shown distance is not a major factor because the high-resolution offered by {\em Spitzer} improves our sensitivity to small structures.  However, infrared-dark clouds are forming star clusters and by nature are highly structured and clustered.  As such, we can not rule out significant line-of-sight structure.  Since independent clumps along the line-of-sight might have distinct characteristic velocities, the addition of kinematical information from high-resolution molecular data (Ragan et al., in prep.) will help the disentanglement.

\section{Mass Function}
\label{mf}

A primary goal of this study is to explore the mass function of clumps in infrared-dark clouds and compare it to that of massive star formation regions, local star formation regions, and the stellar IMF.  We note that there is some ambiguity in the literature about the ``clump'' versus the ``core'' mass functions.  In the following description, a ``core'' mass function refers to the mass spectrum objects with masses in the ``core'' regime (10$^{-1}$-10$^1$~$\msun$, 10$^{-2}$-10$^{-1}$~pc), and a ``clump'' mass function for objects in the ``clump'' regime (10$^{-1}$-10$^1$~$\msun$, 10$^{-2}$-10$^{-1}$~pc), as summarized in \citet{BerginTafalla_ARAA2007}.  Here we present the infrared-dark cloud clump mass function.  We describe the relevance of this result in the context of Galactic star formation and discuss several methods we use to test its validity.

\subsection{Mass Function in Context}
\label{context}

A fundamental property of the star formation process is the mass spectrum of stars, and, more recently, the mass function of pre-stellar objects.  The mass spectrum in either case is most typically characterized by a power law, taking the form $dN/dM \propto M^{-\alpha}$, known as the differential mass function (DMF).  In other contexts, the mass function can be described as a function of the logarithm of mass, which is conventionally presented as $dN/d(log~m) \propto M^{\Gamma}$, in which case $\Gamma=-(\alpha-1)$.  In the results that follow, we present the slope of the clump mass function in terms of $\alpha$.

A commonly used method for studying mass functions of pre-stellar cores is observation of dust thermal continuum emission in nearby star-forming clouds.  Cold dust emission is optically thin at millimeter and sub-millimeter wavelengths, and can therefore be used as a direct tracer of mass.  A number of surveys of local clouds \citep[e.g.][]{Johnstone_Oph, motte_rhooph} have been performed with single-dish telescopes covering large regions in an effort to get a complete picture of the mass distribution of low-mass clouds.  This is an extremely powerful technique, but as \citet{Goodman_col} demonstrate, this technique suffers from some limitations, chief among them poor spatial resolution (in single-dish studies), required knowledge of dust temperatures \citep{Pavlyuchenkov_2007}, and the insensitivity to diffuse extended structures.  

Another technique that has been employed to map dust employs near-infrared extinction mapping \citep{alves_cmfimf, Lombardi_pipe}, which is a way of measuring $A_V$ due to dark clouds by probing the color excesses of background stars \citep{Lombardi_NICER}.  This method is restricted to nearby regions of the Galaxy because of sensitivity limitations and the intervention of foreground stars, both of which worsen with greater distance. Also, the dynamic range of $A_V$ in such studies is limited to $\sim$1-60 \citep{Lombardi_NICER}, while our technique probes from $A_V$ of a few to $\sim$100.

The dust-probing methods mentioned above, both thermal emission from the grains and extinction measures using background stars, often find a core mass function (CMF) that is similar in shape to the stellar initial mass function (IMF), as described by \citet{Salpeter_imf}, where $\alpha=2.35$ ($\Gamma=-1.35$), or \citet{Kroupa_imf}.  This potentially suggests a one-to-one mapping between the CMF and IMF, perhaps scaled by a constant ``efficiency'' factor \citep[e.g.][]{alves_cmfimf}.  Also, both techniques are difficult to apply to regions such as infrared-dark clouds due to their much greater distance.  As we show in $\S$\ref{structure}, absorbing structure exists below the spatial resolution limit of single-dish surveys.  Sensitivity limitations and foreground contamination preclude use of extinction mapping to probe IRDCs.

Structural analysis using emission from CO isotopologues find a somewhat different character to the distribution of mass in molecular clouds.  \citet{Kramer_CMF} determined that the clump mass function in molecular clouds follows a power law with $\alpha$ between 1.4 and 1.8 ($-0.8 < \Gamma < -0.4 $).  This is significantly shallower than the Salpeter-like slope for clumps found in works using dust as a mass probe.  This disagreement may be due to an erroneous assumption about one technique or the other, or it may be that the techniques are finding information about how the fragmentation process takes place from large scale, probed by CO, to small scales, probed by dust.  Another possible explanation is that most of the objects in  \citet{Kramer_CMF} are massive star forming regions, and star formation in these regions may be intrinsically different than tyical regions studied in the local neighborhood (e.g. Taurus, Serpens).

Sub-millimeter observations of more distant, massive star-formation regions have been undertaken \citep[e.g.][]{reid_2, Li_orion, mookerjea2004, rathborne2006} with a mixture of results regarding the mass function shape.  \citet{rathborne2006}, for example, performed IRAM observations of a large sample of infrared-dark clouds.  Each cloud in that sample is comprised of anywhere from 2 to 18 cores with masses ranging from 8 to 2000$\msun$.  They find a Salpeter-like ($\alpha$~$\sim$2.35) mass function for IRDC cores.  However, our Spitzer observations reveal significant structure below the spatial resolution scales of \citet{rathborne2006}.  As we will show (see Section~\ref{mf}), the mass function within a fragmenting IRDC is shallower than Salpeter and closer to the mass function derived from CO emission.

Given the strong evidence for fragmentation, it is clear that IRDCs are the precursors to massive clusters.  We then naturally draw comparisons between the characteristics of fragmenting IRDCs and the nearest region forming massive stars, Orion.  At $\sim$500~pc, it is possible to resolve what are likely to be pre-stellar objects in Orion individually with current observational capabilities.  With the high-resolution of our study, we can examine star formation regions (IRDCs) at a similar level of detail as single-dish telescopes can survey Orion.  For example, we detect structures on the same size scale ($\sim$0.03~pc) as the quiescent cores found by \citet{Li_orion} in the Orion Molecular Cloud, however the most massive core in their study is $\sim$50~$\msun$.  These cores account for only a small fraction of the total mass in Orion.

\begin{figure}
\begin{center}
\includegraphics[scale=0.3]{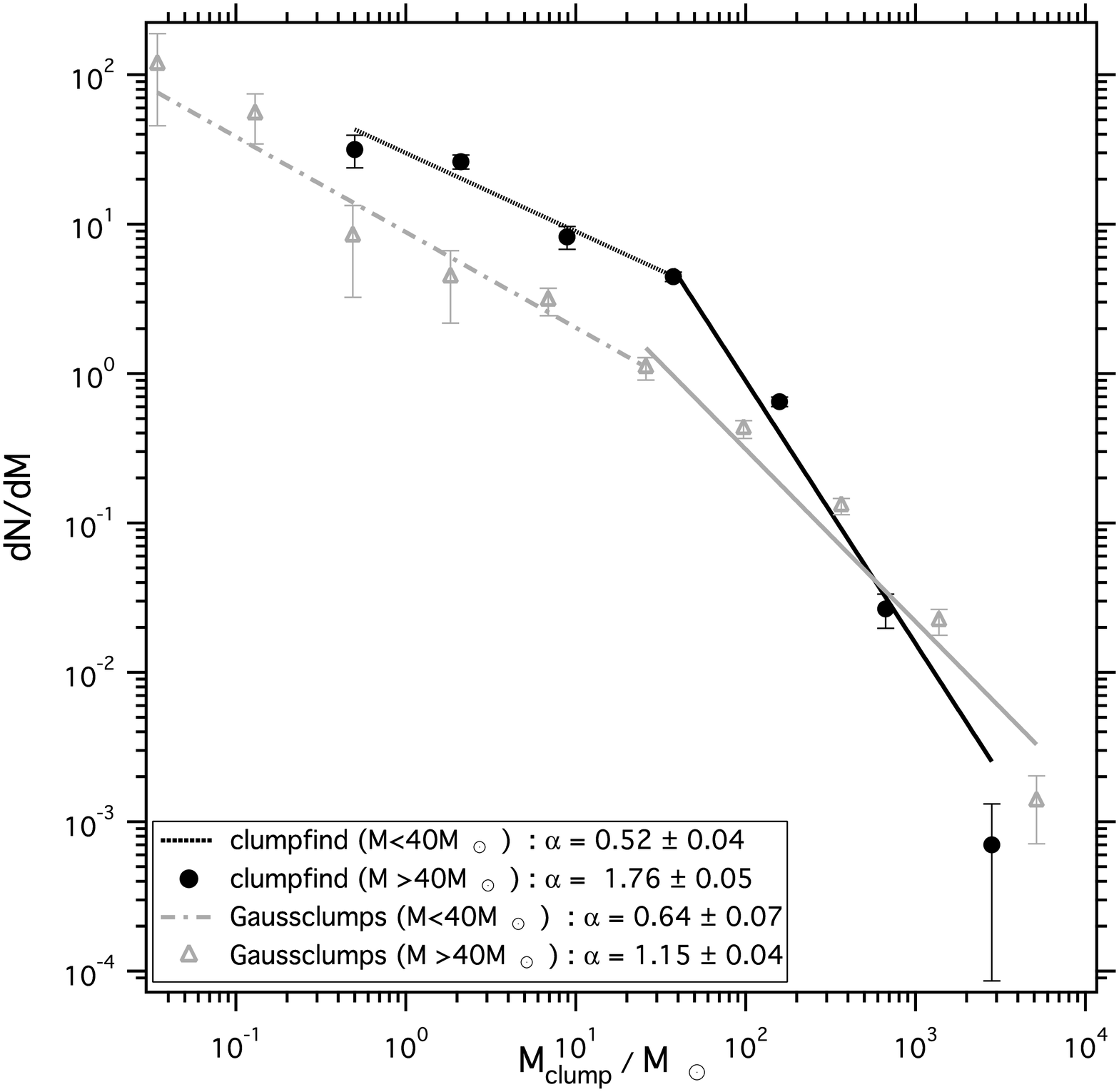}
\end{center}
\caption{\footnotesize{
Differential mass function of ensemble IRDC sample.  Black filled circles indicate results of the {\tt clumpfind} technique, and the green open triangles denote the results of the {\tt gaussclumps} clump-finding method.  The fits are broken power laws.  On the high-mass end, the slope of the {\tt gaussclumps} method mass function ($\alpha=1.15\pm0.04$) is shallower than the slope of hte {\tt clumpfind} mass function ($\alpha=1.76\pm0.05$).}}
\label{fig:cfgccompareMF}
\end{figure}

\subsection{Results: Differential Mass Function}
\label{dmf}

We use the IRDC clump masses calculated in $\S$\ref{structure} (using {\tt clumpfind} and foreground estimation method A) to construct an ensemble mass function in Figure~\ref{fig:cfgccompareMF}.  The mass function that results from using foreground estimation method B is shifted to lower masses by a factor of 2, but the shape is identical.  Because IRDCs appear to be in a roughly uniform evolutionary state over the sample (i.e. they are all likely associated with the Molecular Ring, and they possess similar densities and temperatures), we merge all the clumps listed in Table~6 as ensemble and present a single mass function for all the objects at a range of distances.  This assumes that the character of the mass function is independent of the distance to a given IRDC.  Recall that we see no evidence (see Figure~\ref{fig:masssens}) for the mass distributions to vary significantly with distance.

For the calculation of the errors in the DMF we have separately accounted for the error in the mass calculation and the counting statistics.  We used a method motivated by \citet{reid_1} to calculate the mass error.   We have assumed that the clump mass error is dominated by the systematic uncertainty of 30\% in the optical depth to mass correction.  For each clump we have randomly sampled a Guassian probability function within the 1$\sigma$ envelope defined by the percentage error.  With these new clump masses we have re-determined the differential mass function.   This process is repeated 10$^4$ times, and the standard deviation of the DMF induced by the error in the mass is calculated from the original DMF.  This error is added in quadrature to the error introduced by counting statistics.  The provided errors are 1$\sigma$, with the caveat that the value assumed for the systematic uncertainty is open to debate.  As a result, when there are large numbers in a given mass bin, the error is dominated by the mass uncertainty.  Conversely, when there are few objects in a mass bin, the error is dominated by counting.

The IRDC clump mass function for this sample spans nearly four orders of magnitude in mass.  We fit the mass function with a broken power law weighted by the uncertainties.  At masses greater than $\sim$40$\msun$, the mass function is fit with a power law of slope $\alpha$=1.76$\pm$0.05.  Below $\sim$40$\msun$, the slope becomes much shallower, $\alpha$=0.52$\pm$0.04.  We also include in Figure~\ref{fig:cfgccompareMF} the mass function of clumps found with the {\tt gaussclumps} algorithm, with errors calculated in the identical fashion.  Performing fits in the equivalent mass regimes results in a shallower slope for masses greater than 40$\msun$ ($\alpha$=1.15$\pm$0.04), while the behavior at low masses is similar.  As discussed in $\S$\ref{structure}, the clumps found with {\tt clumpfind} and {\tt gaussclumps} are in good agreement in the central region of each IRDC, but tend to disagree on the outskirts.  This is a consequence of the failure of {\tt gaussclumps} to model the varying background.  Examination of the images reveals that the contribution of the diffuse material varies across the image, thereby setting the background level too high for outer clumps (where the envelope contributes less) to be detected.  In fact, these clumps appear to be preferentially in the 30 to 500$\msun$ range, and a mass function constructed with the {\tt gaussclumps} result is significantly shallower than derived with {\tt clumpfind} (see Figure~\ref{fig:cfgccompareMF}).  We conclude that {\tt gaussclumps} is not suitable to identify structure away from the central region of the IRDC where the envelope level is below the central level.  This is further supported by the wavelet analysis which is capable of accounting for a variable envelope contribution.  It is worth noting that for the one IRDC for which we have the wavelet analysis, that the slope of the derived mass function shows little appreciable change and agrees with the {\tt clumpfind} result.   

\begin{figure}
\begin{center}
\includegraphics[scale=0.3]{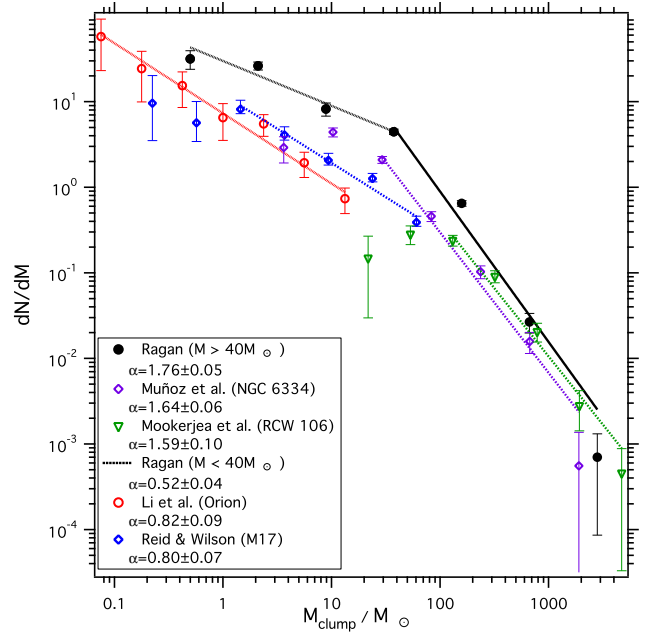}
\end{center}
\caption{
\footnotesize{ 
Differential mass function of this IRDC sample (black filled circles) fit with a single power-law for M$_{clump}>30\msun$ ($\alpha = 1.76\pm0.05$) compared with various star formation regions in the high mass regime and their respective single power-law fit slopes.  At the high mass end, our fit agrees well with that of other studies: {\it Open purple diamonds} from \citet{munoz_ngc6334} ($\alpha = 1.64\pm0.06$); {\it Open green inverted triangles} from \citet{mookerjea2004} ($\alpha = 1.59\pm0.10$).  At the low mass end, we fit a second power law for the bins with M$_{clump}<30\msun$ ($\alpha = 0.52\pm0.04$), which agrees well with other studies in this mass regime: {\it Open blue diamonds} from \citet{reid_2} ($\alpha = 0.80\pm0.07$); {\it Open red circles} from quiescent Orion cores from \citet{Li_orion}($\alpha = 0.82\pm0.09$).  Note that only this study spans the entire range of masses, so the reality of the apparent break at $\sim$30$\msun$ is in question. }}
\label{fig:literature_MF}
\end{figure}

To put the mass function into context with known Galactic star formation, we plot the clump mass function of all clumps in our sample in Figure~\ref{fig:literature_MF} along with the core/clump mass function of a number of other studies probing various mass ranges.  We select four studies, each probing massive star forming regions at different wavelengths and resolutions including quiescent cores in Orion \citep{Li_orion}, clumps in M17 \citep{reid_2}, clumps in RCW 106 \citep{mookerjea2004}, and clumps in massive star formation region NGC 6334 \citep{munoz_ngc6334}.  In their papers, each author presents the mass function in a different way, making it difficult to compare the results directly to one another.  Here, we recompute the mass function for the published masses in each work uniformly (including the treatment of errors, see above).  Each of the mass functions is fit with a power law.  Figure~\ref{fig:literature_MF} highlights the uniqueness of our study in that it spans over a much larger range in masses than any other study to date.  

At the high-mass end, the mass function agrees well with the \citet{mookerjea2004} and \citet{munoz_ngc6334} studies, which probed to lower mass limits of 30$\msun$ and 4$\msun$, respectively.  The fall-off from the steep slope at the high mass end to a shallower slope at the low mass end immediately suggests that completeness, enhanced contribution from the envelope and/or clump blending become an issue.  However, the slope at the low mass end compares favorably with \citet{Li_orion} and \citet{reid_2} which probe mass ranges 0.1 to 46$\msun$ and 0.3 to 200$\msun$, respectively.  In addition to the general DMF shape at both the high mass and low mass end, the ``break'' in the mass function falls in the 10$\msun$ to 50$\msun$ range for the ensemble of studies, including ours.  If this is a real feature of the evolving mass spectrum, this can shed some light on the progression of the fragmentation process from large, massive objects to the numerous low-mass objects like we see in the local neighborhood.  The characteristic ``break'' mass can also be a superficial artifact of differences in binning, mass determination technique, and observational sensitivity.  Our study is the only one that spans both mass regimes, and further such work is needed to explore the authenticity of this feature.  However, in $\S$\ref{conclusion} we speculate that this may be an intrinsic feature. 

It is possible that the slope of the IRDC clump mass function might be an artifact of a limitation in our technique.  With the great distances to these clouds, one would expect the effect of clump blending to play a role in the shape of their mass spectrum.  We have shown in $\S$~\ref{sens} that distance does not dramatically hinder the detection of small clumps.  Our study samples infrared-dark clouds from 2.4~kpc to 4.9~kpc, and we find that the number of clumps does not decrease with greater distance, nor does the median mass tend to be be significantly greater with distance.  Furthermore, with the present analysis, we see no evidence that including clumps from IRDCs at various distances affects the shape of the mass function.

From past studies of local clouds there has been a disparity between mass function slope derived with dust emission and CO \citep[e.g. compare][]{Johnstone_OrionB, Kramer_CMF}.  Our result suggests that massive star forming regions have mass functions with slope in good agreement with CO isotopologues, e.g. $\alpha$=1.8.  This is crucial because CO observations contain velocity information, which allow for the clumps to be decomposed along the line-of-sight.  Still, the authors find a shallow slope in agreement with ours.  We conclude that clump blending, while unavoidable to some extent, does not skew the shape of the mass function as derived from dust emission or absorption.  A close look at \citet{Kramer_CMF} results finds that the majority of objects studied are massive star formation regions.  Given the general agreement of the clump mass function of this sample of IRDCs with other studies of massive star formation regions, we believe this result represents the true character of these objects, not an artifact of the observing technique.  

Several studies of pre-stellar cores in the local neighborhood show a mass distribution that mimics the shape of the stellar IMF.  That the slope of the mass function in infrared-dark clouds is considerably shallower than the stellar IMF should not be surprising.  The masses we estimate for these clumps are unlikely to give rise to single stars.  Instead, the clumps themselves must fragment further and eventually form a star cluster, likely containing multiple massive stars.  Unlike Orion A, for example, which contains $\sim$10$^4$~$\msun$ distributed over a 380 square parsec (6.2 square degrees at 450~pc) region \citep{Carpenter2000}, in IRDCs, a similar amount of mass is concentrated in clumps extending only a 1.5 square parsec area.  Therefore, we posit that IRDCs are not distant analogues to Orion, but more compact complexes capable of star formation on a more massive scale.

Given the high masses estimated for infrared-dark clouds, yet the lack evidence for the massive stars they must form is perhaps indicating that we see them necessarily {\em because} we are capturing them just before the onset of star formation.  Such a selection effect would mean that we preferentially observe these dark objects because massive stars have yet to disrupt their natal cloud drastically in the process of protostar formation.  

\begin{figure*}
\begin{center}
\hbox{
\vspace{1.0cm}
\hspace{-1cm}
\psfig{figure=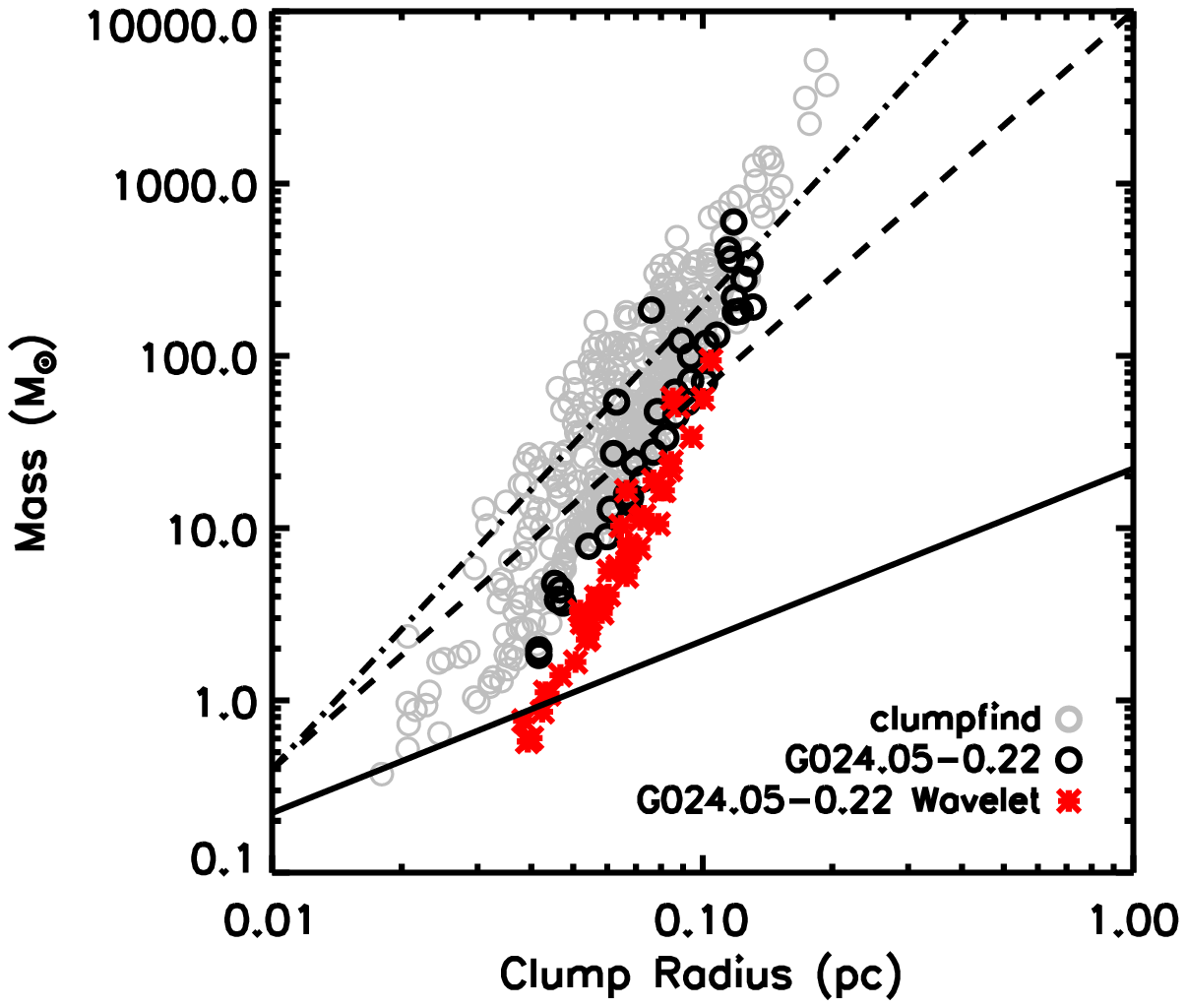,height=9cm}
\hspace{-4cm}
\psfig{figure=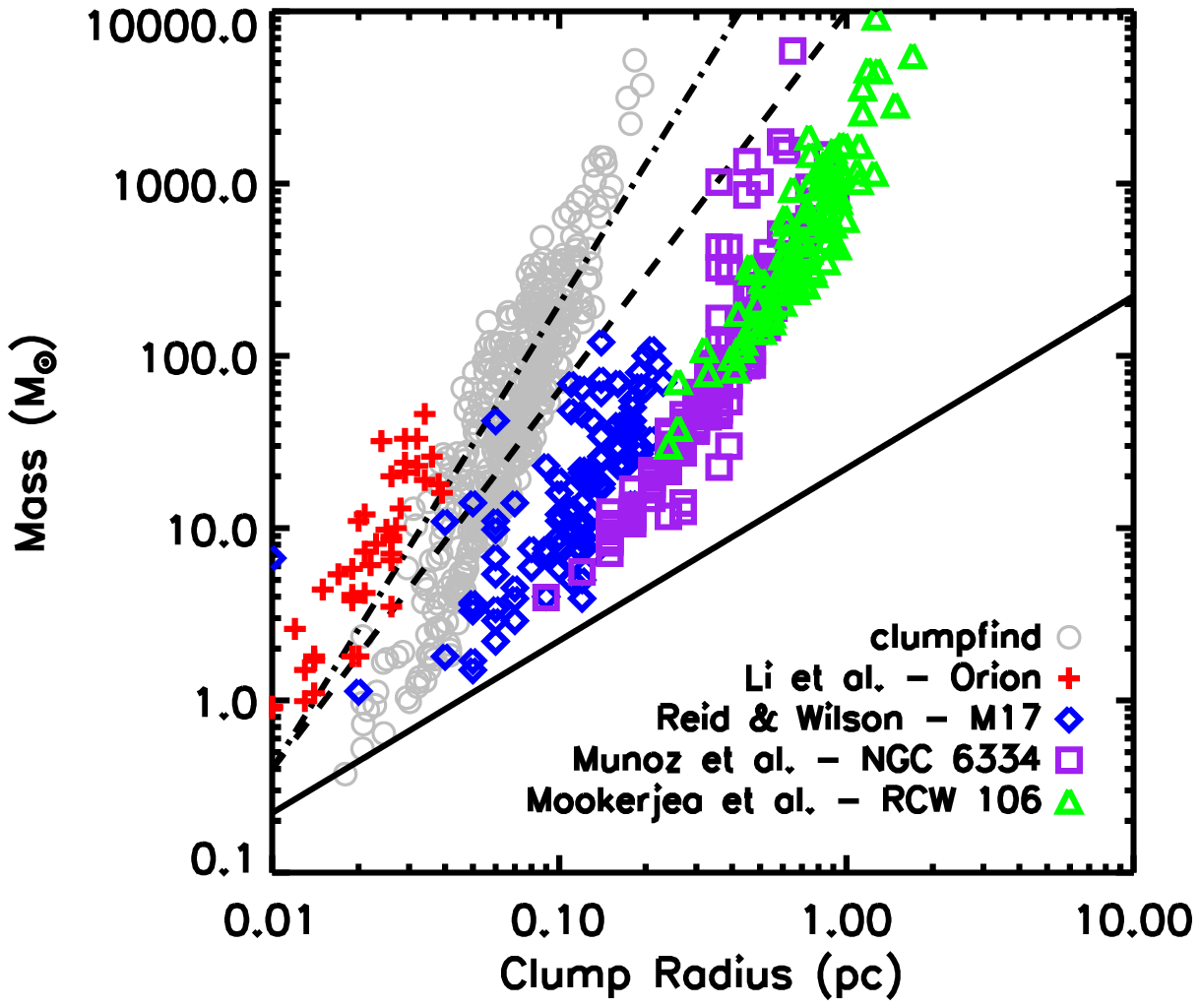,height=9cm}
}
\vspace{-1.0cm}
\end{center}
\caption{\footnotesize{Left: The mass-radius relationship for {\tt clumpfind} clumps (foreground method A) in the entire sample of IRDCs (gray), with the clumps found only in G024.05$-$0.22 highlighted in black, and the clumps found in the wavelet subtracted image (red).  The solid line denotes the critical Bonnor-Ebert mass-radius relation for T$_{internal}$=15~K.  The dashed line is the M~$\propto$~R$^{2.2}$ from the \citet{Kramer1996} CO multi-line study of Orion.  The dash-dotted line is taken from \citet{williams_clumpfind}, which finds M~$\propto$~R$^{2.7}$.  Right: The mass-radius relationship for IRDC clumps, including a comparison to all the studies of massive star forming regions included in Figure~\ref{fig:literature_MF}.}}
\label{fig:massrad}
\end{figure*}

\subsection{The Contribution from the IRDC Envelope}
\label{envelope}

Like nearby clouds, infrared-dark clouds are structured hierarchically, consisting of dense condensations embedded in a more diffuse envelope.  Here we present various attempts to estimate the fraction of the total cloud mass resides in dense clumps compared to the extended clouds.  First, we use archival $^{13}$CO data to probe the diffuse gas and use it to estimate the envelope mass.  To further explore the contribution of the envelope, we demonstrate that a wavelet analysis, a technique designed to remove extended structures from emission maps, gives a similar relationship between envelope and dense clump mass.  Alternatively, applying the {\tt gaussclumps} algorithm to the data provides an average threshold that describes the diffuse structure.  

We use $^{13}$CO~(1-0) molecular line data from the Galactic Ring Survey \citep{Jackson_GRS} in the area covered by our {\em Spitzer} observations of G024.05$-$0.22 to probe the diffuse material in the field.  The $^{13}$CO emission is widespread, covering the entire area in the IRAC field, thus we are not probing the entire cloud.  Assuming local thermodynamic equilibrium (LTE) at a temperature of 15~K and a $^{13}$CO abundance relative to H$_2$ of $4\times10^{-6}$ \citep{Goldsmith_Taurus}, we find that the the clump mass is $\sim$20\% of the total cloud mass. This demonstrates that IRDCs are the densest regions of much larger molecular cloud complexes;  however, the fraction of mass that we estimate the clumps comprise relative to the cloud is an upper limit because the full extent of the cloud is not probed with these data.

In $\S$\ref{masserr}, we discuss two ways in which we account for the envelope in the clump-finding process.  First, the {\tt gaussclumps} algorithm is an alternative method of identifying clumps, and in $\S$\ref{dmf} we examine the effect this method has on the clump mass function.  The algorithm is insensitive to clumps on the outskirts of the IRDC, thereby flattening the mass function.  While {\tt gaussclumps} may oversimplify the structure of the envelope for the purposes of identifying clumps, it does provide a envelope {\em threshold}, above which optical depth peaks fit as clumps and below which emission is subtracted.  This threshold approximates the level of the envelope, and as a result, {\tt gaussclumps} finds 15-50\% of the optical depth level is from the diffuse envelope.  The wavelet subtraction technique results in clumps that are on average 90\% less massive and smaller in size by 25\% ($\sim$0.02~pc) than those extracted from the unaltered map.

These analyses of the IRDC envelope show us that our technique is only sampling 20-40\% of the clouds total mass and, at the same time, the clump masses themselves include a contribution from the surrounding envelope.  Because of these factors, the different methods for isolating ``clumps'' have varying levels of success.  For example, using {\tt gaussclumps} equips us to parametrically remove the envelope component to the clump, but due to the underlying assumption of the baseline level, it misses many clumps that {\tt clumpfind} identifies successfully.  The mass function that results from using the {\tt gaussclumps} method is shallower than that from {\tt clumpfind}, as {\tt gaussclumps} fails to find clumps on the periphery of the dominant (often central) concentration of clumps, where the envelope level is lower.  

While both the {\tt clumpfind} and {\tt gaussclumps} methods have their drawbacks, it is clear that IRDCs have significant structure on a large range of scales.  The relatively shallow mass function for IRDC clumps and other massive star forming regions shows that there is a great deal of mass in large objects, and future work is needed to understand the detailed relationship between the dense clumps and their surroundings.

\section{Mass-Radius Relation}

Next we investigate the relationship between the mass and size of the clumps found in IRDCs, which informs us of the overall stability of the clump structures.  Figure~\ref{fig:massrad} shows the mass-radius relationship of the {\tt clumpfind}-identified clumps, highlighting the results for G024.05$-$0.22 and the wavelet-subtracted case.  Indeed, the clumps extracted from the wavelet-subtracted map are shifted down in mass by 90\% and down in size by 25\%, but the relationship between the quantities does not change.  We plot the relation of simple self-gravitating Bonnor-Ebert spheres ($M~(R)~=~2.4 R a^2 / G$, where $a$ is the sound speed and set to 0.2~km~s$^{-1}$, solid line) and also the mass-radius relationship observed in a multi-line CO survey of Orion \citep[M~$\propto$~R$^{2.2}$,][(dashed line)]{Kramer1996}.  For comparison, Figure~\ref{fig:massrad} also shows these properties from the other studies of massive star formation regions.  We note that the spatial resolution of the comparison studies is larger than the resolution of this study.  The relationship for Orion \citep{Li_orion}, M17 \citep{reid_2}, NGC 6334 \citep{munoz_ngc6334} and RCW 106 \citep{mookerjea2004} all agree with the \citet{Kramer1996} relationship, which is consistent with the mass function agreement to CO studies (see $\S$\ref{dmf}).  

The IRDC clumps are clearly gravitationally unstable, showing higher densities than their local Bonnor-Ebert sphere counterparts.  The relationship for clumps in IRDCs shows a steeper trend, one closer to the \citet{williams_clumpfind} relationship, $M~\propto~R^{2.7}$.  Also, dust extinction at 8~$\mu$m has greater sensitivity to high-densities than CO, which is known to freeze out at extreme densities.  Hence, while the IRDC clumps are clearly Jeans unstable, the slope of the relation may be simply a reflection of the different mass probe used here.

\section{Discussion \& Conclusion} 
\label{conclusion}

The {\em Spitzer Space Telescope} affords us the ability to probe a spatial regime of massive clouds in the Galactic Ring at comparable resolution as has been applied to the numerous studies of local, low-mass star formation.  In this way, we can extend the frontier of detailed star formation studies to include regions the likes of which are not available in the solar neighborhood.  This study demonstrates a powerful method for characterizing infrared-dark clouds, the precursors to massive stars and star clusters.  These objects provide a unique look at the initial conditions of star formation in the Galactic Ring, the dominant mode of star formation in the Galaxy.

We present new {\em Spitzer} IRAC and MIPS 24~$\mu$m photometric measurements supplemented with 2MASS J, H, K$_s$ photometry of the distributed young stellar population observed in the Spitzer fields.  Rigid color criteria are applied to identify candidate young stellar objects that are potentially associated with the infrared-dark clouds.  In all, 308 young stellar objects were identified (see Table~2), seven of which are classified as embedded protostars.  For those objects, we set lower limits on the infrared luminosities.  One IRDC has an IRAS source in the field, which is the best candidate for an associated massive star.  Otherwise, our observations provide no evidence for massive star formation in IRDCs, though sensitivity limitations do not rule out the presence of low mass stars and heavily extincted stars.  Nebulosity at 8 and 24~$\mu$m was detected in four of the fields, but when these regions were correlated with molecular data, they do not appear to be associated with the IRDCs.  On average, 25\% of clumps are in the vicinity of stars and $\sim$10\% are in near YSOs, which are the most likely sources to be associated with the infrared-dark cloud.  Since most of the mass is not associated with any indicator of star formation.  This leads us to conclude that IRDCs are at in earlier stage than, say, the nearest example of massive star formation, the Orion Nebula, and these results are powerful clues to the initial conditions of star cluster formation.  

We detail our method of probing mass in IRDCs using dust absorption as a direct tracer of column density.  We perform the analysis using two different assumptions (methods A and B) for the foreground contribution to the 8~$\mu$m flux.  The IRDC envelope contribution to the  To validate our method in the context of others, we compare and find good agreement between the 8~$\mu$m absorption and other tracers of dust, such as sub-millimeter emission from dust grains measured with SCUBA and N$_2$H$^+$ molecular line emission measured with FCRAO and BIMA.  We show that distance does not play a role in the effectiveness of the technique.  The high resolution {\em Spitzer} observations allows us to probe the absorbing structures in infrared-dark clouds at sub-parsec spatial scales.  We apply the {\tt clumpfind} algorithm to identify independent absorbing structures and use the output to derive the mass and size of the clumps.  Tens of clumps are detected in each IRDC, ranging in mass from 0.5 to a few $\times$ 10$^3~\msun$ with sizes from 0.02 to 0.3~pc in diameter.  We also apply the {\tt gaussclumps} algorithm to identify clumps.  The structures in the central region of the IRDC correspond almost perfectly to the {\tt clumpfind} result, but {\tt gaussclumps} misses clumps on the outskirts because it fails to account for a spatially variable background level.  

The existence of substructure -- from 10$^3$~$\msun$ clumps down to 0.5~$\msun$ ``cores'' -- indicates that IRDCs are undergoing fragmentation and will ultimately form star clusters.  The typical densities (n $>$ 10$^5$ cm$^{-3}$) and temperatures (T~$<$~20 K) of IRDCs are consistent with massive star forming regions, but they lack the stellar content seen in more active massive star formation regions, such as the Orion molecular cloud or W49, for example.  The mass available in the most massive clumps, however, leads us to conclude that IRDCs will eventually form multiple massive stars.  

The IRDC clump mass function, with slope $\alpha = 1.76\pm0.05$ for masses greater than $\sim$40$\msun$, agrees with the mass function we calculate based on data from other studies of massive objects.  The mass function for both IRDCs and these massive clump distributions is shallower than the Salpeter-like core mass function reported in local regions.  In fact, the IRDC clump mass function is more consistent with that found when probing molecular cloud structure using CO line emission ($\alpha = 1.6 - 1.8~$), again supporting the assertion that these objects are at an earlier phase of fragmentation.  At the low-mass end ($M < 40\msun$), we find a much shallower slope, $\alpha = 0.52\pm0.04$, which is somewhat flatter than other studies that cover the same range in masses.  This could be due in part to incomplete sampling of the fields.  Alternatively, the apparent flattening of the clumps mass function around 40~$\msun$ could indicate a transition between objects that will generate clustered star formation and those that give rise to more distributed star formation \citep{AdamsMyers2001}.

IRDC clumps are generally not in thermodynamic equilibrium, but rather are undergoing turbulent fragmentation.  The mass spectrum is consistent with the predictions of gravoturbulent fragmentation of molecular clouds \citep{Klessen2001}.  The dynamic Molecular Ring environment could naturally be conducive for producing concentrated cluster-forming regions.

Just as in all surveys of IRDCs to date, this study is subject to the blending of clumps, which could alter the shape of the mass function to over-represent the most massive clumps at the expense of clumps of all masses and sizes.  To the extent that this sample allows, we find that this does not drastically effect the shape of the mass function.  Other studies of cloud fragmentation that have the advantage of a third dimension of information also find a shallower clump mass function slope \citep{Kramer_CMF}.  We therefore conclude that this result is a true reflection of the structure in IRDCs and nature of massive star formation. 

Infrared-dark clouds are already well-established candidates for the precursors to stellar clusters and exhibit significant structures down to 0.02~pc scales.  The properties of IRDCs provide powerful constraints on the initial conditions of massive and clustered star formation.  We suggest that the mass function is an evolving entity, with infrared-dark clouds marking one of the earliest stages of cluster formation.  The mass distribution is top-heavy, with most of the mass in the largest structures.  As the massive clumps fragment further, the mass function will evolve and become steeper.  The clumps will ultimately fragment to the stellar scale and {\em then} take on the Salpeter core mass function that has been observed so prevalently in local clouds.  For example, following the (mostly) starless IRDC phase of cluster evolution, the mass spectrum will evolve into its steeper form, aligning with the mass function of local embedded clusters \citep{Lada_araa03} or star clusters in the Large and Small Magellanic Clouds \citep{Hunter2003}, with a slope $\alpha\sim$~2.  As fragmentation proceeds on smaller scales, the mass function would take on yet a steeper character observed in core mass functions \citep[e.g.][]{alves_cmfimf} and, ultimately, stars \citep[e.g.][]{Kroupa_imf}.  

\acknowledgements
SR is indebted to Doug Johnstone, Fabian Heitsch, Lori Allen and Lee Hartmann for useful suggestions on this work.  SR and EB thank Darek Lis, Carsten Kramer and Joseph Weingartner for their invaluable assistance in the analysis.  This research was supported by {\em Spitzer} under program ID 3434.  This work was also supported by the National Science Foundation under Grant 0707777.

\bibliographystyle{apj}  
\bibliography{master}  

\begin{center}
\begin{deluxetable}{lccc}
\label{tab:targets}
\tablecolumns{4}
\tablecaption{Spitzer targets}
\tablewidth{9cm}
\tablehead{
\colhead{IRDC} &
\colhead{$\alpha$} &
\colhead{$\delta$} &
\colhead{distance $\tablenotemark{a}$} \\
\colhead{} &
\colhead{(J2000)} &
\colhead{(J2000)} &
\colhead{(kpc)} \\
}
\startdata
G05.85$-$0.23 &
17:59:53 &
$-$24:00:10 &
3.14$^{+0.66}_{-0.76}$
\\
G06.26$-$0.51 &
18:01:50 &
$-$23:47:11 &
3.78$^{+0.59}_{-0.67}$
\\
G09.16$+$0.06 &
18:05:50 &
$-$20:59:12 &
3.8$^{+0.61}_{-0.69}$
\\
G09.28$-$0.15 &
18:06:54 &
$-$20:58:51 &
4.48$^{+0.54}_{-0.61}$
\\
G09.86$-$0.04 &
18:07:40 &
$-$20:25:25 &
2.36$^{+0.78}_{-0.88}$
\\
G12.50$-$0.22 & 
18:13:45 &
$-$18:11:53 &
3.55$^{+0.67}_{-0.75}$
\\
G23.37$-$0.29 & 
18:34:51 &
$-$08:38:58 &
4.70$^{+0.90}_{-0.88}$
\\
G23.48$-$0.53 &
18:35:57 &
$-$08:39:46 &
4.10$^{+0.88}_{-0.90}$
\\
G24.05$-$0.22 &  
18:35:52 & 
$-$08:00:38 &
4.82$^{+0.96}_{-0.90}$
\\
G34.74$-$0.12 &
18:55:14 & 
$+$01:33:42 &
4.86$_{-1.45}$
\\
G37.44$+$0.14 &  
18:59:08 & 
$+$04:03:31 &
2.59$^{+1.47}_{-1.34}$
\\
\enddata
\tablenotetext{a}{Distances calculated from Galactic rotation curve, as presented in \citet{ragan_msxsurv}.}
\end{deluxetable}

\end{center}

\clearpage
\begin{landscape}
\begin{center}
\begin{deluxetable}{lclccccccccccc}
\tabletypesize{\scriptsize}
\tablecolumns{14}
\tablewidth{10.5in}
\tablecaption{Spitzer-identified YSOs: 2MASS, IRAC, and MIPS photometry}
\tablehead{
\colhead{IRDC} &
\colhead{Index} &
\colhead{$\alpha$ (J2000)} &
\colhead{$\delta$ (J2000)} &
\colhead{J} &
\colhead{H} &
\colhead{K$_s$} &
\colhead{[3.6]} &
\colhead{[4.5]} &
\colhead{[5.8]} &
\colhead{[8]} &
\colhead{[24]} &
\colhead{A$_K$} &
\colhead{classification\tablenotemark{a}}  
}
\vspace{2in}
\startdata
G05.85$-$0.23  & & & & & & & & & & & & \\
 &  1 & 17:59:41.27 & -24:03:25.8 & \nodata & \nodata & \nodata & \nodata & 12.13$\pm$0.06 & 10.88$\pm$0.04 &  9.08$\pm$0.02 &  3.58$\pm$0.02 & \nodata & EP \\ 
 &  2 & 17:59:49.14 & -24:03:50.6 & 14.70$\pm$0.04 & 11.27$\pm$0.03 &  9.30$\pm$0.02 &  7.27$\pm$0.01 &  6.63$\pm$0.01 &  6.03$\pm$0.01 &  5.60$\pm$0.01 & \nodata & 2.806 & CII\\ 
 &  3 & 17:59:49.88 & -24:03:44.9 & \nodata & 13.97$\pm$0.06 & 12.91$\pm$0.05 & 11.87$\pm$0.03 & 11.48$\pm$0.04 & 11.41$\pm$0.05 & \nodata & \nodata & 0.963 & CII \\ 
 &  4 & 17:59:51.83 & -24:02:04.2 & 15.41$\pm$0.07 & 11.62$\pm$0.03 &  9.63$\pm$0.02 &  8.37$\pm$0.01 &  7.98$\pm$0.01 &  7.43$\pm$0.01 &  7.30$\pm$0.01 &  5.99$\pm$0.06 & 3.348 & CII \\ 
 &  5 & 17:59:47.68 & -24:01:33.0 & \nodata & \nodata & 13.16$\pm$0.06 & 11.71$\pm$0.02 & 11.02$\pm$0.01 & 10.52$\pm$0.01 & 10.09$\pm$0.05 & \nodata & \nodata & CII \\ 
 &  6 & 17:59:35.96 & -24:00:43.8 & 14.23$\pm$0.05 & 10.82$\pm$0.03 &  8.86$\pm$0.02 &  6.89$\pm$0.01 &  6.49$\pm$0.01 &  6.01$\pm$0.01 &  5.35$\pm$0.01 & \nodata & 2.777 & CII \\ 
 &  7 & 17:59:55.30 & -24:00:39.3 & 15.44$\pm$0.06 & 14.45$\pm$0.05 & 13.63$\pm$0.05 & 12.54$\pm$0.04 & 12.31$\pm$0.04 & 12.28$\pm$0.12 & \nodata & \nodata & \nodata & CII \\ 
 &  8 & 17:59:42.43 & -24:00:29.5 & 10.28$\pm$0.02 &  7.61$\pm$0.03 &  6.28$\pm$0.02 &  5.68$\pm$0.01 &  5.50$\pm$0.01 &  5.14$\pm$0.01 &  4.96$\pm$0.01 &  3.40$\pm$0.01 & 2.176 & CII \\ 
 &  9 & 17:59:46.05 & -24:00:15.1 & \nodata & \nodata & 12.73$\pm$0.04 & 10.25$\pm$0.02 &  9.56$\pm$0.01 &  8.72$\pm$0.01 &  7.74$\pm$0.02 &  4.09$\pm$0.03 & \nodata & CII \\ 
 & 10 & 17:59:48.08 & -24:00:12.6 & 15.30$\pm$0.07 & 11.35$\pm$0.03 &  9.23$\pm$0.02 &  7.28$\pm$0.01 &  6.63$\pm$0.01 &  6.15$\pm$0.01 &  5.57$\pm$0.01 &  4.07$\pm$0.03 & 3.482 & CII \\ 
 & 11 & 18:00:02.83 & -24:00:07.7 & \nodata & \nodata & \nodata & 12.65$\pm$0.05 & 12.33$\pm$0.04 & 11.90$\pm$0.11 & 11.01$\pm$0.18 & \nodata & \nodata & CII \\ 
 & 12 & 17:59:54.07 & -23:59:42.9 & \nodata & 14.39$\pm$0.09 & 11.48$\pm$0.03 &  8.89$\pm$0.01 &  8.48$\pm$0.01 &  7.94$\pm$0.01 &  7.76$\pm$0.02 &  5.85$\pm$0.09 & 4.924 & CII \\ 
 & 13 & 18:00:02.08 & -23:59:41.2 & 15.74$\pm$0.09 & 14.26$\pm$0.06 & 13.29$\pm$0.06 & 12.23$\pm$0.03 & 11.91$\pm$0.04 & 11.67$\pm$0.10 & 11.00$\pm$0.28 & \nodata & 0.627 & CII \\ 
 & 14 & 18:00:02.14 & -23:59:34.7 & \nodata & \nodata & 13.83$\pm$0.05 & 12.38$\pm$0.03 & 12.07$\pm$0.03 & 11.49$\pm$0.07 & 10.59$\pm$0.12 & \nodata & \nodata & CII \\ 
 & 15 & 17:59:58.51 & -23:59:25.3 & 15.52$\pm$0.06 & 14.42$\pm$0.05 & 13.16$\pm$0.04 & 12.01$\pm$0.03 & 11.79$\pm$0.04 & 11.56$\pm$0.07 & 11.75$\pm$0.32 & \nodata & \nodata & CII \\ 
 & 16 & 17:59:57.86 & -23:59:12.0 & 15.35$\pm$0.08 & 11.26$\pm$0.03 &  9.17$\pm$0.03 &  7.54$\pm$0.01 &  7.32$\pm$0.01 &  6.87$\pm$0.01 &  6.74$\pm$0.01 &  4.61$\pm$0.04 & 3.673 & CII \\ 
 & 17 & 18:00:02.31 & -23:58:56.1 & 15.85$\pm$0.09 & 12.86$\pm$0.04 & 11.43$\pm$0.03 & 10.40$\pm$0.01 & 10.22$\pm$0.01 &  9.76$\pm$0.01 &  9.50$\pm$0.03 & \nodata & 2.509 & CII \\ 
 & 18 & 17:59:39.24 & -23:58:31.8 & \nodata & \nodata & 11.99$\pm$0.05 &  9.77$\pm$0.01 &  8.95$\pm$0.01 &  8.29$\pm$0.01 &  7.45$\pm$0.01 &  5.33$\pm$0.04 & \nodata & CII \\ 
 & 19 & 18:00:04.04 & -23:58:04.8 & 14.58$\pm$0.03 & 13.78$\pm$0.05 & 13.35$\pm$0.06 & 12.72$\pm$0.05 & 12.54$\pm$0.04 & 12.57$\pm$0.14 & \nodata & \nodata & 0.059 & CII \\ 
 & 20 & 17:59:51.58 & -23:57:42.7 & 15.20$\pm$0.06 & 13.88$\pm$0.11 & 12.92$\pm$0.09 & 11.73$\pm$0.04 & 11.33$\pm$0.03 & 11.18$\pm$0.05 & 10.63$\pm$0.07 & \nodata & 0.400 & CII \\ 
 & 21 & 17:59:53.59 & -23:57:40.0 & 14.56$\pm$0.03 & 13.67$\pm$0.05 & 13.02$\pm$0.07 & 12.25$\pm$0.06 & 12.02$\pm$0.05 & 12.20$\pm$0.20 & \nodata & \nodata & 0.001 & CII \\ 
 & 22 & 17:59:50.39 & -23:56:59.8 & 14.07$\pm$0.07 & 11.13$\pm$0.05 &  9.23$\pm$0.03 &  7.55$\pm$0.01 &  7.30$\pm$0.01 &  6.71$\pm$0.01 &  6.17$\pm$0.01 & \nodata & 2.102 & CII \\ 
 & 23 & 17:59:46.44 & -23:56:53.6 & 15.26$\pm$0.05 & 11.49$\pm$0.02 &  9.45$\pm$0.02 &  7.67$\pm$0.01 &  7.27$\pm$0.01 &  6.71$\pm$0.01 &  6.30$\pm$0.01 & \nodata & 3.278 & CII \\ 
 & 24 & 17:59:52.22 & -23:59:03.5 & \nodata & 13.83$\pm$0.05 & 12.00$\pm$0.04 & 10.51$\pm$0.01 & 10.13$\pm$0.01 &  9.77$\pm$0.02 &  9.66$\pm$0.06 &  6.90$\pm$0.13 & 2.849 & TD \\ 

\enddata
\tablecomments{Table 2 is published in its entirety in the electronic edition of the Astrophysical Journal.  An excerpt is included here to guide the reader in its content and format.}
\tablenotetext{a}{CI=class I protostar, CII=class II pre-main sequence star, TD=transition disk, EP=embedded protostar}
\end{deluxetable}

\end{center}


\clearpage
\end{landscape}
\clearpage
\begin{center}
\begin{deluxetable}{lccccc}
\tablecolumns{6}
\tablecaption{YSO Summary}
\tablewidth{15cm}
\tablehead{
\colhead{IRDC} &
\colhead{Class I} &
\colhead{Class II} &
\colhead{Transition} &
\colhead{Embedded} & 
\colhead{Total} \\
\colhead{} &
\colhead{Protostars} &
\colhead{PMS stars} &
\colhead{Disks} &
\colhead{Objects} &
\colhead{Number} 
}
\startdata
G05.85$-$0.23 & 0 & 22(2) & 1 & 0 & 23(2) \\
G06.26$-$0.51 & 3(1) & 26(4) & 0 & 0 & 29(5) \\
G09.16$+$0.06 &1 & 12(1) & 0 & 0 & 13(1) \\
G09.28$-$0.15 & 2(1) &15(2) &0 &0& 17(3) \\
G09.86$-$0.04 &5(3) & 21(1) & 3 &2(1) & 31(5)\\
G12.50$-$0.22 & 4(1) &22(1) &1 &1(1) & 28(3) \\
G23.37$-$0.29 & 8 &36(2) &0& 2(1) & 46(3) \\
G23.48$-$0.53 & 5(4) &16 &0& 0 &21(4) \\
G24.05$-$0.22 & 0 &24(4) & 0 &1 &25(4) \\
G34.74$-$0.12 & 4(1) &28(4) & 2(1) &1& 35(6) \\
G37.44$+$0.14 & 5(1) &33(1) &2(1) & 0 &40(3) \\

\hline

Total & 37(12) & 255(22) & 9(2) & 7(3) & 308(39) \\
\enddata
\tablecomments{The number of objects classified as YSOs for each IRDC field.  In parentheses, we indicate the number from each classification that are associated with a ``clump,'' which are determined in Section 4.2 and tabulated in Table 5.}
\end{deluxetable}

\end{center}

\begin{center}
\begin{deluxetable}{lllccccccccc}
\label{startable}
\tabletypesize{\scriptsize}
\tablecolumns{10}
\tablewidth{6.7in}
\tablecaption{Spitzer-identified Embedded Protostars: Flux and Luminosity Estimates}
\tablehead{
\colhead{IRDC} &
\colhead{Index} &
\colhead{$\alpha$} &
\colhead{$\delta$} &
\colhead{3.6$\mu$m} &
\colhead{4.5$\mu$m} &
\colhead{5.8$\mu$m} &
\colhead{8$\mu$m} &
\colhead{24$\mu$m} &
\colhead{L$_{MIR}$} \\
\colhead{} & \colhead{number\tablenotemark{a}} & \colhead{(J2000)} & \colhead{(J2000)} &
\colhead{(mJy)} & \colhead{(mJy)} & \colhead{(mJy)} & \colhead{(mJy)} & \colhead{(mJy)} & \colhead{(L$_{\sun}$)} 
}
\startdata
G09.86$-$0.04 & 6 & 18:07:36.99 & -20:26:03.9 & \nodata &   \nodata & \nodata & \nodata &  7.00$\pm$2.47 & $>$0.05  \\
			& 7 &  18:07:42.12 & -20:23:34.3 & \nodata &  \nodata &  \nodata & \nodata &  43.64$\pm$5.45 & $>$0.33  \\
G12.50$-$0.22 & 5 & 18:13:41.71 & -18:12:29.6 & \nodata &    0.02$\pm$ 0.01 & \nodata &  \nodata & 42.93$\pm$12.81 & $>$2.1  \\
G23.37$-$0.29 & 9 & 18:34:54.12 & -08:38:25.5& \nodata &  \nodata &   \nodata &  \nodata &    33.94$\pm$ 8.18 & $>$1.0  \\
			&10 & 18:35:00.04 & -08:36:57.4  &    0.02$\pm$ 0.01 &   \nodata &   \nodata & \nodata &  18.15$\pm$4.56 & $>$1.5  \\
G24.05$-$0.22 & 1 & 18:35:54.73 & -08:01:30.2 & \nodata &  \nodata &   \nodata &   \nodata & 5.88$\pm$1.27 & $>$0.2  \\
G34.74$-$0.12 & 5 & 18:55:05.20 & +01:34:36.2 & \nodata &  0.02$\pm$ 0.01 &   \nodata & \nodata &    36.13$\pm$4.84 & $>$3.3  \\
\enddata
\tablenotetext{a}{In Table 2.}
\end{deluxetable}
\end{center}
 

\begin{center}
\begin{deluxetable}{lcc}
\tablecolumns{3}
\tablecaption{{\tt clumpfind} parameter summary}
\tablewidth{6.5cm}
\tablehead{
\colhead{IRDC} &
\colhead{Lower $\tau$} &
\colhead{$\Delta \tau$} \\ 
\colhead{} &
\colhead{Threshold} &
\colhead{ } 
}
\startdata
G005.85$-$0.23 & 0.27 & 0.20 \\
G006.26$-$0.51 & 0.27 & 0.11 \\
G009.16$+$0.06 & 0.19 & 0.10 \\
G009.28$-$0.15 & 0.35 & 0.06 \\
G009.86$-$0.04 &0.32 & 0.11 \\
G012.50$-$0.22 & 0.31 & 0.16 \\
G023.37$-$0.29 & 0.36 & 0.11 \\
G023.48$-$0.53 & 0.39 & 0.09 \\
G024.05$-$0.22 & 0.22 & 0.09 \\
G034.74$-$0.12 & 0.27 & 0.07 \\
G037.44$+$0.14 & 0.29 & 0.17 \\
\enddata
\end{deluxetable}

\end{center}

\clearpage
\LongTables
\begin{center}
\begin{deluxetable}{llrrcccc}
\label{tab:allclumps}
\tabletypesize{\scriptsize}
\tablecolumns{8}
\tablecaption{CLUMPFIND original fg/bg estimation run}
\tablewidth{11cm}

\tablehead{
\colhead{IRDC} &
\colhead{} &
\colhead{$\Delta\alpha$} &
\colhead{$\Delta\delta$} &
\colhead{Clump Mass} &
\colhead{$\tau_{max}$} &
\colhead{Clump Size} &
\colhead{Notes\tablenotemark{a}}  \\
\colhead{} &
\colhead{} & 
\colhead{$('')$} &
\colhead{$('')$} &
\colhead{(M$_{\odot}$)} &
\colhead{} &
\colhead{(pc)} &
\colhead{}
}
\startdata
G05.85$-$0.23 & & & & & & & \\
 & C1 &   23 &  -23 &   348.5 & 0.14 & 0.11 & \\ 
 & C2 &  -22 &  -75 &   342.5 & 0.55 & 0.09 & \\ 
 & C3 &  -33 &  -50 &   320.7 & 0.40 & 0.09 & \\ 
 & C4 &  -60 &  -83 &   305.8 & 0.25 & 0.10 & \\ 
 & C5 &  -16 &  -54 &   299.0 & 0.75 & 0.08 & \\ 
 & C6 &   -6 &  -61 &   211.8 & 0.46 & 0.08 & \\ 
 & C7 &  -42 &  -81 &   178.4 & 0.29 & 0.08 & fg \\ 
 & C8 &  -29 &  -63 &   165.3 & 0.93 & 0.07 & \\ 
 & C9 &   78 &   18 &   108.3 & 0.17 & 0.08 & fg \\ 
 & C10 &  -21 &  -52 &    72.4 & 0.58 & 0.06 & \\ 
 & C11 &  -22 &  -61 &    64.9 & 1.09 & 0.05 & \\ 
 & C12 &  -86 &  -93 &    47.1 & 0.23 & 0.07 & 5 - CII \\ 
 & C13 &   15 &   44 &    32.9 & 0.17 & 0.06 & 12 - CII \\ 
 & C14 & -105 & -123 &     3.3 & 0.14 & 0.04 & \\ 
 & C15 & -100 &  -94 &     0.6 & 0.13 & 0.02 & \\ 
\hline

G06.26$-$0.51 & & & & & & & \\
 & C1 &  -71 &   97 &  2226.8 & 0.26 & 0.18 & \\ 
 & C2 & -179 &  -25 &   963.5 & 0.15 & 0.15 & 17 - CII \\ 
 & C3 &  -99 &   93 &   820.9 & 0.15 & 0.15 & \\ 
 & C4 &  -51 &  -45 &   683.1 & 1.58 & 0.11 & \\ 
 & C5 &   37 &  -82 &   641.2 & 0.21 & 0.14 & \\ 
 & C6 &  -91 &   78 &   331.7 & 0.21 & 0.11 & \\ 
 & C7 & -154 &  -64 &   255.4 & 0.17 & 0.11 & \\ 
 & C8 &   70 & -112 &   240.6 & 0.13 & 0.11 & \\ 
 & C9 &  -62 &  -64 &   233.8 & 0.55 & 0.10 & \\ 
 & C10 &  -59 &  -35 &   221.8 & 0.75 & 0.08 & \\ 
 & C11 & -186 &  -52 &   210.7 & 0.17 & 0.10 & \\ 
 & C12 &  -21 &  110 &   176.8 & 0.15 & 0.10 & \\ 
 & C13 &   66 & -117 &   175.5 & 0.13 & 0.11 & \\ 
 & C14 &   47 &  -75 &   169.1 & 0.20 & 0.09 & \\ 
 & C15 &   91 &  -80 &   161.1 & 0.36 & 0.10 & 1 - CI \\ 
 & C16 &  -67 &  -61 &   145.0 & 0.49 & 0.08 & \\ 
 & C17 &  -70 &  -19 &   144.6 & 0.45 & 0.09 & \\ 
 & C18 &  -58 &  -40 &   119.1 & 0.88 & 0.06 & \\ 
 & C19 &   70 &  -77 &   112.5 & 0.18 & 0.09 & \\ 
 & C20 &  -71 &   56 &   109.3 & 0.19 & 0.09 & \\ 
 & C21 &  -70 &  -30 &   107.5 & 0.52 & 0.07 & \\ 
 & C22 &  -75 &  -39 &   106.9 & 0.47 & 0.08 & \\ 
 & C23 & -173 &  -44 &   102.1 & 0.14 & 0.09 & \\ 
 & C24 &  106 &  -53 &    83.8 & 0.14 & 0.09 &13 - CII  \\ 
 & C25 & -207 &  -38 &    77.5 & 0.11 & 0.09 & \\ 
 & C26 &  -83 &  -86 &    77.0 & 0.13 & 0.09 & \\ 
 & C27 &   61 &  -72 &    71.8 & 0.19 & 0.08 & \\ 
 & C28 & -133 &   83 &    65.4 & 0.10 & 0.09 & \\ 
 & C29 & -105 &  -88 &    59.6 & 0.38 & 0.08 & \\ 
 & C30 &  -61 &  -84 &    56.3 & 0.14 & 0.08 & \\ 
 & C31 & -167 &  -62 &    56.0 & 0.14 & 0.08 & 14 - CII \\ 
 & C32 &  -90 & -173 &    52.7 & 0.48 & 0.07 & \\ 
 & C33 & -110 &  -99 &    49.4 & 0.20 & 0.07 & \\ 
 & C34 & -125 &   83 &    42.5 & 0.10 & 0.08 & \\ 
 & C35 & -135 &  -75 &    41.8 & 0.12 & 0.08 & \\ 
 & C36 & -110 &   70 &    37.7 & 0.12 & 0.07 & \\ 
 & C37 &   -5 &   99 &    36.9 & 0.13 & 0.07 & \\ 
 & C38 &  -27 &  117 &    36.6 & 0.15 & 0.07 & \\ 
 & C39 &   55 &  102 &    34.0 & 0.14 & 0.07 & 26 - CII\\ 
 & C40 &  103 &  -62 &    32.3 & 0.15 & 0.07 & \\ 
 & C41 & -102 &   86 &    29.9 & 0.18 & 0.06 & \\ 
 & C42 & -178 &  -60 &    29.7 & 0.16 & 0.07 & 14 - CII \\ 
 & C43 &   65 &  -75 &    19.8 & 0.16 & 0.06 & \\ 
 & C44 &  -25 &  121 &    18.7 & 0.13 & 0.06 & \\ 
 & C45 &  -91 &  -49 &    17.3 & 0.15 & 0.06 & \\ 
 & C46 &  -51 &  -84 &    15.6 & 0.11 & 0.06 & \\ 
 & C47 & -154 &  -12 &    14.4 & 0.13 & 0.06 & \\ 
 & C48 &   46 & -164 &    12.7 & 0.10 & 0.06 & \\ 
 & C49 &   -4 &   89 &    11.0 & 0.13 & 0.05 & \\ 
 & C50 & -129 &   80 &    10.0 & 0.13 & 0.05 & \\ 
 & C51 & -122 &  -86 &     9.5 & 0.12 & 0.05 & \\ 
 & C52 &   82 &  -62 &     8.8 & 0.13 & 0.05 & \\ 
 & C53 &   65 &  117 &     8.5 & 0.11 & 0.05 & \\ 
 & C54 &  103 &   -9 &     7.8 & 0.13 & 0.05 & \\ 
 & C55 & -175 &  -58 &     7.6 & 0.15 & 0.04 & \\ 
 & C56 & -159 &   -9 &     7.3 & 0.10 & 0.05 & \\ 
 & C57 & -106 &   72 &     5.5 & 0.11 & 0.05 & \\ 
 & C58 &  -71 & -177 &     4.0 & 0.18 & 0.04 & \\ 
 & C59 &   53 & -170 &     1.4 & 0.10 & 0.03 & \\ 
 & C60 & -131 &  -47 &     1.3 & 0.10 & 0.03 & \\ 
 & C61 &   77 &  111 &     1.2 & 0.10 & 0.03 & \\ 
 & C62 &  -77 &   -3 &     1.0 & 0.13 & 0.03 & \\ 
 & C63 & -102 &  -63 &     1.0 & 0.10 & 0.03 & \\ 
\hline

G09.16$+$0.06 & & & & & & & \\
 & C1 &  -63 &  -80 &  3732.1 & 0.52 & 0.19 & \\ 
 & C2 &  -25 &  -66 &   741.2 & 0.43 & 0.14 & \\ 
 & C3 &  -50 &  -82 &   359.0 & 0.76 & 0.10 & \\ 
 & C4 & -180 &  -87 &   207.6 & 0.17 & 0.12 & \\ 
 & C5 &  -36 &  -81 &   162.5 & 0.52 & 0.08 & \\ 
 & C6 &    1 &  -97 &   153.6 & 0.23 & 0.10 & 4 - CII \\ 
 & C7 & -213 &  -96 &   118.6 & 0.09 & 0.10 & \\ 
 & C8 & -106 &  -61 &    69.7 & 0.08 & 0.09 & \\ 
 & C9 &  -24 &  -83 &    68.3 & 0.42 & 0.07 & \\ 
 & C10 & -170 &   17 &    64.1 & 0.08 & 0.09 & \\ 
 & C11 &  -20 &  -87 &    57.8 & 0.32 & 0.07 & \\ 
 & C12 & -164 &  -10 &    54.2 & 0.11 & 0.09 & \\ 
 & C13 &  -11 & -111 &    51.9 & 0.22 & 0.07 & \\ 
 & C14 & -203 &   -1 &    38.0 & 0.11 & 0.08 & \\ 
 & C15 & -185 &   11 &    36.8 & 0.11 & 0.08 & \\ 
 & C16 & -191 &   10 &    33.2 & 0.11 & 0.08 & \\ 
 & C17 & -123 &  -75 &    26.0 & 0.12 & 0.07 & \\ 
 & C18 &   32 & -118 &    23.2 & 0.19 & 0.07 & \\ 
 & C19 &   -5 & -130 &    16.3 & 0.16 & 0.06 & \\ 
 & C20 & -194 &  -31 &    10.8 & 0.08 & 0.06 & \\ 
 & C21 &   -2 &  -82 &    10.7 & 0.19 & 0.05 & 4 - CII \\ 
 & C22 & -119 &  -38 &     4.4 & 0.08 & 0.05 & \\ 
 & C23 & -116 &  -45 &     4.0 & 0.09 & 0.04 & \\ 
 & C24 &   87 &  -92 &     2.9 & 0.11 & 0.04 & \\ 
 & C25 & -100 &  -40 &     1.7 & 0.08 & 0.04 & \\ 
 & C26 &  -69 &  -23 &     1.5 & 0.11 & 0.04 & \\ 
 & C27 & -104 &  -39 &     1.3 & 0.08 & 0.03 & \\ 

\hline

G09.28$-$0.15 & & & & & & & \\
 & C1 &  -77 &    1 &  1036.9 & 0.55 & 0.13 & fg \\ 
 & C2 &  -64 &  -21 &   636.1 & 0.77 & 0.10 & fg\\ 
 & C3 &  -59 &  -14 &   492.4 & 0.61 & 0.11 & \\ 
 & C4 &  -55 &  -43 &   343.3 & 0.71 & 0.10 & \\ 
 & C5 &  -81 &  -12 &   339.3 & 0.50 & 0.10 & \\ 
 & C6 &  -49 &  -56 &   283.5 & 0.49 & 0.09 & fg\\ 
 & C7 &  -50 &  -32 &   238.3 & 0.71 & 0.09 & \\ 
 & C8 &  -89 &   36 &   234.6 & 0.21 & 0.11 &  14 - CII \\ 
 & C9 & -127 &  -25 &   169.2 & 0.31 & 0.10 & \\ 
 & C10 &  -80 &   23 &   119.8 & 0.28 & 0.09 & \\ 
 & C11 &  -36 & -125 &   119.3 & 0.29 & 0.09 & fg\\ 
 & C12 &  -55 &  -98 &   118.8 & 0.23 & 0.09 & \\ 
 & C13 &  -54 &  -38 &   115.2 & 0.67 & 0.07 & \\ 
 & C14 &  -37 & -112 &   110.2 & 0.27 & 0.09 & fg \\ 
 & C15 &  -51 &  -70 &   106.8 & 0.40 & 0.08 & \\ 
 & C16 &  -43 &  -39 &    96.3 & 0.41 & 0.08 & \\ 
 & C17 &  -59 &  -76 &    87.0 & 0.24 & 0.08 & \\ 
 & C18 &  -34 &  -98 &    73.0 & 0.30 & 0.08 & \\ 
 & C19 &  -16 & -147 &    72.5 & 0.20 & 0.08 & \\ 
 & C20 & -139 &  -25 &    70.0 & 0.23 & 0.08 & fg\\ 
 & C21 &  -38 & -104 &    65.5 & 0.29 & 0.07 & fg\\ 
 & C22 &    8 &  -34 &    60.8 & 0.20 & 0.08 & \\ 
 & C23 & -101 &   -9 &    60.2 & 0.22 & 0.08 & \\ 
 & C24 &  -97 &  -20 &    59.0 & 0.21 & 0.07 & \\ 
 & C25 & -102 &  -23 &    58.2 & 0.19 & 0.08 & fg \\ 
 & C26 &  -96 &   27 &    51.5 & 0.27 & 0.07 & \\ 
 & C27 &  -34 & -137 &    48.4 & 0.20 & 0.08 & 10 - CII \\ 
 & C28 &  -37 &  -83 &    46.8 & 0.38 & 0.07 & \\ 
 & C29 &  -25 &  -65 &    45.8 & 0.23 & 0.07 & \\ 
 & C30 &  -57 &  -92 &    45.3 & 0.21 & 0.07 & \\ 
 & C31 &  -40 &  -90 &    44.2 & 0.38 & 0.06 & \\ 
 & C32 &  -80 &   27 &    43.6 & 0.29 & 0.07 & \\ 
 & C33 &    4 &  -26 &    41.5 & 0.19 & 0.07 & 2 - CI \\ 
 & C34 &  -76 &  -33 &    37.6 & 0.16 & 0.07 & \\ 
 & C35 &  -26 &  -45 &    31.7 & 0.20 & 0.07 & fg\\ 
 & C36 &  -41 &  -69 &    31.6 & 0.28 & 0.06 & \\ 
 & C37 &  -32 &  -74 &    30.9 & 0.23 & 0.06 & \\ 
 & C38 &  -36 &  -61 &    28.1 & 0.28 & 0.06 & \\ 
 & C39 & -102 &  -15 &    27.1 & 0.22 & 0.06 & \\ 
 & C40 &  -53 &  -80 &    26.2 & 0.26 & 0.06 & \\ 
 & C41 &  -49 &  -49 &    25.8 & 0.41 & 0.05 & \\ 
 & C42 &  -32 &  -55 &    19.1 & 0.22 & 0.06 & \\ 
 & C43 &  -51 &  -88 &    18.1 & 0.20 & 0.06 & \\ 
 & C44 &  -19 &  -54 &    13.6 & 0.25 & 0.05 & fg\\ 
 & C45 &  -20 &  -86 &    13.6 & 0.23 & 0.05 & \\ 
 & C46 &  -40 &  -16 &    10.7 & 0.13 & 0.05 & \\ 
 & C47 &  -76 & -134 &    10.5 & 0.20 & 0.05 & \\ 
 & C48 & -107 &  -50 &    10.0 & 0.14 & 0.05 & fg \\ 
 & C49 &  -17 & -161 &     9.6 & 0.16 & 0.05 & \\ 
 & C50 &   -4 & -146 &     6.7 & 0.16 & 0.05 & \\ 
 & C51 &  -12 &   31 &     4.6 & 0.13 & 0.04 & \\ 
 & C52 &  -58 &    7 &     4.5 & 0.11 & 0.04 & \\ 
 & C53 & -110 &  -39 &     4.2 & 0.15 & 0.04 & fg \\ 
 & C54 &  -12 &  -93 &     2.6 & 0.14 & 0.04 & \\ 
 & C55 & -139 &   12 &     2.6 & 0.16 & 0.04 & fg\\ 
 & C56 & -141 &   44 &     2.0 & 0.12 & 0.04 & fg\\ 
 & C57 &  -32 & -157 &     1.8 & 0.13 & 0.03 & \\ 
 & C58 & -113 &   45 &     1.8 & 0.14 & 0.03 & fg\\ 

\hline

G09.86$-$0.04 & & & & & & & \\
 & C1 &  -15 &  -68 &   299.0 & 0.42 & 0.08 & 2 - CI, 3 - CI\\ 
 & C2 &  -62 &  -82 &   185.7 & 0.16 & 0.08 & \\ 
 & C3 &  -37 &  -71 &   174.3 & 0.20 & 0.07 & fg\\ 
 & C4 & -101 &  -57 &   167.9 & 0.50 & 0.07 &fg \\ 
 & C5 &  -74 &  -44 &   156.9 & 1.57 & 0.06 & \\ 
 & C6 &  -92 &  -42 &   120.3 & 0.45 & 0.06 & \\ 
 & C7 &  -52 &  -58 &   117.4 & 0.27 & 0.06 & \\ 
 & C8 &  -42 &  -38 &   116.5 & 0.41 & 0.06 & 6 - EP \\ 
 & C9 &    9 &  -67 &   112.3 & 0.32 & 0.07 &fg \\ 
 & C10 &  -33 &  -44 &   109.9 & 0.42 & 0.06 & \\ 
 & C11 & -106 &  -35 &    92.4 & 0.52 & 0.05 & \\ 
 & C12 &  -85 &  -61 &    82.7 & 0.30 & 0.06 & \\ 
 & C13 & -126 &  -28 &    80.4 & 0.59 & 0.06 & \\ 
 & C14 &  -65 &  -39 &    80.0 & 0.77 & 0.05 & \\ 
 & C15 &  -24 &  -11 &    76.3 & 0.18 & 0.06 & \\ 
 & C16 & -114 & -106 &    70.6 & 0.18 & 0.06 & \\ 
 & C17 &  -72 &  -27 &    65.9 & 0.26 & 0.06 & \\ 
 & C18 & -115 &  -38 &    63.9 & 0.64 & 0.05 & \\ 
 & C19 &  -15 &  -40 &    56.1 & 0.17 & 0.06 & \\ 
 & C20 &  -26 &  -61 &    48.5 & 0.38 & 0.05 & 3 - CI \\ 
 & C21 &    7 &  -87 &    43.2 & 0.13 & 0.06 & \\ 
 & C22 &  -15 &  -56 &    40.8 & 0.22 & 0.05 &fg \\ 
 & C23 & -114 & -123 &    36.4 & 0.17 & 0.05 & \\ 
 & C24 & -139 &   -7 &    35.9 & 0.24 & 0.05 & 19 - CII \\ 
 & C25 & -137 & -121 &    27.7 & 0.25 & 0.05 &fg \\ 
 & C26 & -163 & -122 &    26.6 & 0.22 & 0.05 & \\ 
 & C27 & -121 &  -31 &    23.8 & 0.56 & 0.04 & \\ 
 & C28 &   27 &  -65 &    21.9 & 0.32 & 0.05 & fg\\ 
 & C29 &   -5 &  -39 &    19.1 & 0.15 & 0.04 & 5 - CI \\ 
 & C30 &  -52 &  -45 &    18.0 & 0.28 & 0.04 & \\ 
 & C31 &  -45 &  -52 &    17.8 & 0.26 & 0.04 & \\ 
 & C32 & -184 & -125 &    12.9 & 0.17 & 0.04 & \\ 
 & C33 & -160 & -131 &     6.5 & 0.13 & 0.04 & \\ 
 & C34 & -157 &   -1 &     4.7 & 0.12 & 0.03 & \\ 
 & C35 &   -3 & -116 &     0.7 & 0.12 & 0.02 & \\ 
 & C36 & -201 &   -7 &     0.4 & 0.11 & 0.02 & \\ 

\hline

G12.50$-$0.22 & & & & & & & \\
 & C1 &  -70 &  -30 &  1418.6 & 0.39 & 0.14 & \\ 
 & C2 &  -51 &  -61 &  1293.3 & 0.44 & 0.14 & 2 - CI \\ 
 & C3 &  -41 &  -59 &   838.6 & 0.62 & 0.12 & \\ 
 & C4 &  -50 &  -41 &   488.8 & 0.98 & 0.09 & 5 - EP, 12 - CII\\ 
 & C5 & -196 &   39 &   385.4 & 0.46 & 0.10 & \\ 
 & C6 & -104 &  -31 &   373.2 & 0.23 & 0.11 & fg \\ 
 & C7 &  -51 &  -50 &   333.5 & 0.95 & 0.08 & 2 - CI \\ 
 & C8 &  -78 &  -12 &   204.6 & 0.40 & 0.09 & 14 - CII\\ 
 & C9 &  -42 &  -43 &   179.0 & 1.16 & 0.07 & 5 - EP, 12 - CII \\ 
 & C10 & -190 &   49 &   165.1 & 0.32 & 0.08 & \\ 
 & C11 & -131 &  -47 &    99.9 & 0.26 & 0.08 & \\ 
 & C12 &   80 & -146 &    36.2 & 0.21 & 0.07 & fg\\ 
 & C13 &  -32 &  -43 &    31.1 & 0.25 & 0.06 & \\ 
 & C14 &  -32 & -112 &    21.9 & 0.22 & 0.06 & \\ 
 & C15 &  -35 & -101 &     3.6 & 0.20 & 0.04 & \\ 

\hline

G23.37$-$0.29 & & & & & & & \\
 & C1 &   45 &   43 &  5199.6 & 1.64 & 0.18 & \\ 
 & C2 &   45 &   37 &  3143.6 & 1.28 & 0.17 & 9 - EP \\ 
 & C3 &  -18 &  183 &   417.3 & 0.32 & 0.13 & \\ 
 & C4 &  -44 &  139 &   317.3 & 0.82 & 0.10 & \\ 
 & C5 &  -16 &  197 &   231.2 & 0.58 & 0.11 & \\ 
 & C6 &  -44 &  153 &   188.9 & 0.28 & 0.10 & \\ 
 & C7 &   87 &   24 &   188.2 & 0.21 & 0.11 & \\ 
 & C8 &   48 &   70 &   167.5 & 0.24 & 0.10 & \\ 
 & C9 &  -79 &  107 &   128.0 & 0.18 & 0.10 & \\ 
 & C10 &  192 &  209 &   127.5 & 1.19 & 0.09 & fg\\ 
 & C11 &  -50 &  167 &    94.0 & 0.26 & 0.09 & fg\\ 
 & C12 &  -18 &  191 &    81.4 & 0.31 & 0.08 & \\ 
 & C13 &   39 &  112 &    78.8 & 0.27 & 0.08 & \\ 
 & C14 &    7 &  -86 &    78.2 & 0.33 & 0.08 & \\ 
 & C15 &  -26 &  206 &    67.3 & 0.17 & 0.09 & \\ 
 & C16 &    0 &  151 &    63.4 & 0.20 & 0.09 & 39 - CII\\ 
 & C17 &  -36 &  194 &    61.0 & 0.18 & 0.08 & \\ 
 & C18 &  124 &  -13 &    30.8 & 0.14 & 0.07 & \\ 
 & C19 &  -39 &  126 &    29.3 & 0.21 & 0.07 & \\ 
 & C20 &  171 &  131 &    28.0 & 0.28 & 0.06 & \\ 
 & C21 &   10 &  104 &    27.7 & 0.24 & 0.07 & \\ 
 & C22 &  -52 &  186 &    24.6 & 0.28 & 0.06 & fg\\ 
 & C23 &   44 &  102 &    24.2 & 0.21 & 0.06 & \\ 
 & C24 &    5 &  -64 &    22.2 & 0.21 & 0.06 & fg\\ 
 & C25 &   48 &  122 &    21.1 & 0.22 & 0.06 & \\ 
 & C26 &  -76 &  118 &    20.3 & 0.17 & 0.06 & \\ 
 & C27 &  192 &  202 &    18.6 & 1.40 & 0.05 &fg \\ 
 & C28 &    0 &  -64 &    18.2 & 0.20 & 0.06 & \\ 
 & C29 &  -39 &  115 &    15.9 & 0.21 & 0.06 & \\ 
 & C30 &  -47 &  199 &    15.1 & 0.18 & 0.06 & \\ 
 & C31 &   74 &  -65 &    14.9 & 0.16 & 0.06 & \\ 
 & C32 &  -34 &  211 &    14.7 & 0.18 & 0.06 & \\ 
 & C33 &   49 &   98 &    13.9 & 0.25 & 0.05 & 9 - EP \\ 
 & C34 &   53 &   88 &    13.7 & 0.20 & 0.06 & \\ 
 & C35 &   15 &  114 &    12.4 & 0.18 & 0.06 & fg\\ 
 & C36 & -100 &  172 &    12.1 & 0.16 & 0.06 & \\ 
 & C37 &   26 &  114 &    12.0 & 0.15 & 0.06 &fg \\ 
 & C38 &  196 &  133 &    10.6 & 0.15 & 0.06 & \\ 
 & C39 &   36 &  -70 &     8.2 & 0.17 & 0.05 & \\ 
 & C40 &   71 &  -43 &     8.1 & 0.19 & 0.05 & \\ 
 & C41 &  -62 &  188 &     6.2 & 0.17 & 0.05 & \\ 
 & C42 &  -63 &  199 &     6.0 & 0.16 & 0.05 & \\ 
 & C43 &  164 &  140 &     5.4 & 0.16 & 0.05 & \\ 
 & C44 &   58 &   86 &     5.4 & 0.13 & 0.05 &fg \\ 
 & C45 &  -83 &  185 &     4.3 & 0.15 & 0.04 & \\ 
 & C46 & -109 &   70 &     4.2 & 0.14 & 0.04 & \\ 
 & C47 &   50 &  -58 &     4.0 & 0.15 & 0.04 & \\ 
 & C48 &  -29 &   38 &     3.8 & 0.18 & 0.04 &33 - CII \\ 
 & C49 &  -77 &   81 &     2.6 & 0.13 & 0.04 & fg\\ 
 & C50 &  -56 &  202 &     1.8 & 0.14 & 0.04 & \\ 

\hline

G23.48$-$0.53 & & & & & & & \\
 & C1 &   78 &   26 &  1274.2 & 0.82 & 0.13 & \\ 
 & C2 &   68 &   32 &   775.0 & 0.71 & 0.12 & \\ 
 & C3 &  104 &   29 &   436.3 & 0.33 & 0.12 & \\ 
 & C4 &   52 &   37 &   363.6 & 0.76 & 0.09 & \\ 
 & C5 &   43 &   33 &   249.7 & 0.91 & 0.08 & 4 - CI\\ 
 & C6 &   50 &   56 &   247.5 & 0.58 & 0.09 & \\ 
 & C7 &  -43 &  -45 &   206.2 & 0.41 & 0.09 & \\ 
 & C8 &   33 &   40 &   204.1 & 0.72 & 0.08 & 4 - CI \\ 
 & C9 &   12 &   42 &   202.3 & 0.58 & 0.09 & \\ 
 & C10 &   60 &   51 &   186.1 & 0.37 & 0.09 & \\ 
 & C11 &  -79 &  -72 &   182.4 & 0.31 & 0.09 &1 - CI, 2 - CI, 3 - CI \\ 
 & C12 &   19 &   59 &   178.0 & 0.45 & 0.09 & \\ 
 & C13 &  -62 &  -62 &   154.1 & 0.36 & 0.09 & \\ 
 & C14 &   91 &    0 &   142.7 & 0.20 & 0.09 & \\ 
 & C15 &  -45 &  -57 &   129.6 & 0.49 & 0.08 & \\ 
 & C16 &  -34 &  -30 &   129.0 & 0.31 & 0.09 & \\ 
 & C17 &   23 &   51 &   101.2 & 0.61 & 0.06 & \\ 
 & C18 &  -12 &  -24 &    96.7 & 0.28 & 0.08 & \\ 
 & C19 &   34 &   59 &    96.6 & 0.46 & 0.07 & \\ 
 & C20 &    8 &   75 &    91.3 & 0.28 & 0.08 & \\ 
 & C21 &   44 &   44 &    83.3 & 0.59 & 0.06 & 4 - CI \\ 
 & C22 &  -24 &  -29 &    79.2 & 0.26 & 0.08 & \\ 
 & C23 &  -54 &  -64 &    69.7 & 0.38 & 0.07 & \\ 
 & C24 &  116 &   30 &    64.0 & 0.19 & 0.08 & \\ 
 & C25 &   -1 &   57 &    47.5 & 0.26 & 0.07 & \\ 
 & C26 &   28 &   54 &    46.4 & 0.52 & 0.05 & \\ 
 & C27 &    4 &   37 &    40.4 & 0.46 & 0.06 & \\ 
 & C28 &  -91 &  101 &    39.0 & 0.23 & 0.07 & \\ 
 & C29 &   -9 &  -12 &    38.6 & 0.19 & 0.07 & \\ 
 & C30 &  105 &  -10 &    38.0 & 0.29 & 0.07 & \\ 
 & C31 &   36 &   40 &    27.1 & 0.72 & 0.04 & 4 - CI\\ 
 & C32 &   68 &   68 &    18.0 & 0.25 & 0.06 & \\ 
 & C33 &  123 &   42 &    14.5 & 0.17 & 0.05 & \\ 
 & C34 &   52 &   80 &    12.9 & 0.17 & 0.05 & \\ 
 & C35 &   59 &   70 &     9.6 & 0.21 & 0.05 & fg\\ 
 & C36 &  -59 &  -44 &     5.0 & 0.16 & 0.04 & fg\\ 
 & C37 &   66 &    3 &     4.5 & 0.17 & 0.04 & \\ 
 & C38 &  -21 &   89 &     2.4 & 0.17 & 0.03 & \\ 

\hline

G24.05$-$0.22 & & & & & & & \\
 & C1 &   44 &   55 &   598.9 & 0.70 & 0.12 & \\ 
 & C2 &   32 &   37 &   411.5 & 0.58 & 0.11 &7 - CII \\ 
 & C3 &   26 &   56 &   363.0 & 0.37 & 0.12 & \\ 
 & C4 &   31 &   83 &   343.8 & 0.29 & 0.13 &fg \\ 
 & C5 &   44 &   20 &   277.8 & 0.28 & 0.12 & \\ 
 & C6 &   40 &   84 &   217.8 & 0.21 & 0.12 & fg\\ 
 & C7 &   36 &  183 &   192.1 & 0.10 & 0.13 & 13 - CII\\ 
 & C8 &   38 &   46 &   184.5 & 1.10 & 0.08 & 10 - CII\\ 
 & C9 &   47 &  128 &   181.7 & 0.14 & 0.12 & \\ 
 & C10 &   40 &  165 &   179.6 & 0.14 & 0.12 & \\ 
 & C11 &  173 &  181 &   131.7 & 0.17 & 0.11 & \\ 
 & C12 &   52 &   42 &   122.2 & 0.36 & 0.09 & \\ 
 & C13 &   56 &  119 &   118.1 & 0.19 & 0.10 & \\ 
 & C14 &   45 &   73 &    99.4 & 0.23 & 0.09 & fg\\ 
 & C15 &   69 &  124 &    71.9 & 0.16 & 0.09 & \\ 
 & C16 &   28 &  160 &    70.7 & 0.11 & 0.10 & \\ 
 & C17 &   66 &  113 &    61.5 & 0.20 & 0.09 & \\ 
 & C18 &   44 &   45 &    54.0 & 0.44 & 0.06 & \\ 
 & C19 &  144 &  205 &    53.6 & 0.11 & 0.09 & \\ 
 & C20 &   59 &   37 &    47.4 & 0.23 & 0.08 & \\ 
 & C21 &  130 &  208 &    45.2 & 0.13 & 0.09 &fg \\ 
 & C22 &  152 &  173 &    33.5 & 0.12 & 0.08 & \\ 
 & C23 &  136 &  177 &    27.6 & 0.11 & 0.08 & \\ 
 & C24 &   28 &   45 &    27.1 & 0.31 & 0.06 & \\ 
 & C25 &  -90 &  173 &    23.9 & 0.19 & 0.07 & fg \\ 
 & C26 &  128 &  186 &    19.3 & 0.11 & 0.07 & fg\\ 
 & C27 &  184 &  172 &    15.6 & 0.11 & 0.07 & \\ 
 & C28 &   24 &  127 &    15.1 & 0.13 & 0.07 &8 - CII \\ 
 & C29 &   71 &   59 &    14.6 & 0.13 & 0.07 & 8 - CII\\ 
 & C30 &  191 &  179 &    14.1 & 0.10 & 0.07 & \\ 
 & C31 &   68 &   32 &    12.8 & 0.19 & 0.06 & \\ 
 & C32 &  148 &  175 &     8.9 & 0.10 & 0.06 & \\ 
 & C33 &  -83 &  -24 &     7.8 & 0.14 & 0.05 & \\ 
 & C34 &  -94 &  173 &     4.8 & 0.19 & 0.05 & fg \\ 
 & C35 & -100 &  167 &     4.4 & 0.14 & 0.05 & fg \\ 
 & C36 &   76 &   37 &     3.8 & 0.14 & 0.05 & \\ 
 & C37 & -106 &  176 &     3.7 & 0.11 & 0.05 & fg\\ 
 & C38 &  -75 &  -42 &     2.0 & 0.10 & 0.04 & \\ 
 & C39 &    7 &  184 &     1.8 & 0.09 & 0.04 & \\ 

\hline

G34.74$-$0.12 & & & & & & & \\
 & C1 &  -62 &  -26 &  1411.1 & 0.66 & 0.14 & \\ 
 & C2 &  -59 &  -49 &   353.2 & 0.42 & 0.11 & \\ 
 & C3 &  -58 &  -39 &   309.7 & 0.42 & 0.10 & \\ 
 & C4 &   -5 &  -83 &   282.1 & 0.19 & 0.13 & 14 - CII\\ 
 & C5 &  -23 &  -60 &   223.0 & 0.49 & 0.10 & \\ 
 & C6 &  -11 &  -62 &   185.3 & 0.32 & 0.10 &14 - CII \\ 
 & C7 &    0 & -108 &   157.0 & 0.09 & 0.12 & \\ 
 & C8 &  -36 &  -54 &   150.8 & 0.35 & 0.10 & 35 - TD\\ 
 & C9 &  -39 &  -40 &   147.7 & 0.50 & 0.08 & \\ 
 & C10 &  -19 &  -40 &   146.8 & 0.35 & 0.09 & \\ 
 & C11 &  -26 &  -76 &   144.9 & 0.26 & 0.10 & \\ 
 & C12 &  -59 &  -10 &   134.7 & 0.22 & 0.10 & \\ 
 & C13 &  -37 &  -17 &   113.7 & 0.22 & 0.10 & \\ 
 & C14 &  -47 &  -26 &    92.3 & 0.29 & 0.08 & \\ 
 & C15 &  -89 &  -24 &    89.8 & 0.14 & 0.10 & fg\\ 
 & C16 &  -39 &  -83 &    86.6 & 0.12 & 0.10 & \\ 
 & C17 &  -44 &   -5 &    83.4 & 0.21 & 0.09 & fg\\ 
 & C18 &  -32 &  -39 &    74.7 & 0.29 & 0.08 & \\ 
 & C19 &  -42 &  -28 &    69.4 & 0.27 & 0.08 & \\ 
 & C20 &    7 &  -59 &    63.1 & 0.16 & 0.09 & \\ 
 & C21 &   -6 &  -38 &    57.6 & 0.42 & 0.07 & 3 - CI \\ 
 & C22 &  -15 &  -89 &    54.6 & 0.15 & 0.09 & fg\\ 
 & C23 & -122 &    0 &    54.4 & 0.19 & 0.09 & 18 - CII \\ 
 & C24 &  -26 &  -48 &    54.3 & 0.34 & 0.07 & \\ 
 & C25 & -108 &  -15 &    52.5 & 0.17 & 0.09 & \\ 
 & C26 &  -48 &  -15 &    50.2 & 0.22 & 0.07 & \\ 
 & C27 &  -28 & -106 &    46.2 & 0.11 & 0.09 & fg\\ 
 & C28 &   -8 &  -43 &    42.6 & 0.38 & 0.06 & 3 - CI, 16 - CII\\ 
 & C29 &   -5 &  -55 &    41.7 & 0.26 & 0.07 & \\ 
 & C30 &  -36 &  -44 &    41.0 & 0.50 & 0.06 & \\ 
 & C31 &  -40 &    5 &    40.5 & 0.12 & 0.08 & \\ 
 & C32 &   -8 &  -33 &    34.3 & 0.22 & 0.07 & \\ 
 & C33 &  -28 &  -91 &    34.0 & 0.15 & 0.08 & \\ 
 & C34 &  -14 &  -49 &    32.8 & 0.30 & 0.06 & 16 - CII\\ 
 & C35 &   18 &  -73 &    30.5 & 0.09 & 0.08 & \\ 
 & C36 &  -81 & -114 &    29.8 & 0.13 & 0.08 & \\ 
 & C37 & -141 &   19 &    28.6 & 0.17 & 0.07 & 22 - CII \\ 
 & C38 &  -29 & -113 &    28.3 & 0.09 & 0.08 & \\ 
 & C39 &    1 &  -53 &    28.0 & 0.20 & 0.07 & \\ 
 & C40 &   12 & -103 &    25.7 & 0.14 & 0.07 & \\ 
 & C41 &  -12 &  -83 &    25.2 & 0.14 & 0.07 & \\ 
 & C42 &  -84 & -106 &    23.5 & 0.16 & 0.07 &fg \\ 
 & C43 &  -42 &   63 &    23.4 & 0.15 & 0.07 &fg \\ 
 & C44 & -127 &  -60 &    18.1 & 0.16 & 0.07 & \\ 
 & C45 &  -56 &   10 &    16.7 & 0.13 & 0.07 & \\ 
 & C46 &  -92 &  -16 &    15.4 & 0.12 & 0.06 & fg \\ 
 & C47 &  -60 &  -62 &    13.8 & 0.14 & 0.06 & \\ 
 & C48 &  -47 &  -94 &    13.0 & 0.11 & 0.06 & \\ 
 & C49 & -130 &    9 &    12.4 & 0.11 & 0.06 & \\ 
 & C50 &  -21 &  -86 &    11.4 & 0.13 & 0.06 & \\ 
 & C51 &    6 & -113 &     8.7 & 0.10 & 0.06 & \\ 
 & C52 &  -48 &   57 &     8.6 & 0.13 & 0.06 & \\ 
 & C53 &  -25 &  -92 &     5.8 & 0.14 & 0.05 & fg \\ 
 & C54 & -131 &  -81 &     2.8 & 0.09 & 0.04 & \\ 
 & C55 & -110 &   -5 &     2.1 & 0.09 & 0.04 & \\ 

\hline

G37.44$+$0.14 & & & & & & & \\
 & C1 &   52 &  159 &   352.7 & 0.19 & 0.10 & \\ 
 & C2 &   64 &   48 &   258.3 & 0.45 & 0.08 & 2 - CII\\ 
 & C3 &   77 &   43 &   218.6 & 0.40 & 0.08 & \\ 
 & C4 &  -86 &   36 &   203.7 & 0.30 & 0.08 & \\ 
 & C5 &   45 &  141 &   170.1 & 0.17 & 0.08 & \\ 
 & C6 &   70 &  196 &   130.4 & 0.17 & 0.08 & fg\\ 
 & C7 &   35 &  206 &   119.3 & 0.14 & 0.08 & \\ 
 & C8 &   61 &  211 &   117.1 & 0.29 & 0.07 &fg \\ 
 & C9 &   30 &   47 &   116.4 & 0.80 & 0.06 & \\ 
 & C10 & -101 &   23 &   110.0 & 0.35 & 0.07 & \\ 
 & C11 &   18 &  203 &    95.7 & 0.16 & 0.07 & \\ 
 & C12 &   24 &   58 &    89.8 & 0.92 & 0.06 & 40 - TD\\ 
 & C13 &   51 &   56 &    88.6 & 0.59 & 0.06 & \\ 
 & C14 &   47 &  205 &    68.6 & 0.23 & 0.06 & \\ 
 & C15 &   52 &  210 &    60.4 & 0.25 & 0.06 & \\ 
 & C16 & -111 &   31 &    53.4 & 0.81 & 0.05 & \\ 
 & C17 & -102 &   38 &    52.7 & 0.38 & 0.05 & \\ 
 & C18 &   51 &   72 &    44.8 & 0.59 & 0.05 & \\ 
 & C19 &   44 &   16 &    35.4 & 0.15 & 0.06 & \\ 
 & C20 &    5 &   16 &    34.6 & 0.32 & 0.05 & \\ 
 & C21 &   42 &   56 &    26.8 & 0.60 & 0.04 & \\ 
 & C22 & -114 &   40 &    25.5 & 0.98 & 0.04 & \\ 
 & C23 &   17 &   42 &    24.7 & 0.23 & 0.05 & \\ 
 & C24 &   41 &   77 &    22.5 & 0.64 & 0.04 & \\ 
 & C25 &   91 &   39 &    20.9 & 0.18 & 0.05 & \\ 
 & C26 &   33 &  131 &    17.8 & 0.13 & 0.05 & \\ 
 & C27 &   28 &   14 &    17.5 & 0.16 & 0.05 & \\ 
 & C28 &   97 &   33 &    16.2 & 0.42 & 0.04 & \\ 
 & C29 &   42 &   23 &    15.8 & 0.13 & 0.05 & 20 - CII \\ 
 & C30 &   13 &   48 &    15.3 & 0.13 & 0.05 & \\ 
 & C31 &   36 &   61 &    14.2 & 0.60 & 0.03 & \\ 
 & C32 &  105 &   34 &    13.3 & 0.32 & 0.04 & \\ 
 & C33 &   31 &   60 &    13.0 & 0.87 & 0.03 & \\ 
 & C34 &   34 &    3 &    12.6 & 0.14 & 0.04 & \\ 
 & C35 &   11 &   18 &    11.3 & 0.23 & 0.04 & \\ 
 & C36 &   37 &   55 &    10.3 & 0.55 & 0.03 & \\ 
 & C37 &    5 &   34 &    10.0 & 0.13 & 0.04 & \\ 
 & C38 &  112 &   33 &    10.0 & 0.28 & 0.04 & \\ 
 & C39 &   53 &   82 &     8.7 & 0.21 & 0.04 & \\ 
 & C40 &   33 &  105 &     7.2 & 0.13 & 0.04 & fg\\ 
 & C41 &   29 &   29 &     6.7 & 0.14 & 0.04 & \\ 
 & C42 &  -52 &   88 &     5.9 & 0.83 & 0.03 & \\ 
 & C43 &   37 &   20 &     5.0 & 0.15 & 0.03 & \\ 
 & C44 & -108 &  -19 &     4.4 & 0.12 & 0.03 &fg \\ 
 & C45 &   68 &   66 &     3.7 & 0.12 & 0.03 & \\ 
 & C46 & -112 &   37 &     2.4 & 0.96 & 0.02 & \\ 
 & C47 &   85 &   54 &     1.9 & 0.12 & 0.03 & \\ 
 & C48 &    8 &   37 &     1.8 & 0.13 & 0.03 & \\ 
 & C49 & -111 &  134 &     1.7 & 0.26 & 0.03 & \\ 
 & C50 &  -46 &   82 &     1.7 & 0.27 & 0.02 & \\ 
 & C51 &  -59 &   88 &     1.1 & 0.19 & 0.02 & \\ 
 & C52 &  110 &   39 &     1.0 & 0.30 & 0.02 & \\ 
 & C53 &  136 &   -6 &     0.9 & 0.15 & 0.02 & \\ 
 & C54 & -103 &  121 &     0.9 & 0.21 & 0.02 & fg \\ 
 & C55 &   90 &   16 &     0.5 & 0.11 & 0.02 & \\ 

\enddata
\tablenotetext{a}{Numbers indicate which stars from Table~2 are associated with a given clump.  The YSO type (CI = Class I; CII = Class II; EP = Embedded Protostar; TD = Transition Disk) is also listed. The "fg" denotation indicates that a foreground (or background) star in the field may contaminate the properties listed for that clump. }
\end{deluxetable}
\end{center}


\end{document}